\documentclass[twocolumn,article,amsmath,amssymb,floatfix,aps]{revtex4}
\usepackage{graphicx}
\usepackage{epsfig}
\usepackage{multirow}
\usepackage{color}
\usepackage{cancel}
\begin{document}

\title{A new approach for the wobbling motion in the even-odd isotopes $^{161,163,165,167}$Lu}

\author{A. A. Raduta$^{a), b)}$,  R. Poenaru $^{a), c)}$ and C. M. Raduta $^{a)}$ }

\affiliation{$^{a)}$ Department of Theoretical Physics, Institute of Physics and
  Nuclear Engineering, Bucharest, POBox MG6, Romania}

\affiliation{$^{b)}$Academy of Romanian Scientists, 54 Splaiul Independentei, Bucharest 050094, Romania}

\affiliation{$^{c)}$Doctoral School of Physics, Bucharest University, 405 Atomistilor Str., Bucharest-Magurele, Romania}

\begin{abstract}
 A new interpretation for the wobbling bands in the even-odd Lu isotopes is given within a particle-triaxial rotor semi-classical formalism. While in the previous papers the  bands TSD1, TSD2, TSD3 and TSD4 are viewed as the  ground, one, two and three  phonon wobbling bands, here the corresponding experimental results are described as the ground band with  spin equal to I=R+j, for R=0,2,4,...(TSD1), the  ground band with I=R+j and R=1,3,5,...(TSD2), the one phonon excitations of TSD2 (TSD3), with the odd proton moving in the orbit  $j=i_{13/2}$, and the ground band of I=R+j, with R=1,3,5,... and $j=h_{9/2}$ (TSD4). The moments of inertia (MoI) of the core for the first three bands are the same, and considered to be free parameters. Due to the core polarization effect caused by the particle-core coupling, the MoI's for TSD4 are different. The energies and the e.m. transitions are quantitatively well described. Also, the phase diagram of the odd system is drawn. In the parameter space, one indicates where the points associated with the fitted parameters are located, which is the region where the transversal wobbling mode might be possible, as well as where the wobbling motion is forbidden.
\end{abstract} 
\maketitle

\renewcommand{\theequation}{1.\arabic{equation}}
\setcounter{equation}{0}
\section{Introduction}

The wobbling motion consists in a precession of the total angular momentum of a triaxial system combined with an oscillation of its projection on the quantization axis around a steady position. Bohr and Mottelson described the wobbling motion within a triaxial rotor  model for high spin states, where the total angular momentum almost aligns to the principal axis with the largest moment of inertia \cite{BMott}. This pioneering paper was followed by a fully microscopic description due to Marshalek \cite{Marsh}. Since then a large volume of experimental and theoretical results has been accumulated
\cite{Odeg,Jens,Ikuko,Scho,Amro,Gorg,Ham,Matsu,Ham1,Jens1,Hage,Tana3,Oi,Bring,Hart,Cast,Alme,MikIans,Rad016,Rad017,Rad018}. Also, the concept of wobbling motion has been extended to even-odd nuclei. Experimentally, the excited wobbling states in triaxial strongly deformed (TSD) bands are known in several even-odd nuclei like $^{161,163,165,167}$Lu, $^{167}$Ta \cite{Bring,Hart}, $^{135}$Pr  
\cite{Tan017,Frau,Buda} and $^{187}$Au \cite{Sen}.

In a previous publication \cite{Rad017} we formulated a semi-classical formalism as to describe the main features of the wobbling motion for a particle-triaxial-rotor system, which was successfully applied to $^{163}$Lu. The odd particle is a proton in the $j=i_{13/2}$ orbital. Subsequently, the method was applied to $^{165,167}$Lu \cite{Rad018}. Therein, each state of the TSD1 band is determined by a time dependent variational principle  equation under the restriction of small amplitudes. The solution leads to a phonon operator which applied successively to the ground states with the spin I=R+j and R=0,2,4,..., gives rise to the so called TSD2 band. Applying twice the phonon operator on the TSD1 states, one obtains the TSD3 band. The states of the TSD4 have negative parity and are obtained by acting with three phonons, two of positive and one of negative parity, on the TSD1 states. The negative parity wobbling phonon corresponds to a $j=h_{9/2}$ proton coupled to a triaxial rotor. The core's moments of inertia are the same across all four TSD bands. The phonon operator increases the spin of a state by one unit. Also, the e.m. properties of the mentioned isotopes have been well described. The approach is consistent with the experimental result claiming that it provides evidence of multiple wobbling phonon  states \cite{Hage}.

By contradistinction, here the states  I of TSD2 band are obtained variationally by coupling a proton from the single particle orbital $i_{13/2}$ to the triaxial core with the angular momentum
R equal to 1,3,5,.... The  phonon operator defined for each I is then applied to the TSD2 states which results in generating the states of the TSD3 band. In the case of $^{163}$Lu there exists a fourth band called TSD4 of negative parity. This is the collection of the ground states of I=R+j with R=0,2,4,... and $j=h_{9/2}$. Note that TSD1 and TSD2 are formed from ground states of two sets of different angular momenta, but their energies accounts however for the phonon energies due to the so called zero point energy term. Such an effect is also involved in the structure of the wave function used to calculate the reduced e.m. transition probabilities. Indeed, this is expanded in the first order around the classical coordinates which make the energy function minimum. 
The moments of inertia characterizing the core of the first three TSD bands are the same, since they use an unique particle-core interaction. For the TSD4 band the moments of inertia are modified due to the specific particle-core coupling.
In order to characterize the agreement between our results and the corresponding data as well as the specific wobbling features of the considered bands, additional magnitudes are calculated: the alignment, the dynamic moment of inertia (MoI), the relative energies to a reference axial symmetric rotor.

The above mentioned project is achieved according to the following plan. In Section 2 we briefly present the main ingredients of the proposed formalism. The numerical analysis of our calculations
is given in Section 3. The nuclear phases are defined in the parameter space associated to the model Hamiltonian within Section 4, where also a short comment on the transversal wobbling is presented. In Section 5 one summarizes the main results of our calculations, and the final conclusions are drawn.
\renewcommand{\theequation}{2.\arabic{equation}}
\setcounter{equation}{0}
\section{The formalism}
\label{sec:level2}
We study an odd-mass system  consisting of an even-even core described by a triaxial rotor Hamiltonian $H_{rot}$ and a single j-shell proton moving in a quadrupole deformed mean-field:
\begin{equation}
\hat{H}_{sp}=\epsilon_j+\frac{V}{j(j+1)}\left[\cos\gamma(3j_3^2-{\bf j}^2)-\sqrt{3}\sin\gamma(j_1^2-j_2^2)\right].
\label{hassp}
\end{equation}
Here $\epsilon_j$ is the single particle energy and $\gamma$, the deviation from the axial symmetric picture.
In terms of the total angular momentum ${\bf I}(={\bf R}+{\bf j}) $, and the angular momentum carried by the odd particle, ${\bf j}$, the rotor Hamiltonian is written as:
\begin{equation}
\hat{H}_{rot}=\sum_{k=1,2,3}A_k(\hat{I}_k-\hat{j}_k)^2,
\end{equation}
where $A_k$ are half of the reciprocal moments of inertia associated to the principal axes of the inertia ellipsoid, i.e. $A_k=1/(2{\cal I}_k)$, and considered as free parameters. 

The eigenvalues of interest for $\hat{H}(=\hat{H}_{rot}+\hat{H}_{sp})$ are obtained through a time dependent variational principle equation.
Thus, the total Hamiltonian $\hat{H}$ is dequantized through  the time dependent variational principle:
\begin{equation}
\delta\int_{0}^{t}\langle \Psi_{IjM}|{\hat H}-i\frac{\partial}{\partial t'}|\Psi_{IjM}\rangle d t'=0,
\label{minact}
\end{equation}
with the trial function chosen as:
\begin{equation}
|\Psi_{Ij;M}\rangle ={\bf N}e^{z\hat{I}_-}e^{s\hat{j}_-}|IMI\rangle |jj\rangle ,
\end{equation} 
with $\hat{I}_-$ and $\hat{j}_-$ denoting the lowering operators for the intrinsic angular momenta ${\bf I}$ and ${\bf j}$ respectively, while ${\bf N}$ is  the normalization factor.
$|IMI\rangle $ and $|jj\rangle$ are extremal states for the operators ${\hat I}^2, {\hat I}_3$, and ${\hat j}^2, {\hat j}_3$, respectively. Obviously, in Eq.(\ref{minact}) the unit-system, where $\hbar = c =1$, has been used.
The efficiency of the semi-classical procedure was tested in Refs.\cite{Rad017,Rad018}, showing that the resulting energies agree very well with the exact eigenvalues of the model Hamiltonian.

The variables $z$ and $s$ are complex functions of time and play the role of classical phase space coordinates describing the motion of the core, and the odd particle, respectively:
\begin{equation}
z=\rho e^{i\varphi},\;\;s=fe^{i\psi}.
\end{equation}
Changing the variables $\rho$ and $f$ to  $ r$ and $t$, respectively:
\begin{equation}
r=\frac{2I}{1+\rho^2},\;\;0\le r\le 2I;\;\;
t=\frac{2j}{1+f^2},\;\; 0\le t\le 2j,
\end{equation}
the classical equations of motion acquire the canonical form:
\begin{eqnarray}
\frac{\partial {\cal H}}{\partial r}&=&\stackrel{\bullet}{\varphi},\;\;\frac{\partial {\cal H}}{\partial \varphi}=-\stackrel{\bullet}{r}, \nonumber\\ 
\frac{\partial {\cal H}}{\partial t}&=&\stackrel{\bullet}{\psi},\;\;\frac{\partial {\cal H}}{\partial \psi}=-\stackrel{\bullet}{t}, 
\label{eqmot}
\end{eqnarray}
where ${\cal H}$ denotes the average of $\hat{H}$ with the trial function $|\Psi_{IjM}\rangle$, and plays the role of the classical energy having the expression : 
\begin{eqnarray}
&&{\cal H}\equiv\langle \Psi_{IjM}|H|\Psi_{IjM}\rangle\nonumber\\
        &=&\frac{I}{2}(A_1+A_2)+A_3I^2\nonumber\\
&+&\frac{2I-1}{2I}r(2I-r)\left(A_1\cos^2\varphi+A_2\sin^2\varphi -A_3\right)\nonumber\\
        &+&\frac{j}{2}(A_1+A_2)+A_3j^2\nonumber\\
&+&\frac{2j-1}{2j}t(2j-t)\left(A_1\cos^2\psi+A_2\sin^2\psi -A_3\right)\nonumber\\
        &-&2\sqrt{r(2I-r)t(2j-t)}\left(A_1\cos\varphi\cos\psi+A_2\sin\varphi\sin\psi\right)\nonumber\\
        &+&A_3\left(r(2j-t)+t(2I-r)\right)-2A_3Ij+V\frac{2j-1}{j+1}\nonumber\\
        &\times&\left[\cos\gamma-\frac{t(2j-t)}{2j^2}
        \sqrt{3}\left(\sqrt{3}\cos\gamma +\sin\gamma\cos2\psi\right)\right].
\label{classen}
\end{eqnarray} 
 ${\cal H}$ is minimal (${\cal H}_{I,min}(j)$) in the point
$(\varphi,r;\psi,t)=(0,I;0,j)$, when $A_1<A_2<A_3$.  
Linearizing the equations of motion around the minimum point of ${\cal H}$, one obtains a harmonic motion for the system, with the frequency given by the equation:
\begin{equation}
\Omega^4+B\Omega^2+C=0,
\label{equOm}
\end{equation}
where the coefficients B and C have the expressions:
\begin{widetext}
\begin{eqnarray}
-B&=&\left[(2I-1)(A_3-A_1)+2jA_1\right]\left[(2I-1)(A_2-A_1)+2jA_1\right]+8A_2A_3Ij\nonumber\\
 &+&\left[(2j-1)(A_3-A_1)+2IA_1+V\frac{2j-1}{j(j+1)}\sqrt{3}(\sqrt{3}\cos\gamma+\sin\gamma)\right]\nonumber\\
 &\times&\left[(2j-1)(A_2-A_1)+2IA_1+V\frac{2j-1}{j(j+1)}2\sqrt{3}\sin\gamma\right],\\
C&=&\left\{\left[(2I-1)(A_3-A_1)+2jA_1\right]\left[(2j-1)(A_3-A_1)+2IA_1+V\frac{2j-1}{j(j+1)}\sqrt{3}(\sqrt{3}\cos\gamma+\sin\gamma)\right]-4IjA_3^2\right \}\nonumber\\
 &\times&\left\{\left[(2I-1)(A_2-A_1)+2jA_1\right]\left[(2j-1)(A_2-A_1)+2IA_1+V\frac{2j-1}{j(j+1)}2\sqrt{3}\sin\gamma\right]-4IjA_2^2\right\}.\nonumber\\
\label{BandC}
\end{eqnarray}
\end{widetext}
Under certain restrictions for MoI's, the dispersion equation (\ref{equOm}) admits two real and positive solutions. Hereafter, these will be denoted by $\Omega^{I}_1$ and $\Omega^{I}_{1'}$ for $j=i_{13/2}$,
and $\Omega_2$ and $\Omega_{2'}$ for $j=h_{9/2}$. These energies are ordered as: $\Omega^I_1<\Omega^I_{1'}$ and $\Omega^I_2<\Omega^I_{2'}$. Energies of the states in the four bands are defined as:
\begin{eqnarray}
E^{TSD1}_I&=&\epsilon_{13/2}+{\cal H}_{I,min}(13/2)+\frac{1}{2}\left(\Omega^I_1+\Omega^I_{1'}\right),\nonumber\\
&&I=13/2, 17/2, 21/2,.....\nonumber\\
E^{TSD2}_I&=&\epsilon_{13/2}+{\cal H}_{I,min}(13/2)+\frac{1}{2}\left(\Omega^I_1+\Omega^I_{1'}\right),\nonumber\\
&&I=27/2, 31/2, 35/2,.....\nonumber\\
E^{TSD3}_I&=&\epsilon_{13/2}+{\cal H}_{I-1,min}(13/2)+\frac{1}{2}\left(3\Omega^{I-1}_1+\Omega^{I-1}_{1'}\right),\nonumber\\
&&I=33/2, 37/2, 41/2,.....\nonumber\\
E^{TSD4}_I&=&\epsilon_{9/2}+{\cal H}_{I,min}(9/2)+\frac{1}{2}\left(\Omega^I_2+\Omega^I_{2'}\right),\nonumber\\
&&I=47/2, 51/2, 55/2,.....
\label{ener}
\end{eqnarray}
The excitation energies are obtained by subtracting $E^{TSD1}_{13/2}$ from the above expressions. 
\section{Numerical results}
\renewcommand{\theequation}{3.\arabic{equation}}
\setcounter{equation}{0}
\label{sec:level3}
\subsection{Parameters}
\begin{widetext}
\begin{table}[h!]
\begin{tabular}{|c|c|c|c|c|c|c|c|c|c|}
\hline
isotope&j& bands& ${\cal I}_1[\hbar^2/MeV]$& ${\cal I}_2[\hbar^2/MeV]$ &${\cal I}_3[\hbar^2/MeV]$& V[MeV] &$\gamma $ [degrees]&nr states&r.m.s.[MeV]\\
\hline
$^{161}$Lu&13/2&TSD1,TSD2     &87.555&2.773 &22.744 &2.933&20&29&0.168\\
$^{163}$Lu&13/2&TSD1,TSD2,TSD3&63.2  &  20  &  10  & 3.1 &  17&52&0.264\\
          &9/2&TSD4           &67    &  34.5 & 50  & 0.7&  17&10&0.057\\
$^{165}$Lu&13/2&TSD1,TSD2,TSD3&77.295&16.184&4.399&1.673&20&42&0.125\\
$^{167}$Lu&13/2&TSD1,TSD2&87.032&10.895&3.758&8.167&19.48&30&0.165\\
\hline
\end{tabular}
\caption{The MoI's, the strength of the single particle potential (V), and  the triaxial parameter ($\gamma$) as provided by the adopted fitting procedure.}
\label{Table 1}
\end{table}
\end{widetext}
From Eq.(\ref{ener}), one sees that the excitation energies depend on the MoI's, the strength of the particle-core interaction, V, and the triaxial shape parameter $\gamma$. These are considered as free parameters to be fixed by a fitting procedure.
Fitting the experimental data for the excitation energies, one obtains the parameters mentioned above, with the results  collected in Table I. Aiming at appraising the quality of the fit, we also mention the r.m.s. values of results deviation from the experimental excitation energies, and the number of states belonging to the bands under consideration. One may conclude that the energy description is fairly good. In order to evidence the MoI dependence on the isotope mass number A, data from Table I are visualized in Fig. 1, where a change of the MoI's ordering at A=163 is noticed, which may suggest a phase transition taking place at A=163. It is worth noting that the MoI's order  which is specific for $^{161}$Lu, is also valid for $^{163}$Lu, but only for the band TSD4. As for the other bands, the 
${\cal I}_2$ and ${\cal I}_3$ ordering specific for $^{161}$Lu is changed for the other nuclei. In fact this reflects  the critical point dependence on the excited band \cite{Rad05}. Note that for the  bands TSD1, TSD2 and TSD3, the excitation energies do not depend on the single particle energies. On the contrary,  the  excitation energies for the TSD4 states, in $^{163}$Lu, contain the constant term $\epsilon_{9/2}-\epsilon_{13/2}=-0.334$ MeV. 

The adopted option of fitting the MoI's is naturally required by the observation that the experiment indicates that they are neither irrotational nor rigid but satisfy the relation
${\cal I}_{1}^{irr}<{\cal I}_{1}<{\cal I}_{1}^{rig}$.
Note the MoI's provided by the fitting procedure take care of the particle-core interaction. In that respect the fact that the maximal MoI is ${\cal I}_{1}$ does not necessarily imply that the system motion is of longitudinal character. Indeed, even though for the non-interacting core the ordering is ${\cal I}_{2}>{\cal I}_{1}>{\cal I}_{3}$, as the microscopic studies show, the particle-core interaction renormalizes the bare MoI's which results in a strong increasing of ${\cal I}_{1}$ (due to the alignment) and only a moderate decreasing of ${\cal I}_{2}$ (caused by the pairing interaction), ending with the dominance of the new ${\cal I}_{1}$, characterizing the whole system \cite{Matsu}. Then, one can assert that the interaction with the odd proton stabilizes the system into a large deformed shape and moreover drives it to a longitudinal-like motion where the maximal MoI is the normalized ${\cal I}_{1}$. This change in the rotation regime is caused by both the angular momentum alignment and the pairing interaction. The transition from a transversal to a longitudinal wobbling motion is not abruptly achieved, but only at a certain critical angular momentum $I_{cr}$. Note that the MoI's were fixed such that the best agreement with the corresponding experimental data is obtained for energies of the whole spectrum and therefore no angular momentum dependence can be inferred. Furthermore, the study of the phase transition, transversal-longitudinal, cannot be performed with the present formalism. However, due to this feature one may say that the present formalism does not exclude the possible transversal wobbling motion in the low lying spectrum where the alignment is small, but that part of bands cannot be explored because the adopted fitting procedure does not use any I dependence for  MoI's.

\begin{figure}
\includegraphics[width=0.3\textwidth]{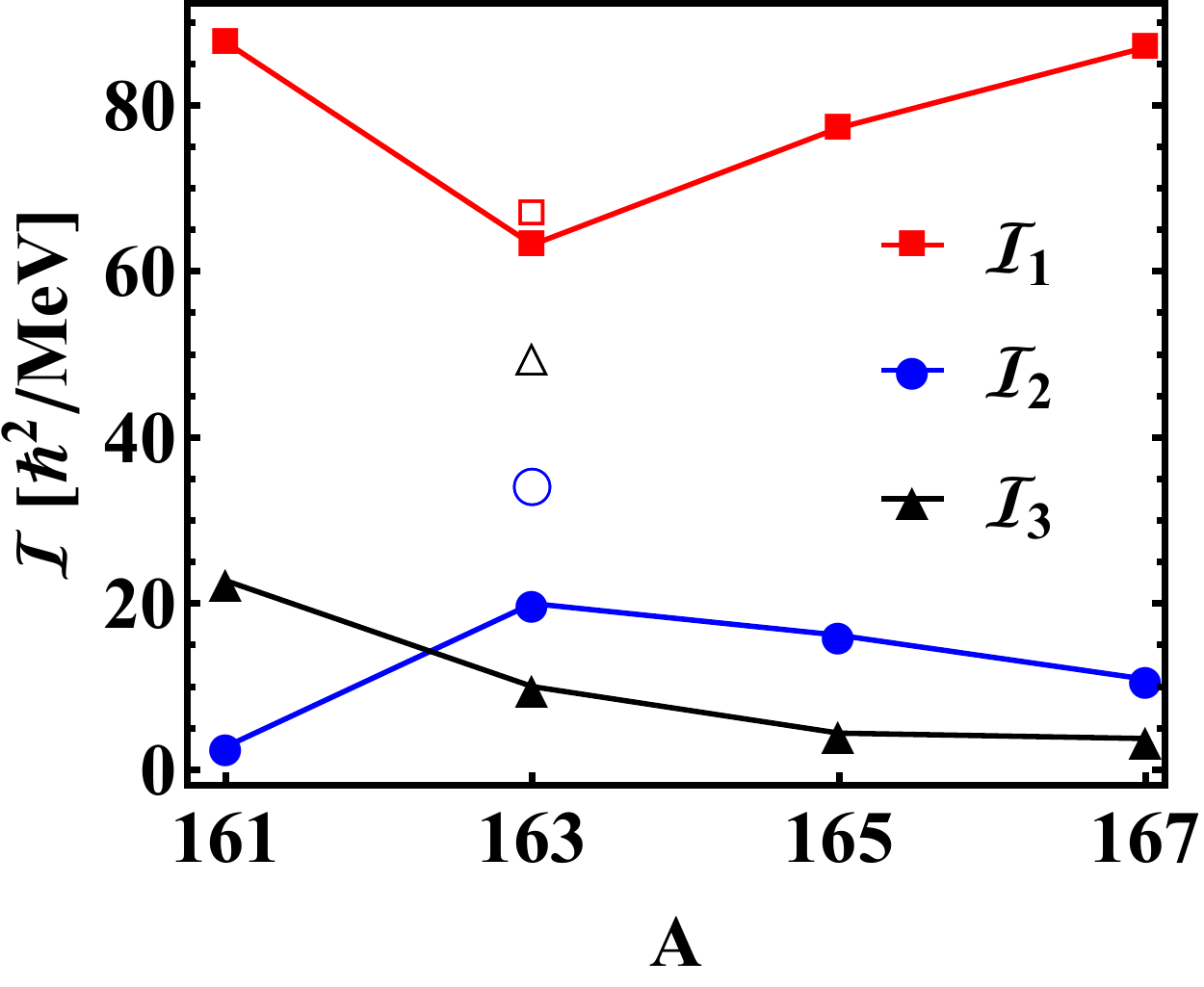}
\caption{(Color online) Results for the MoI's are plotted as function of A. In the case of $^{163}$Lu, one shows also the results for the band TSD4 using the same but open symbols.}
\label{Fig.1}
\end{figure}

\subsection{Energies}
Once the parameters involved in the model Hamiltonian are fixed, Eqs.(\ref{ener}) provide the excitation energies in the considered TSD bands.  
Inspecting Figs.2-5, where the calculated energies are compared with the corresponding experimental data, we may conclude that the energy description is fairly good.  At this stage we may ask ourselves how the energies provided by our approach compare with the exact eigenvalues of the model Hamiltonian. Such a comparison has been already performed in a previous paper \cite{Rad018}, where the exact energies were obtained by diagonalizing the model Hamiltonian within a particle-core basis. The result was that the two sets of energies agree with each other quite well.
 This confirms that the proposed formalism is  appropriate not only for simulating the data, but also  provides a good approximation for the exact results.
We notice that the least square fit predicts that the maximal moment of inertia corresponds to the one-axis, and therefore the system rotates around the short axis. Moreover, the odd proton angular momentum is oriented also along the short axis, and thereby the system motion is of longitudinal wobbling character. The numerical values of MoI's are consistent with the angular momenta orientation corresponding to the minimum point of ${\cal H}$. 
In Ref.\cite{Frau} one states that a signature for a transversal wobbling motion is the decreasing behavior of the wobbling energy with the spin, and moreover that the Lu isotopes would belong to such a category of wobblers. In this context using the standard definition, the wobbling energy  for the one phonon band, i.e.,the TSD3, was plotted in Fig. 6 as function of the angular momentum. As seen from there, the wobbling frequency is slightly increasing with spin, as predicted by our approach. Indeed, the experimental wobbling energy increases from 144 to 170 keV when the spin goes from $33/2$ to $77/2$, and finally decreases for the last two states, with spins 81/2 and 85/2, to 143 keV. On the other hand, the calculated wobbling energy increases faster with angular momentum, from 331 keV, at spin 
$33/2$, up to 570 keV, for spin $85/2$. The effect is more pronounced in $^{165}$Lu, where the increment in the experimental curve is about 100 keV. The agreement between the wobbling energy behavior given by our calculations, and the corresponding experimental data is to be considered as a specific feature of the present approach. Actually, this result is consistent with the microscopic study of Ref.\cite{Matsu}.

\begin{figure}[h!]
\includegraphics[width=0.25\textwidth]{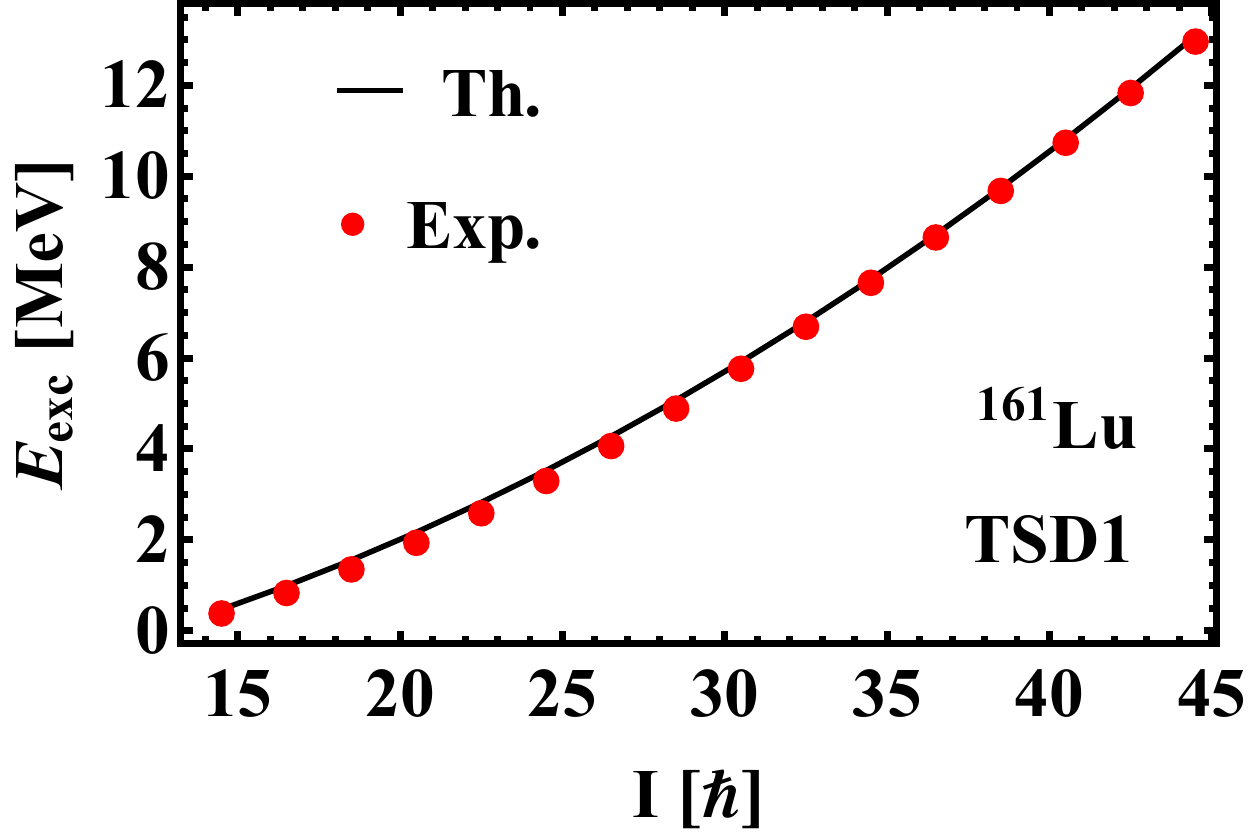}\includegraphics[width=0.25\textwidth]{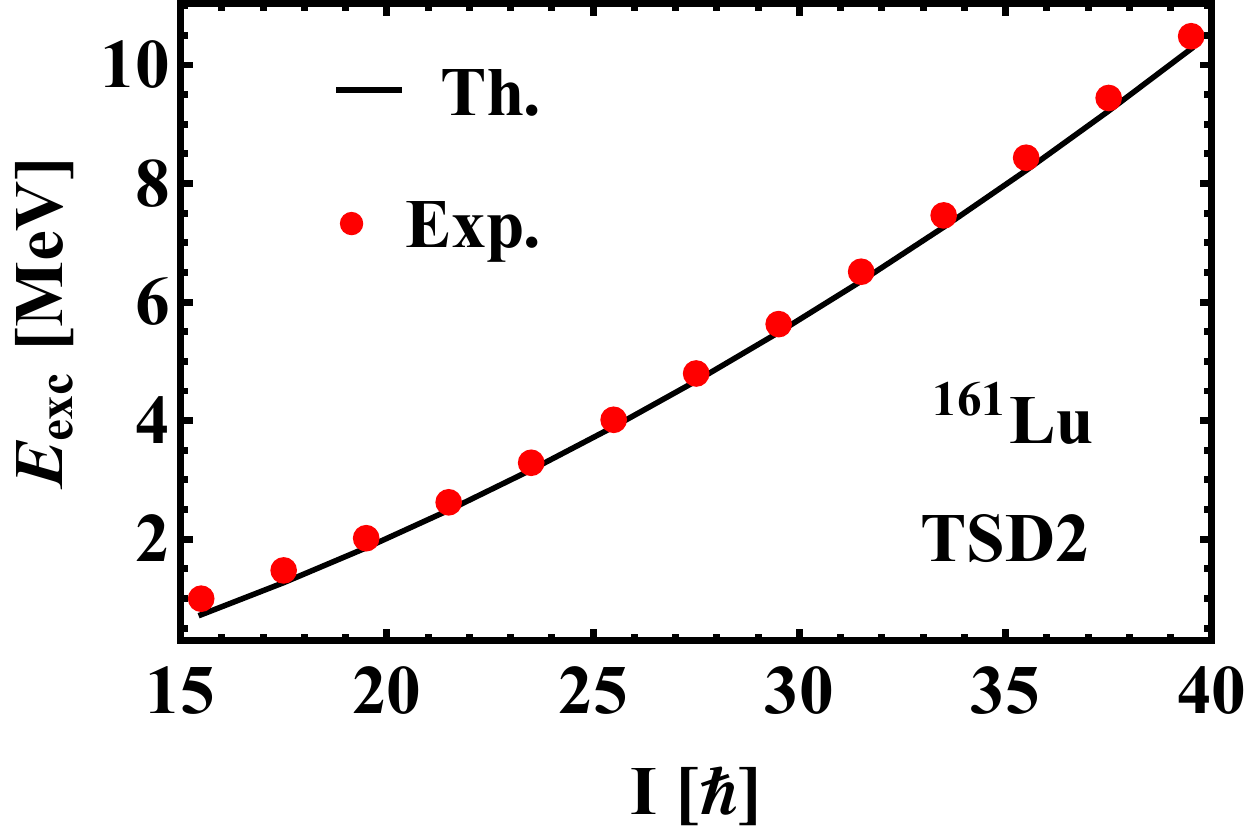}
\caption{(Color online)Calculated energies for the bands TSD1, TSD2 are compared with the corresponding experimental data \cite{Bring} for $^{161}$Lu.}
\label{Fig.2}
\end{figure}

\begin{figure}[h!]
\includegraphics[width=0.25\textwidth]{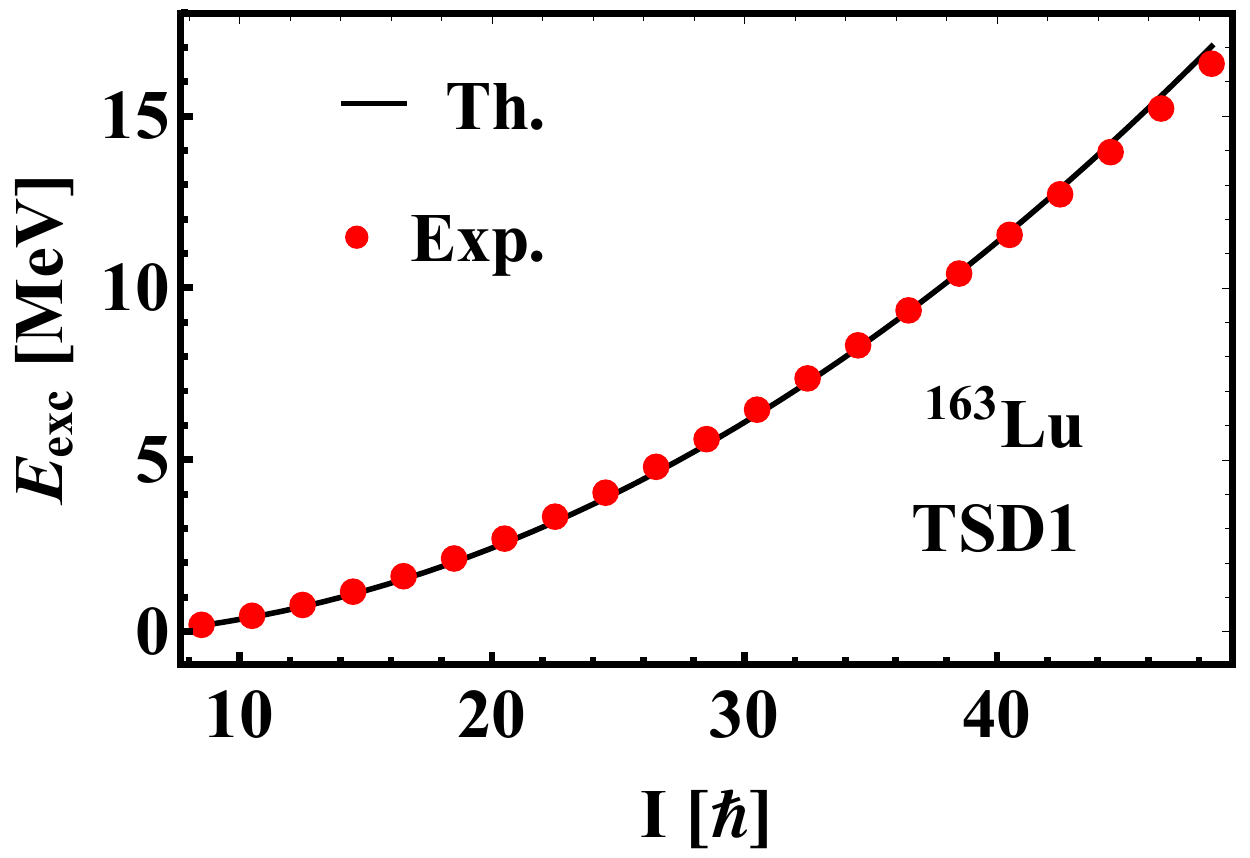}\includegraphics[width=0.25\textwidth]{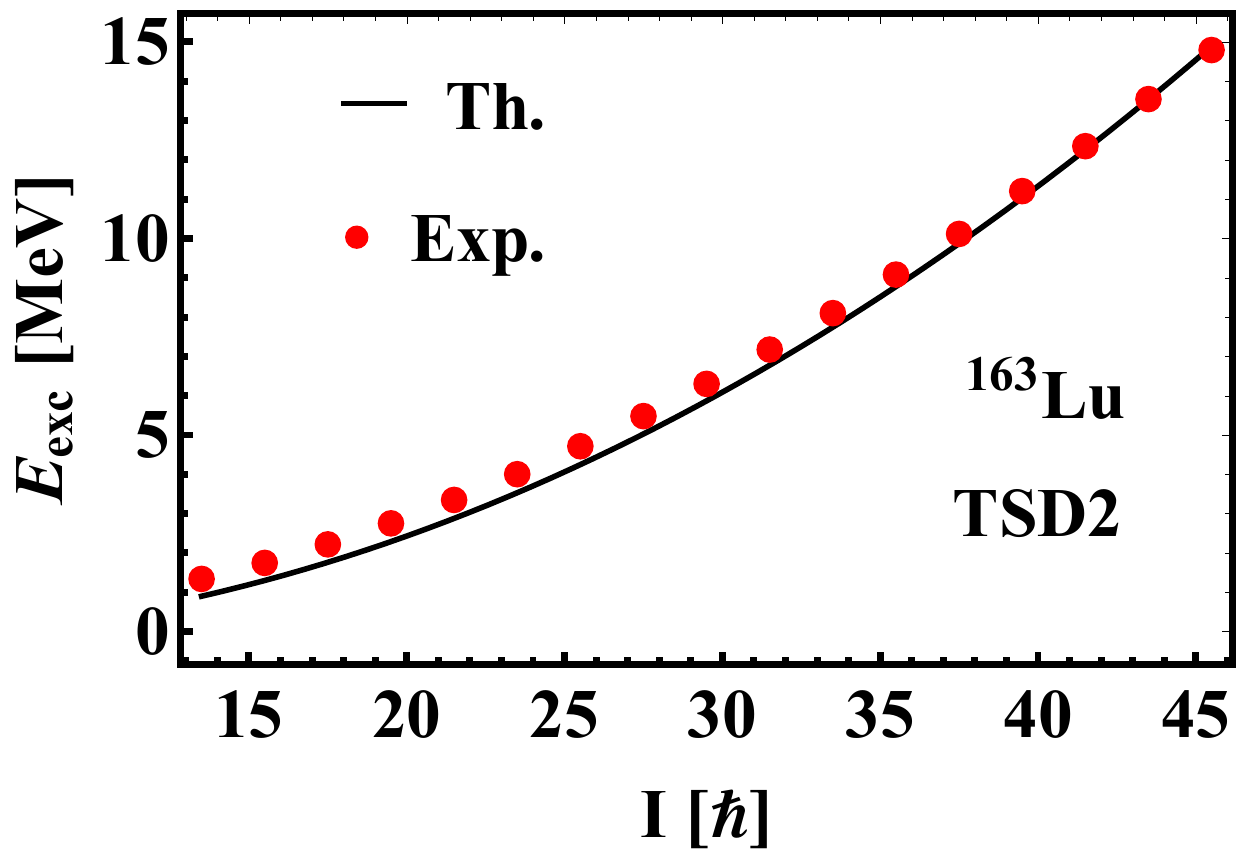}
\includegraphics[width=0.25\textwidth]{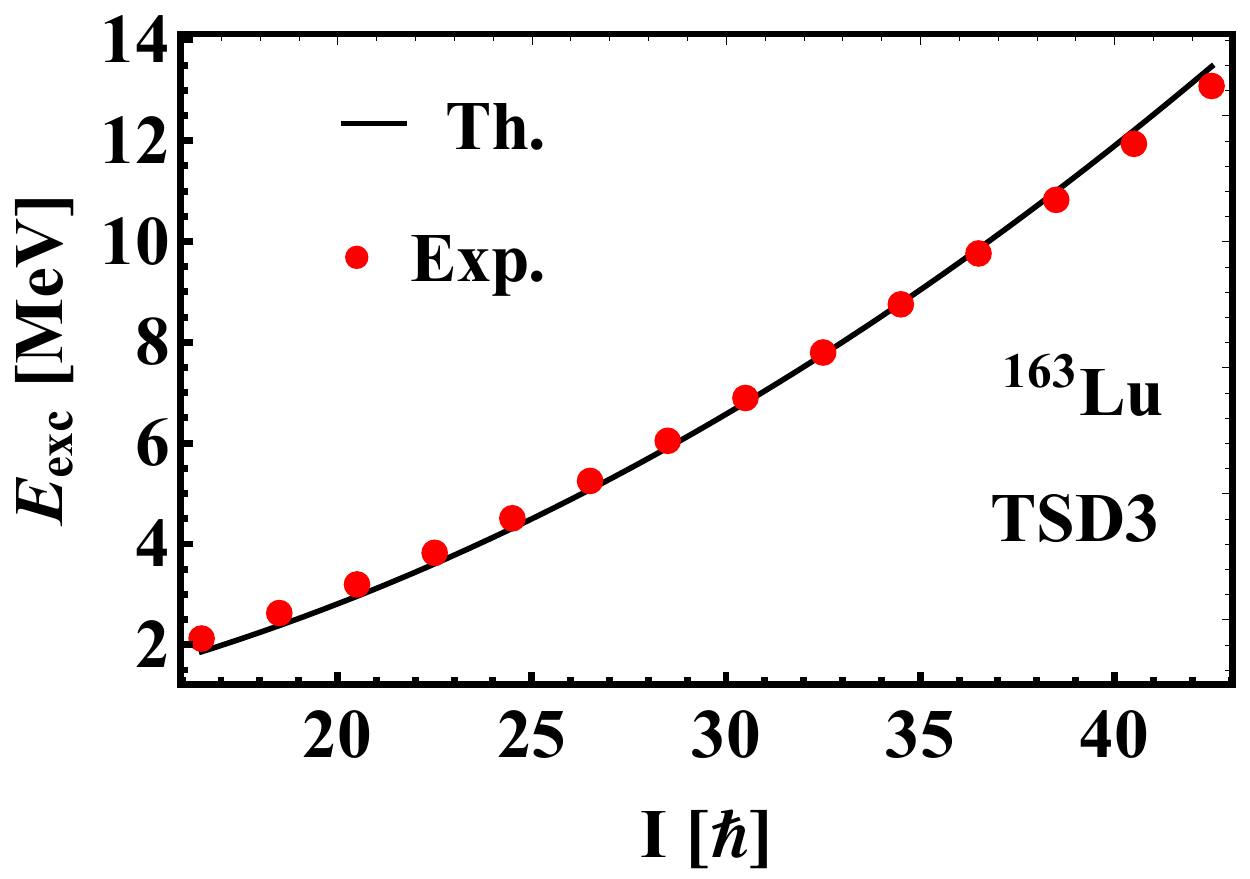}\includegraphics[width=0.25\textwidth]{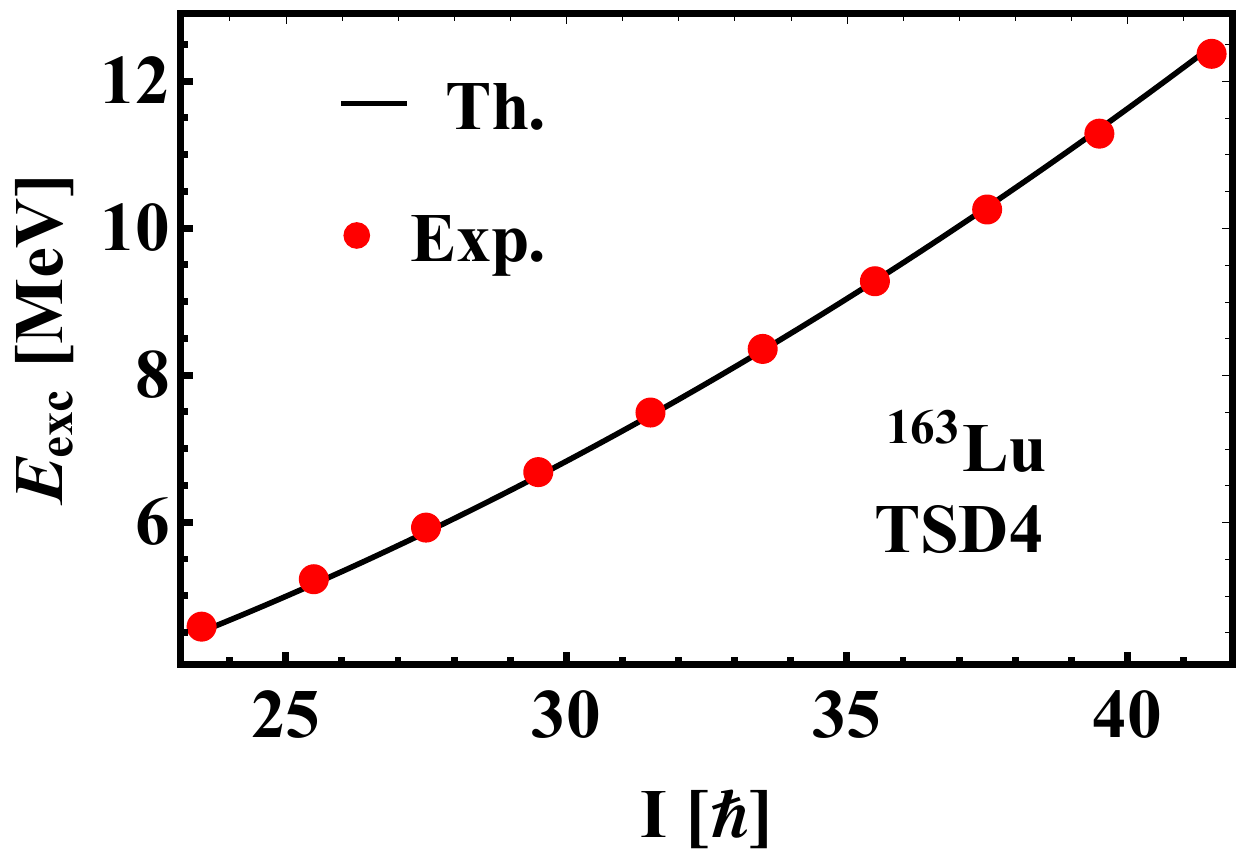}
\caption{(Color online)Calculated energies for the bands TSD1, TSD2, TSD3 and TSD4 are compared with the corresponding experimental data \cite{Jens1,Hage} for $^{163}$Lu.}
\label{Fig.3}
\end{figure}

\begin{figure}[h!]
\includegraphics[width=0.25\textwidth]{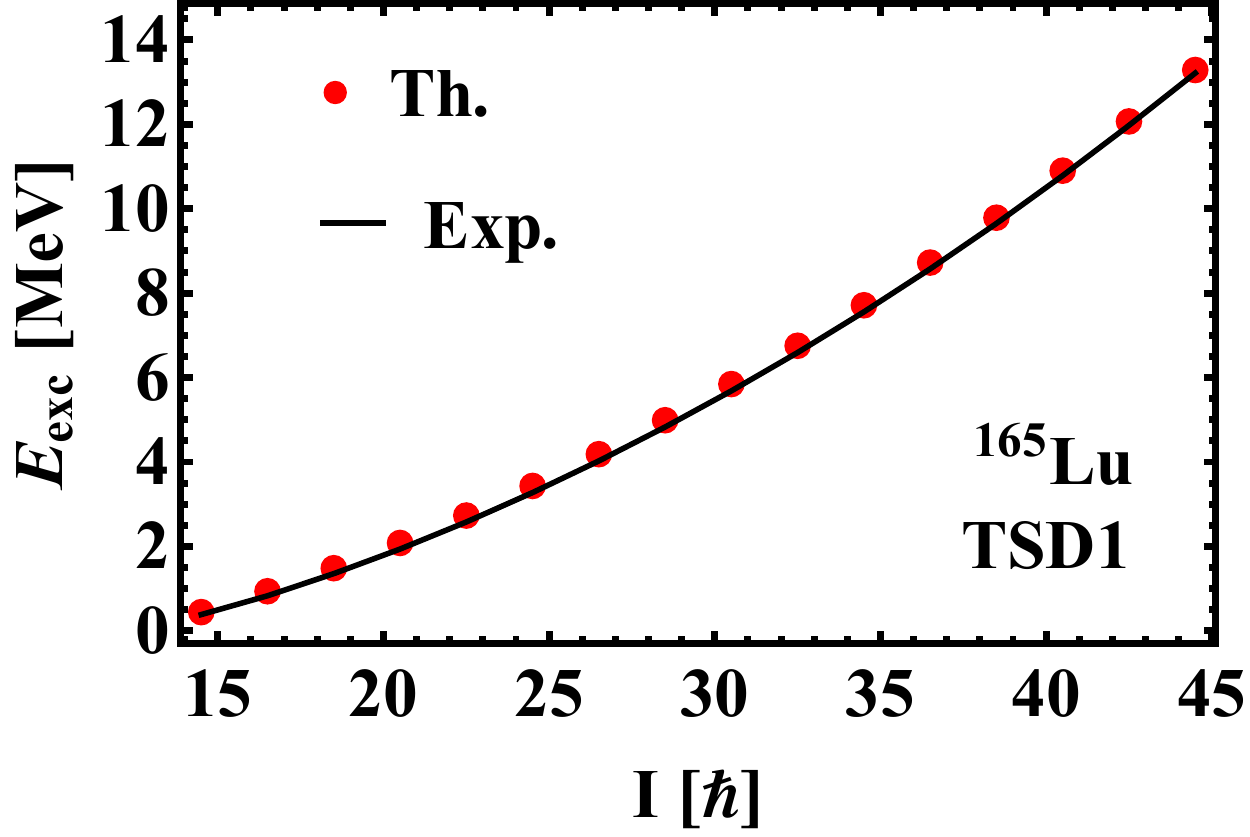}\includegraphics[width=0.25\textwidth]{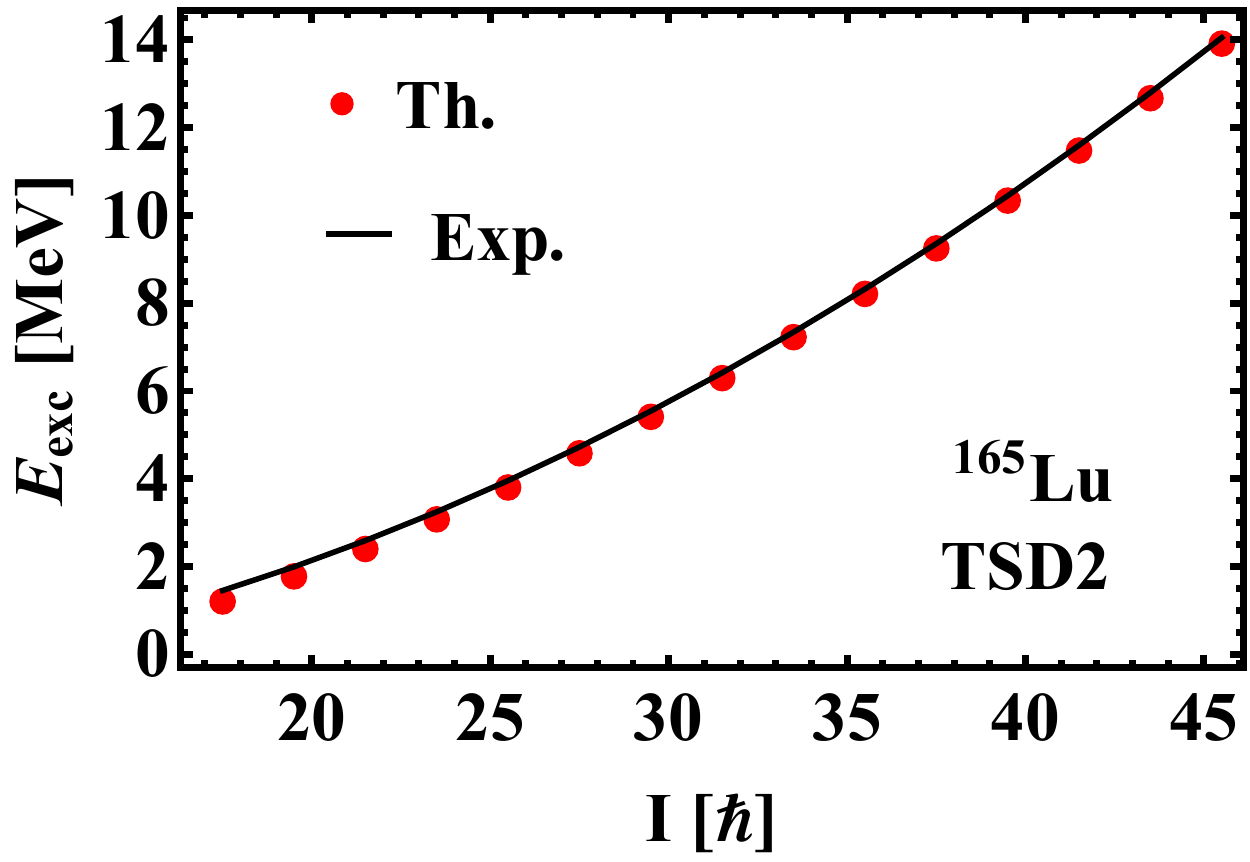}
\includegraphics[width=0.25\textwidth]{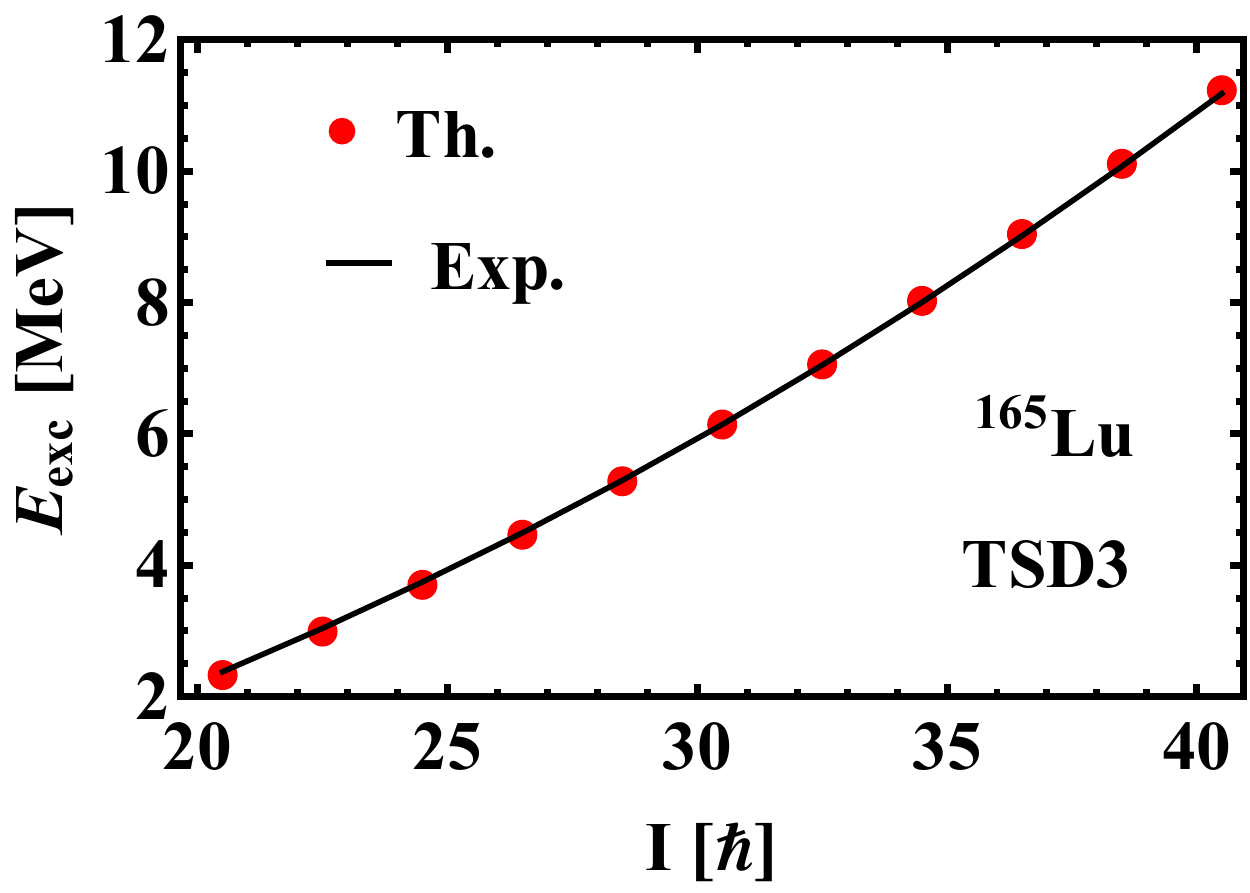}
\caption{(Color online)Calculated energies for the bands TSD1, TSD2 and TSD3 are compared with the corresponding experimental data \cite{Scho} for $^{165}$Lu.}
\label{Fig.4}
\end{figure}

\begin{figure}[h!]
\includegraphics[width=0.25\textwidth]{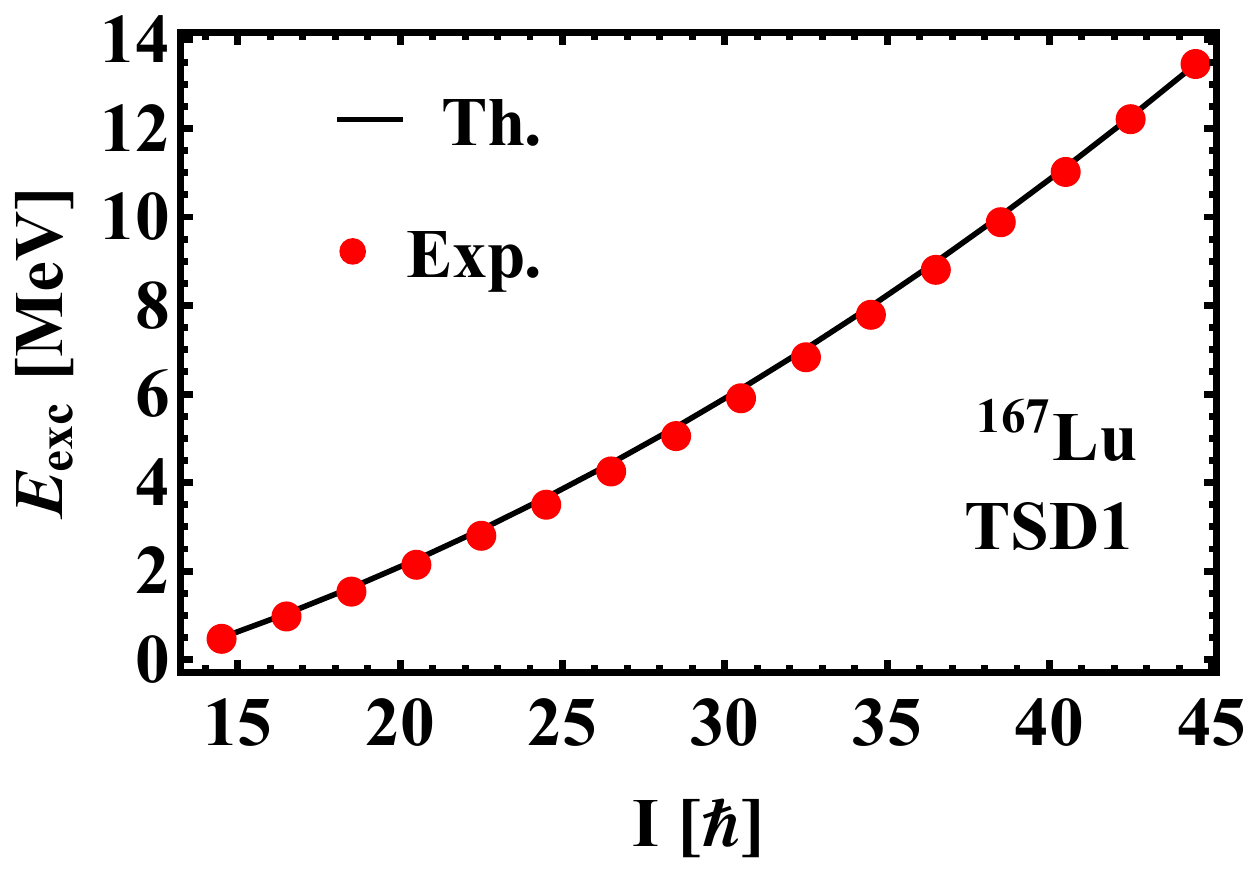}\includegraphics[width=0.25\textwidth]{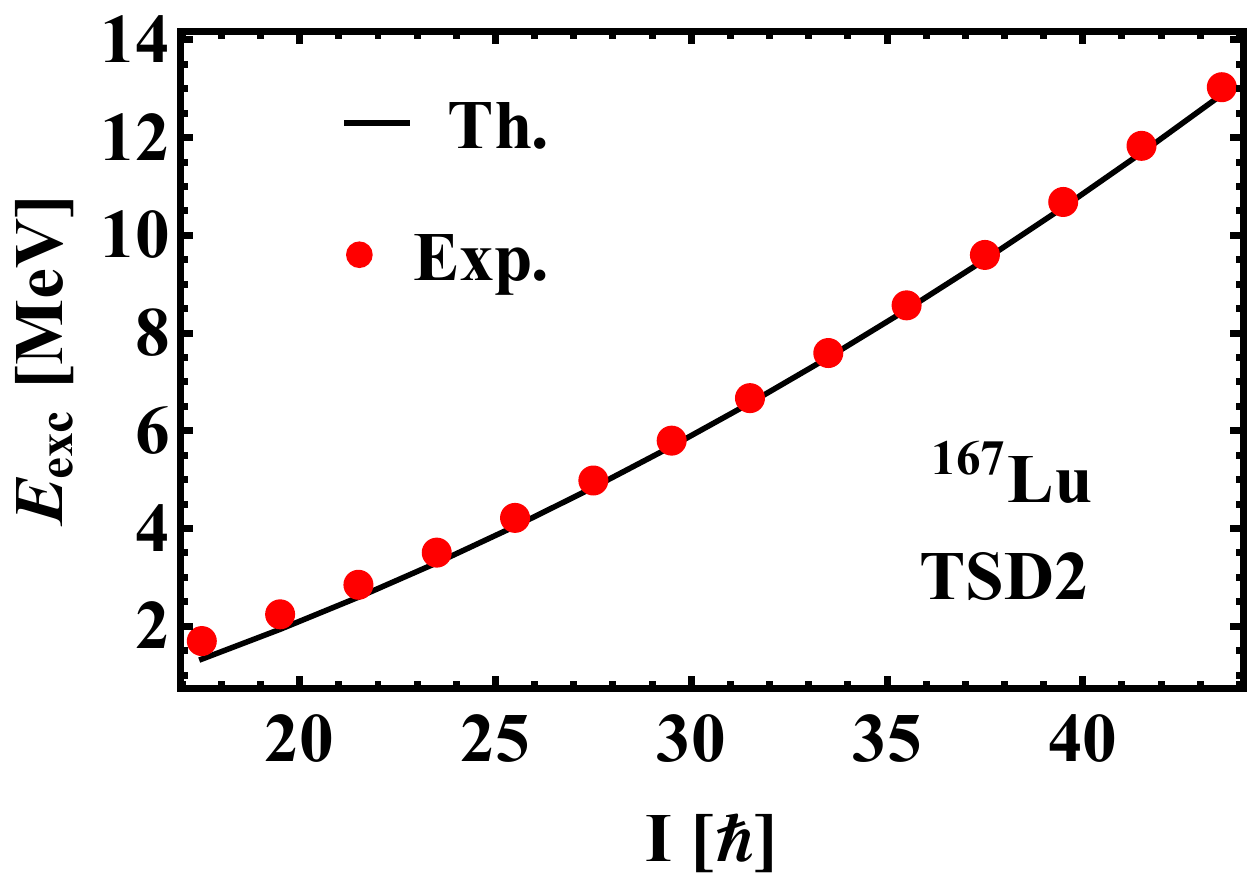}
\caption{(Color online)Calculated energies for the bands TSD1, TSD2 and TSD3 are compared with the corresponding experimental data\cite{Amro} for $^{167}$Lu.}
\label{Fig.5}
\end{figure}
\begin{figure}[ht!]
\includegraphics[width=0.25\textwidth,height=4cm]{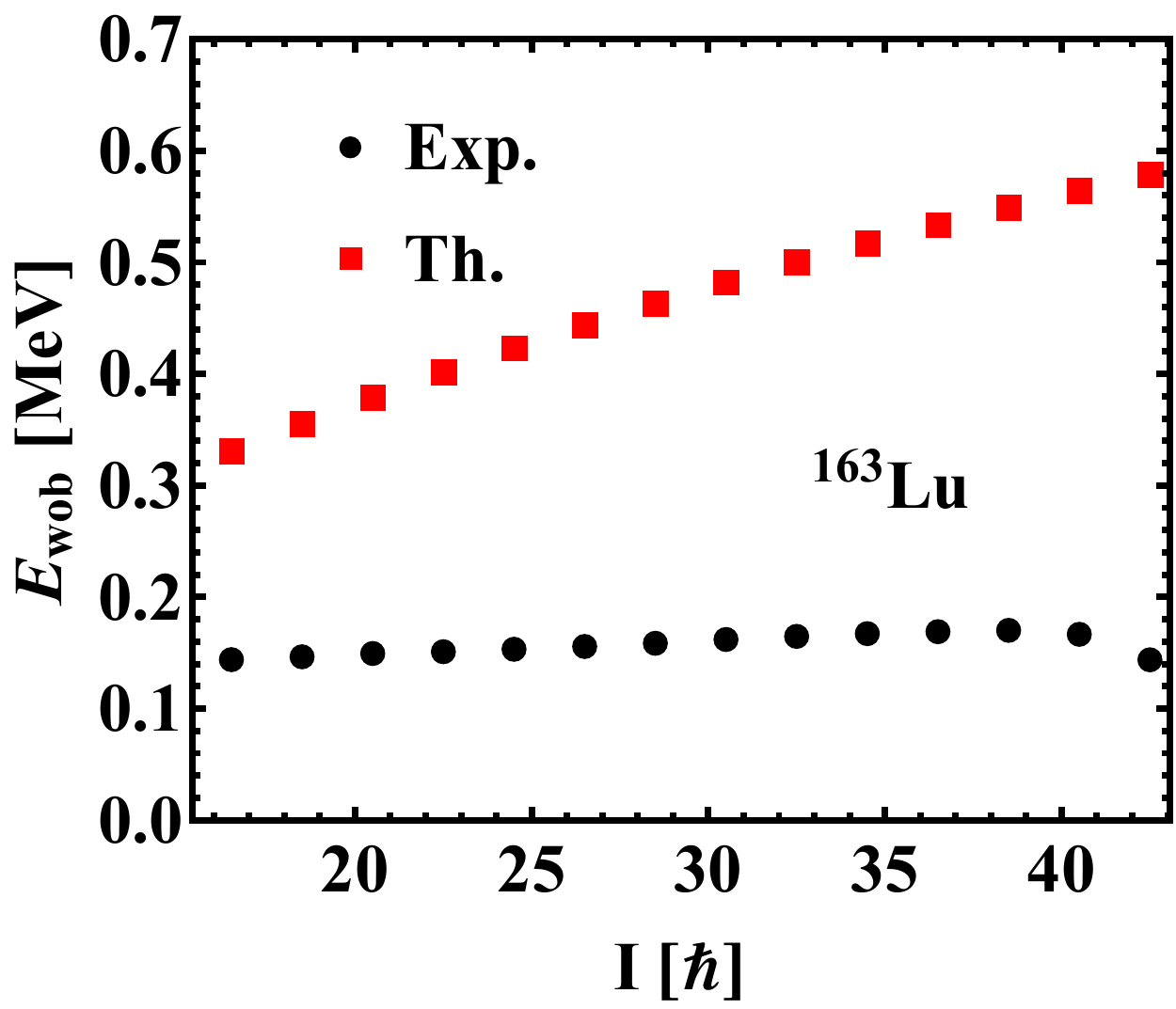}\includegraphics[width=0.25\textwidth,height=4cm]{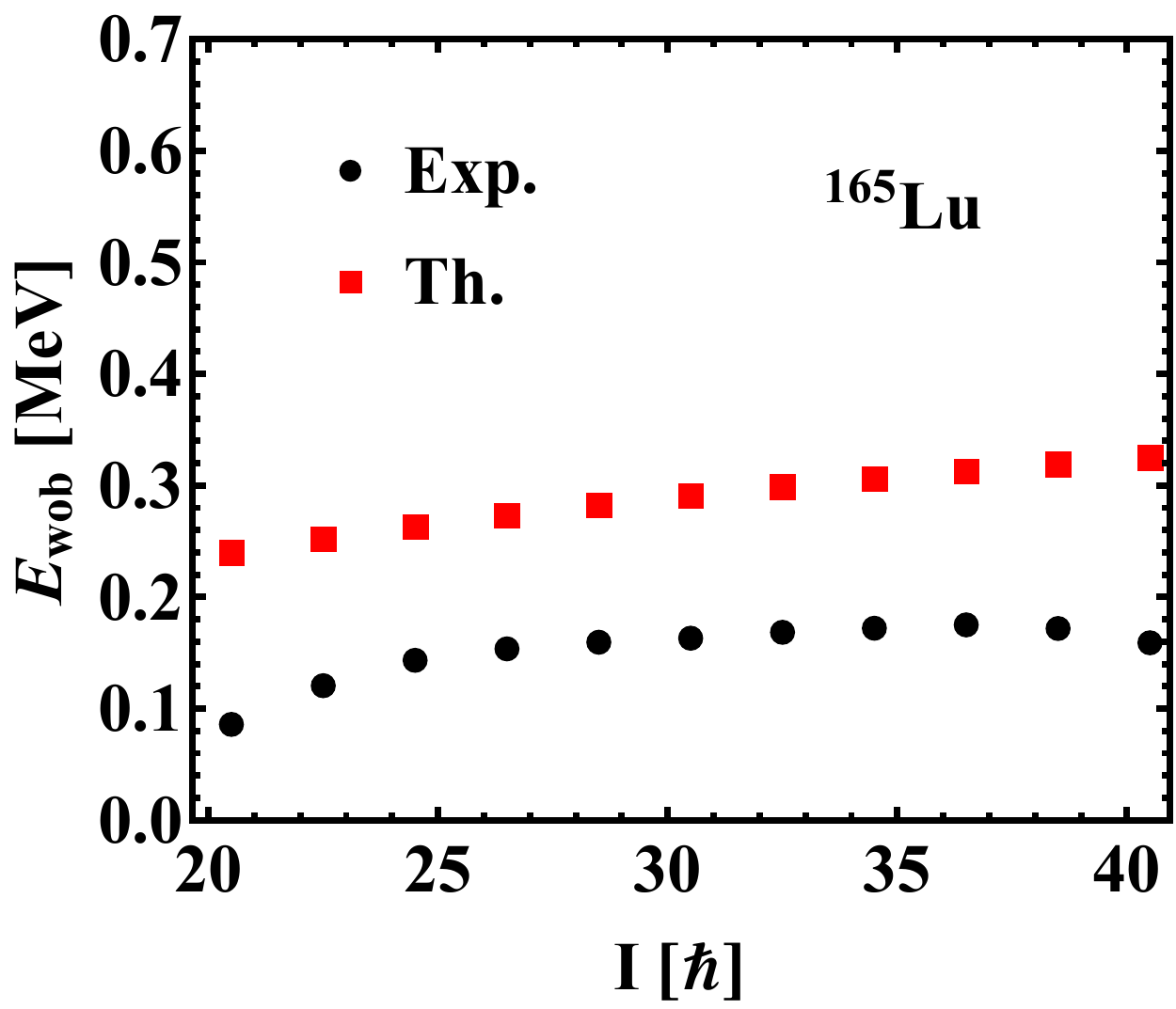}
{\scriptsize
\caption{(Color online)Wobbling energies, $E_{Wob}=E_1(I)-0.5(E_0(I+1)+E_0(I-1))$, with $E_1$ and $E_0$ defined as excitation energies  from TSD3 and TSD2 band, respectively. Experimental data are taken from Ref.\cite{Jens1} for $^{163}$Lu and from\cite{Scho} for $^{165}$Lu.}}
\label{Fig.6}
\end{figure}
\subsection{Alignment}
The alignment in the TSD bands is defined by subtracting from the angular momentum a reference value $I_{ref}={\cal J}_{0}\omega+{\cal J}_{1}\omega^3$, with the coefficients  ${\cal J}_{0}$
and ${\cal J}_{1}$ obtained by a least square procedure fit. Calculation results are compared with the corresponding experimental data in Figs 7-10 for the four isotopes of Lu. The linear term in
 $\omega$ involved in the expression of $I_{ref}$ corresponds to a spherical symmetry, while the second term is determined by the axial symmetry. Thus, the alignment gives a measure of triaxiality effect on the angular momentum.
\begin{figure}[ht!]
\includegraphics[width=0.25\textwidth]{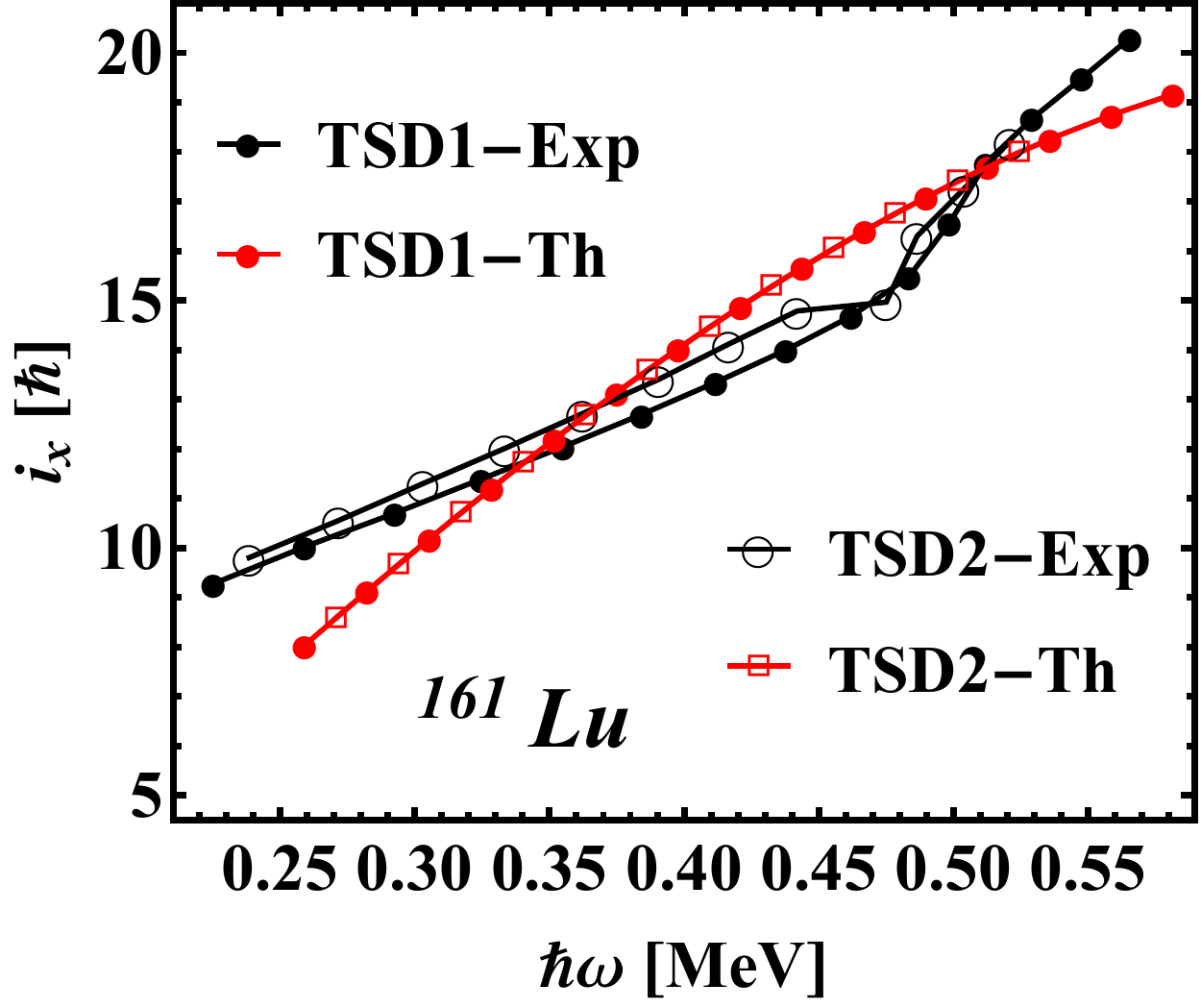}
\caption{(Color online)Results for the aligned angular momenta, $i_x$, relative to a reference $I_{ref}={\cal J}_{0}\omega+{\cal J}_{1}\omega^3$ with ${\cal J}_{0}=30\hbar^2MeV^{-1}$ and 
${\cal J}_{1}=40\hbar^4MeV^{-3}$ are compared with  the corresponding experimental data \cite{Bring}.}
\label{Fig.7}
\end{figure}
\begin{figure}[ht!]
\includegraphics[width=0.25\textwidth]{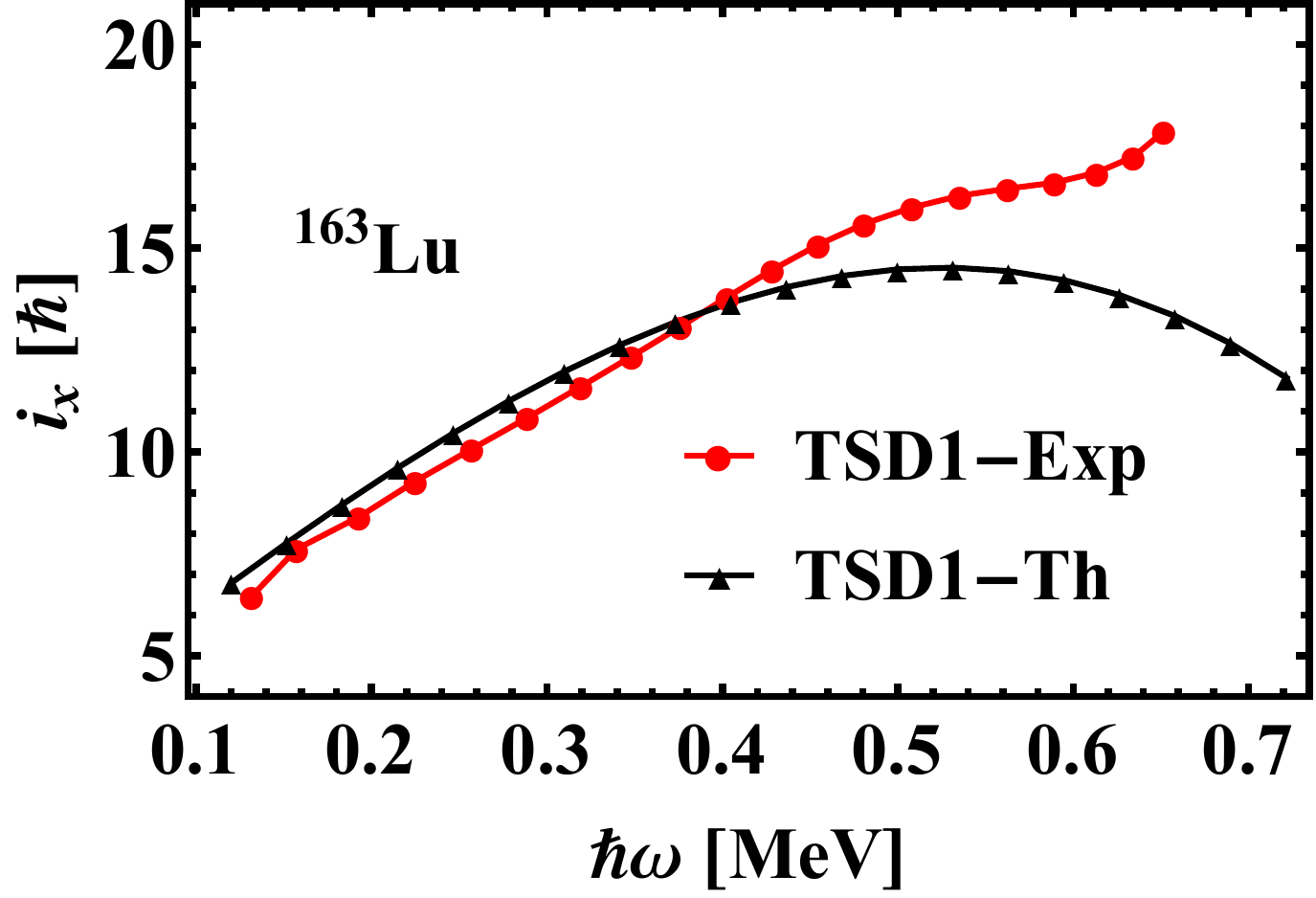}\includegraphics[width=0.25\textwidth]{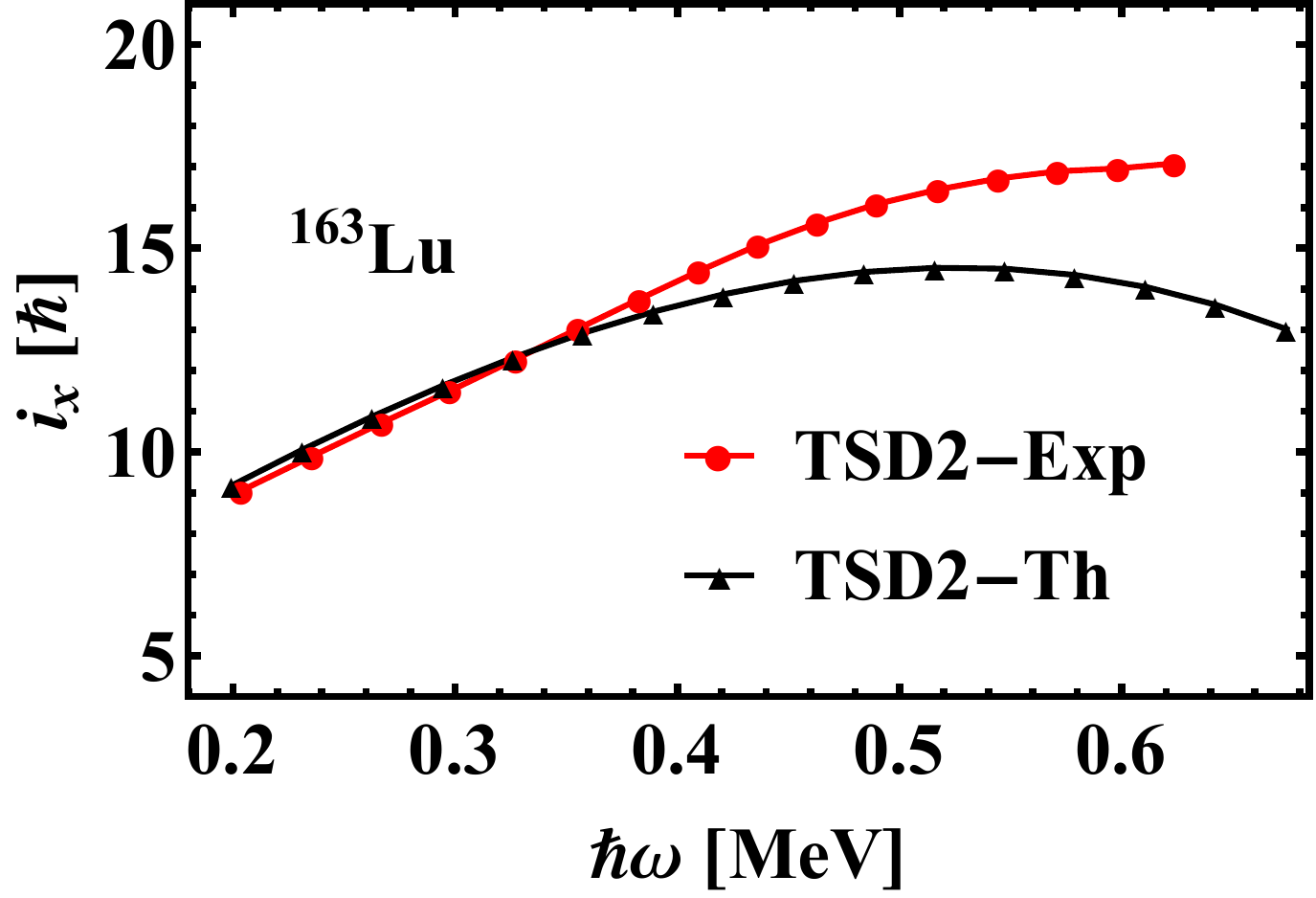}
\includegraphics[width=0.25\textwidth]{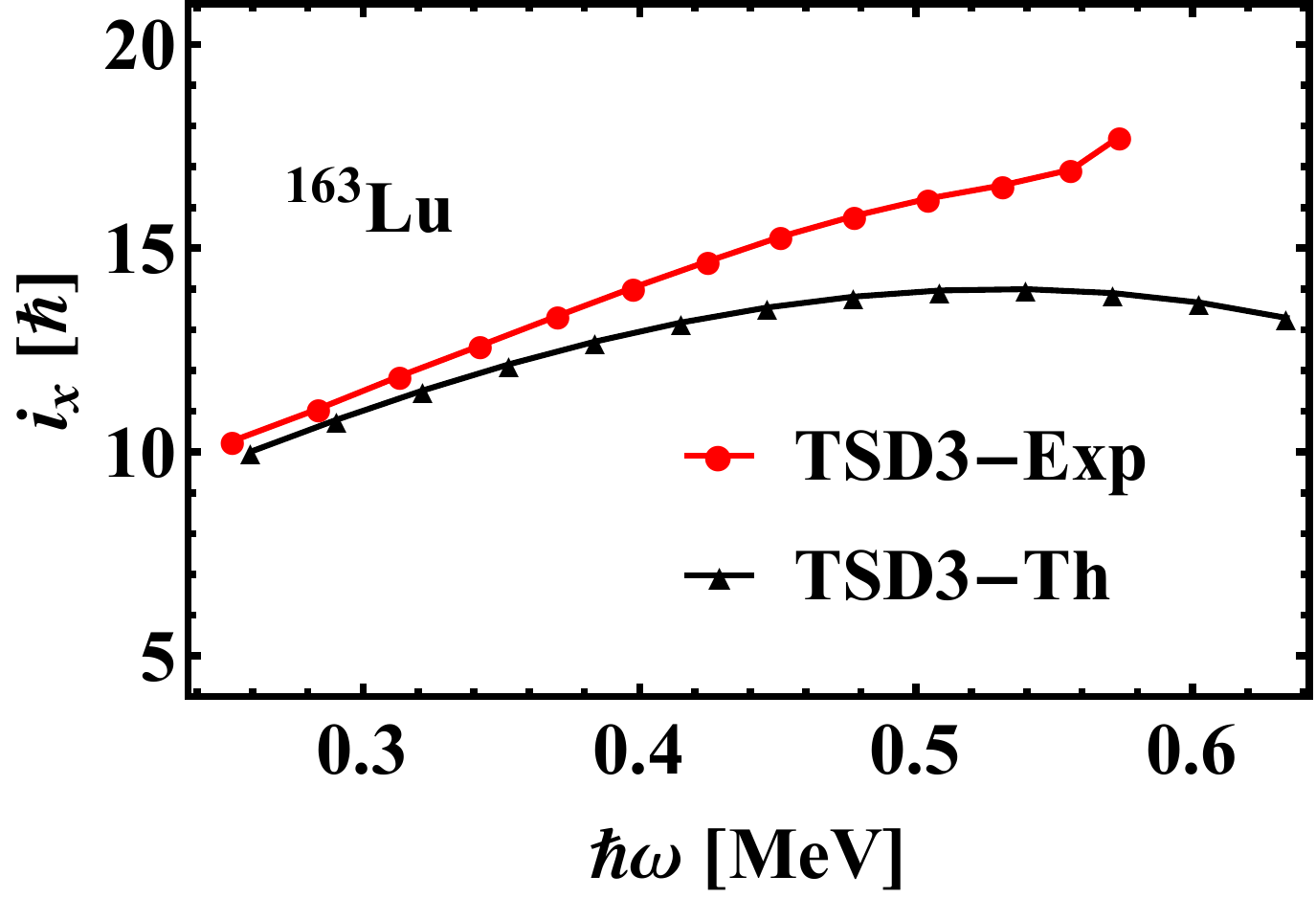}\includegraphics[width=0.25\textwidth]{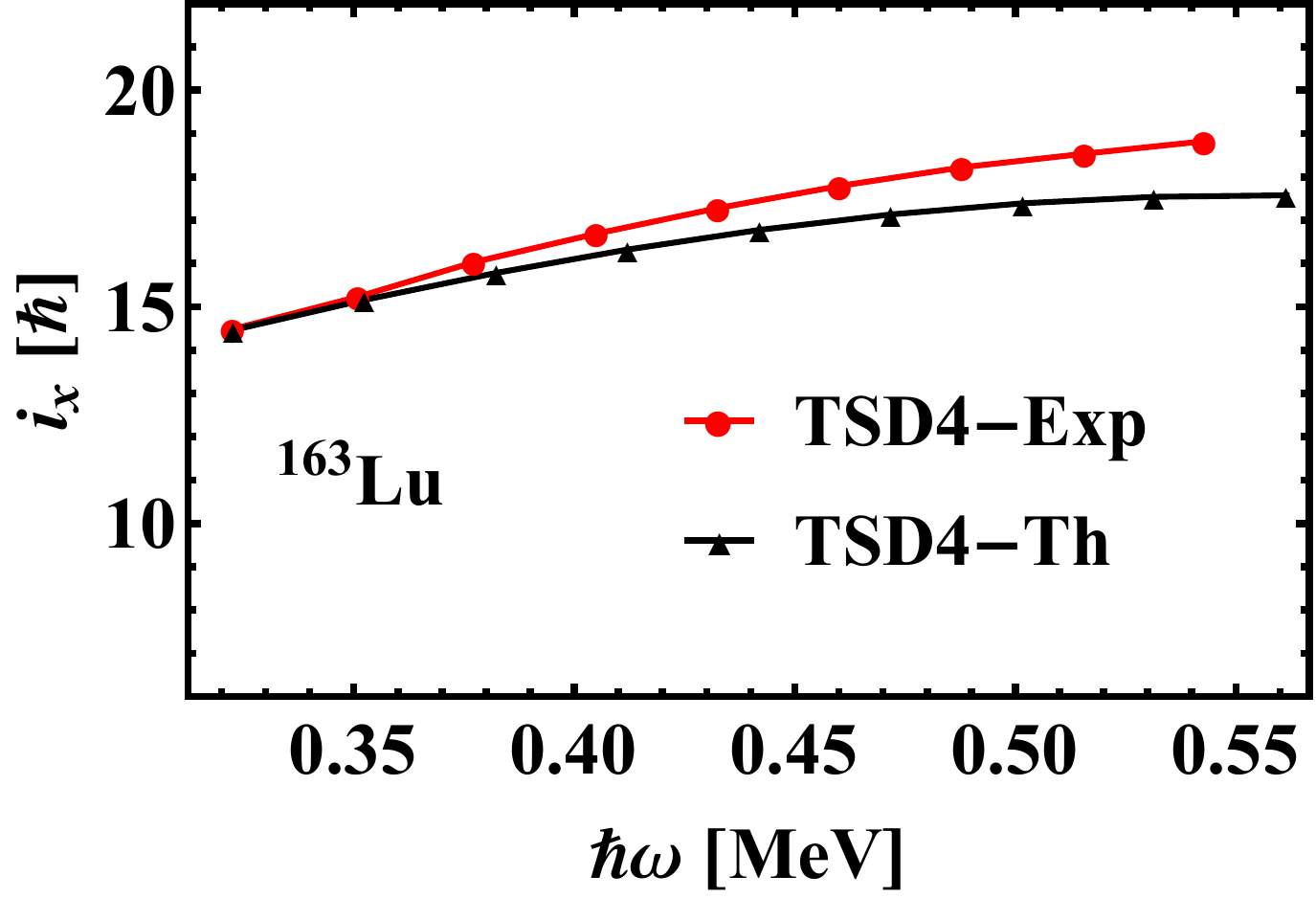}
\caption{(Color online)Results for the aligned angular momenta, $i_x$, relative to a reference $I_{ref}={\cal J}_{0}\omega+{\cal J}_{1}\omega^3$ with ${\cal J}_{0}=30\hbar^2MeV^{-1}$ and 
${\cal J}_{1}=40\hbar^4MeV^{-3}$, in $^{163}$Lu, are compared with  the corresponding experimental data \cite{Jens1,Hage}.}
\label{Fig.8}
\end{figure}
\begin{figure}[ht!]
\includegraphics[width=0.25\textwidth]{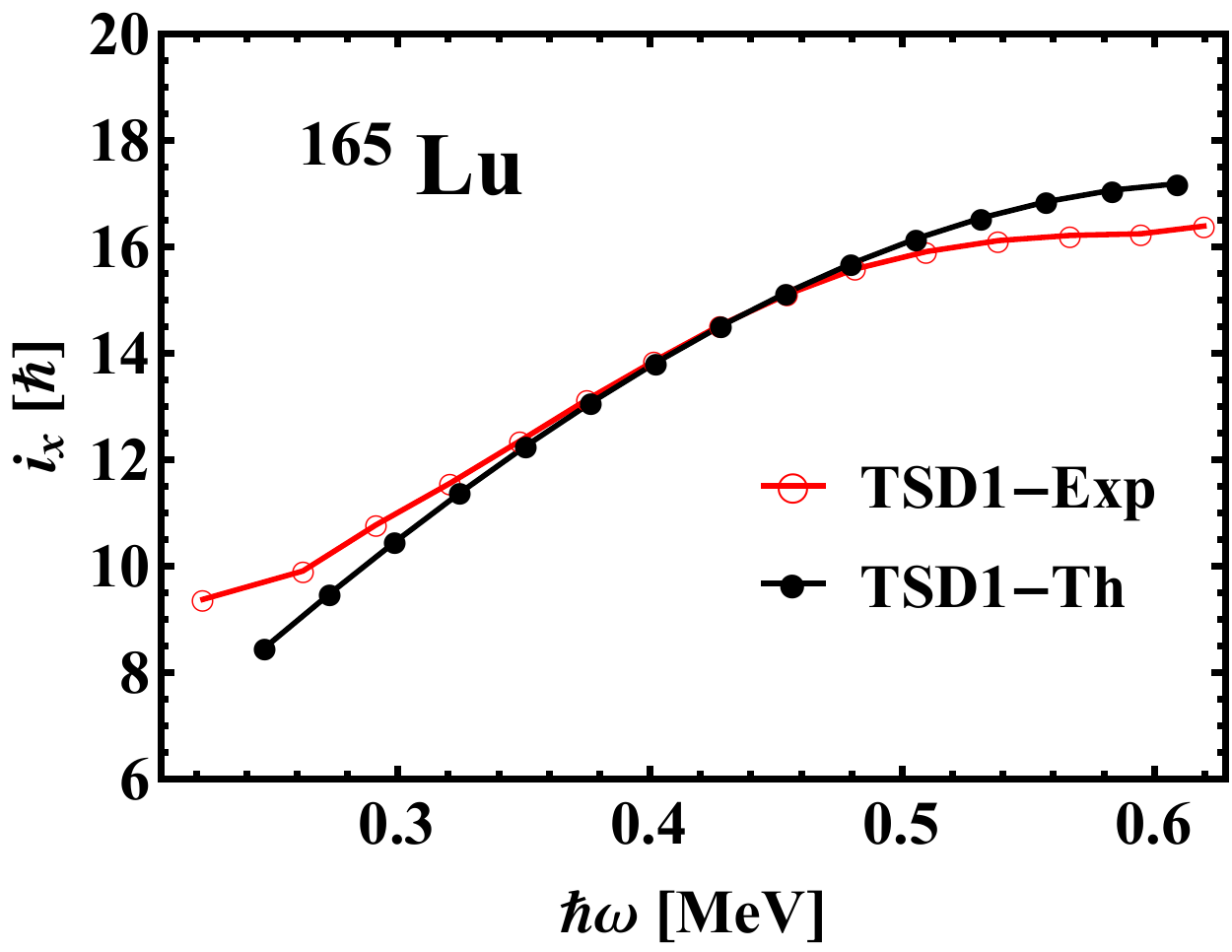}\includegraphics[width=0.25\textwidth]{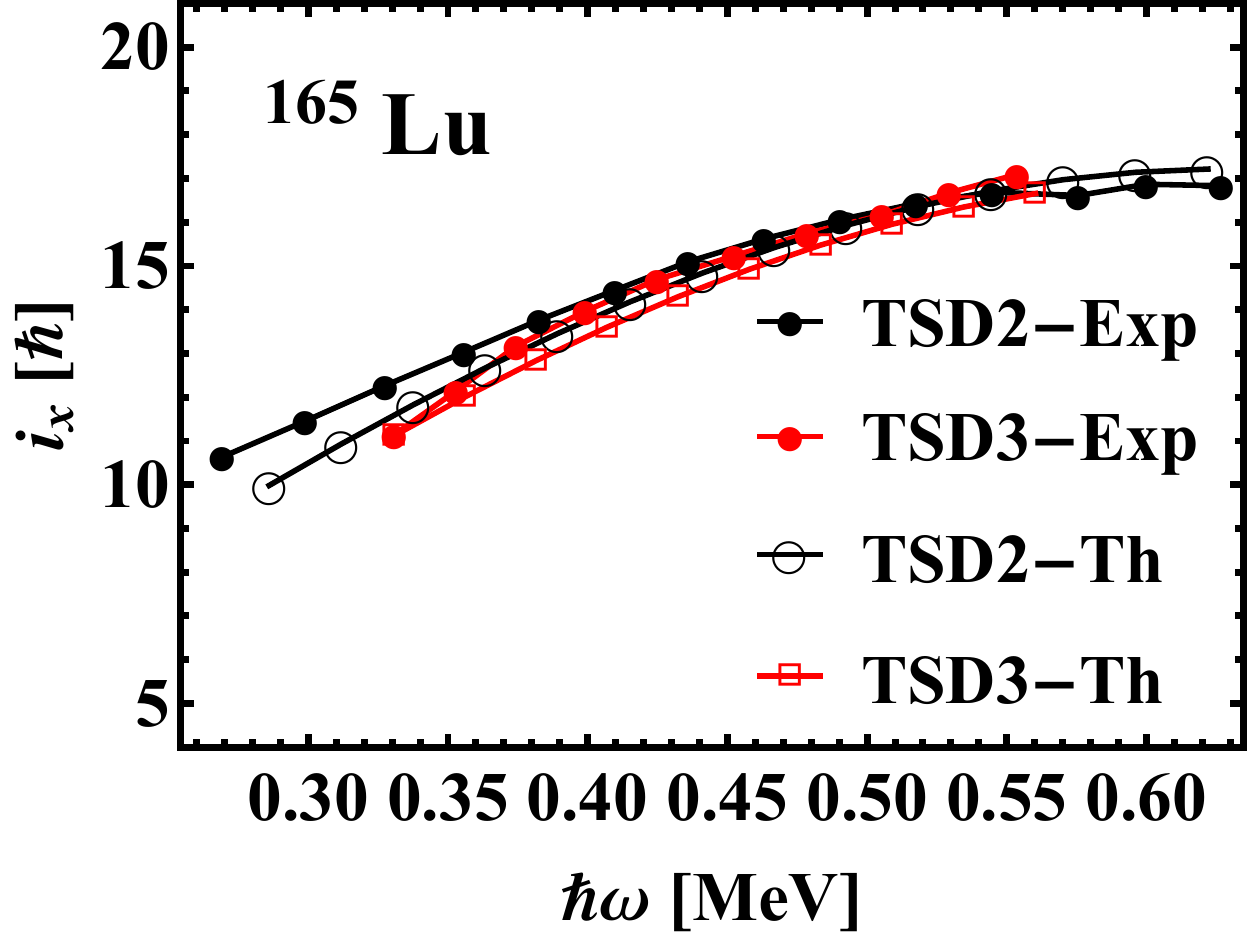}
\caption{(Color online)Results for the aligned angular momenta, $i_x$, relative to a reference $I_{ref}={\cal J}_{0}\omega+{\cal J}_{1}\omega^3$, with ${\cal J}_{0}=30\hbar^2MeV^{-1}$ and 
${\cal J}_{1}=40\hbar^4MeV^{-3}$, are compared with  the corresponding experimental data \cite{Scho}.}
\label{Fig.9}
\end{figure}

\begin{figure}[ht!]
\includegraphics[width=0.25\textwidth]{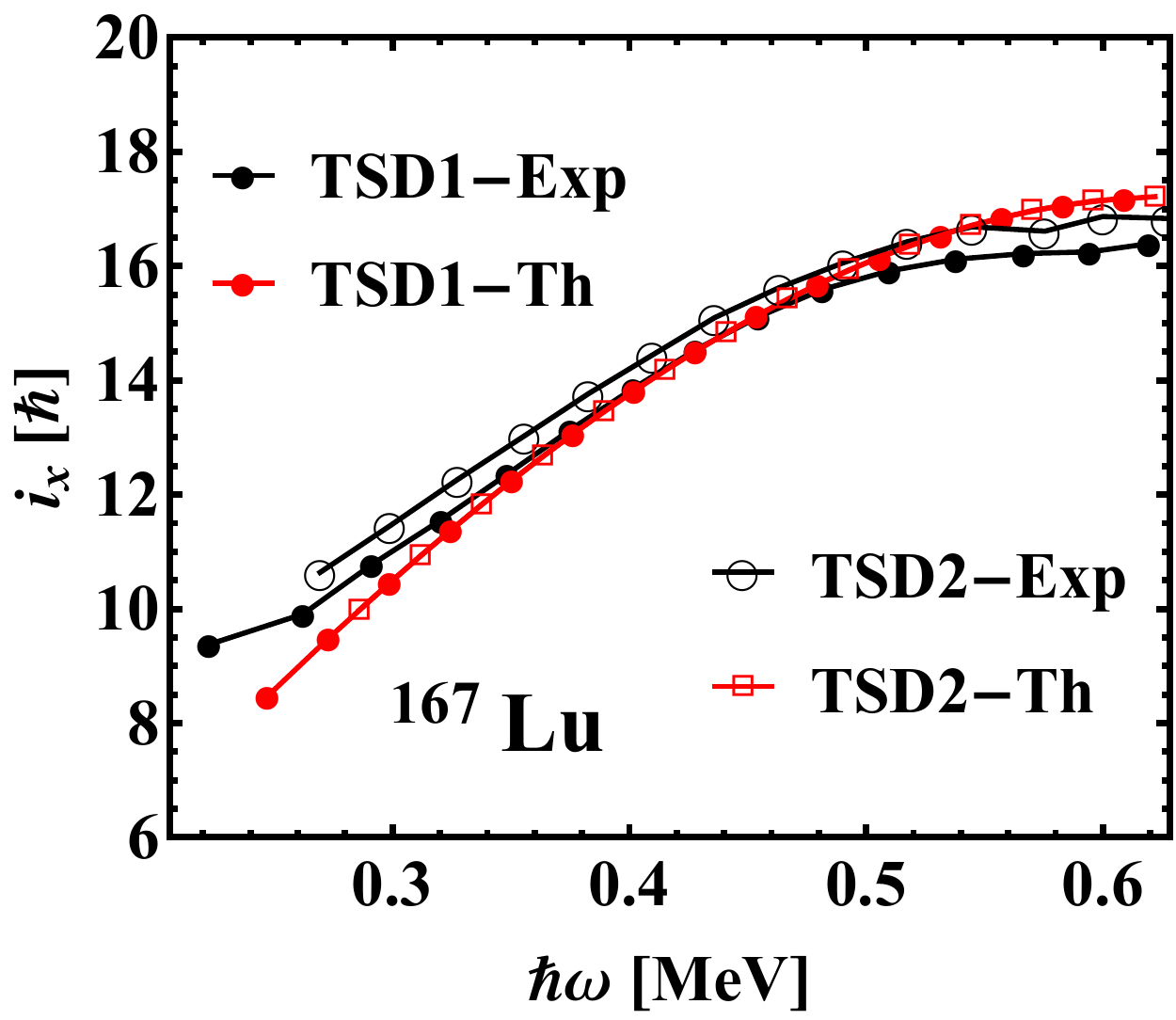}
\caption{(Color online)Results for the aligned angular momenta, $i_x$, relative to a reference $I_{ref}={\cal J}_{0}\omega+{\cal J}_{1}\omega^3$, with ${\cal J}_{0}=30\hbar^2MeV^{-1}$ and 
${\cal J}_{1}=40\hbar^4MeV^{-3}$, are compared with  the corresponding experimental data \cite{Amro}.}
\label{Fig.10}
\end{figure}

From Figs. 7-10 one remarks a quite good agreement between the theoretical results and experimental data. Only for $^{163}$Lu we notice discrepancies in the region of large rotational frequencies.
For a large interval of $\hbar\omega$ the alignment shows a linear increasing behavior, while for very high frequencies an alignment saturation tendency may be observed, which results in a forward and then a slight bending down for the experimental data and results, respectively. In the case of $^{163}$Lu the slope of the linear part is small and the bending of the theoretical curve induces deviations from the experimental one. Accordingly, the angular momentum behavior as function of rotational frequency is similar to that of the reference function but adding some corrections specific to the three mention regions: a) a linear term in $\hbar\omega$ in the first part; b) a constant term, in the second region and a linear term of negative slope in the sector, where one observes a down bending. The fact that the curves associated to different TSD bands are close to each other reflects their wobbling character. 
In a previous publication the alignment \cite{Rad018} for the bands of $^{165}$Lu and $^{167}$Lu where plotted in Figs 7 and 14, respectively. Results were obtained using for TSD2 and TSD3 one and two wobbling phonon states built on top of the states from TSD1. Comparing the results of Ref.\cite{Rad018} with those from the present paper given in Figs 9 and 10 respectively, one notes
an improvement of the agreement  with the data in favor of the present approach.

\subsection{Reference energy}
A similar analysis is performed also for the excitation energy relative to a spherical rigid rotor with an effective moment of inertia, as function of the angular momentum. This is shown in 
Figs. 11-14 for  all TSD bands in the four even-odd isotopes of Lu. The reference energy has the expression $aI(I+1)$, where $a$ is fixed by fitting the experimental energies through a least square procedure. Obviously, the value of $a$ may differ from one isotope to another, and is specified for each of the cases.  The mentioned figure indicate that the deviation of the excitation energies  from the reference values is a decreasing function of the angular momentum. This feature suggests that the effect of triaxiality is diminishing with angular momentum. In Fig. 11,  although the energy follows the general trend, it is  smaller than the reference value. The general  pattern is obtained by amending the relative energy by an amount of about 2.5 MeV. Except for the band TSD3 in $^{163}$Lu, the experimental and theoretical relative energies agree with each other. It is worth mentioning the fact that the experimental and theoretical curves associated to the TSD4 band in $^{163}$Lu and the TSD3 band in $^{165}$Lu band respectively, are almost identical. As in the alignment case, the triaxial features are diminished at large angular momentum and the system tends to rotate around the principal axis with largest MoI. From Figs. 13 and 14  of this paper and Figs.10 and 14 from Ref.\cite{Rad018} one remarks on a better description provided by
the present formalism for $^{165,167}$Lu.

\begin{figure}[ht!]
\includegraphics[width=0.25\textwidth]{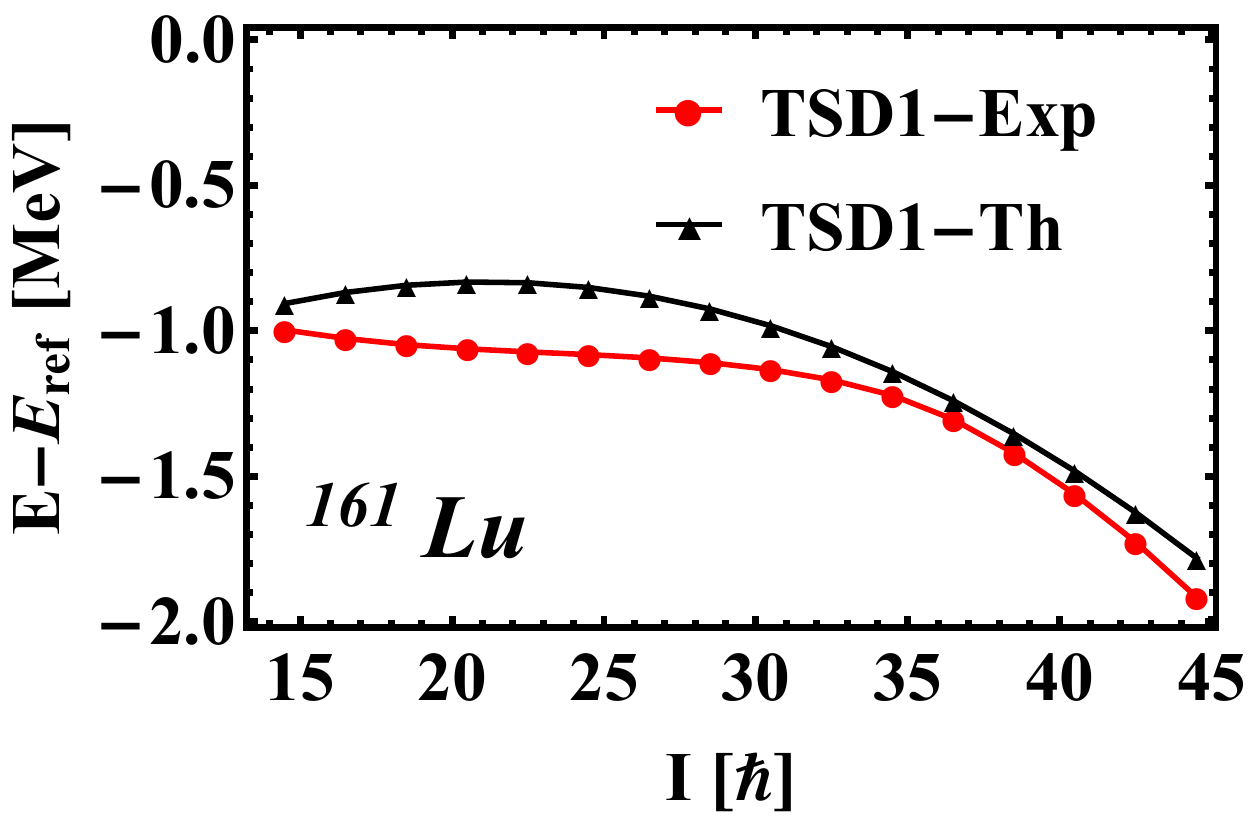}\includegraphics[width=0.25\textwidth]{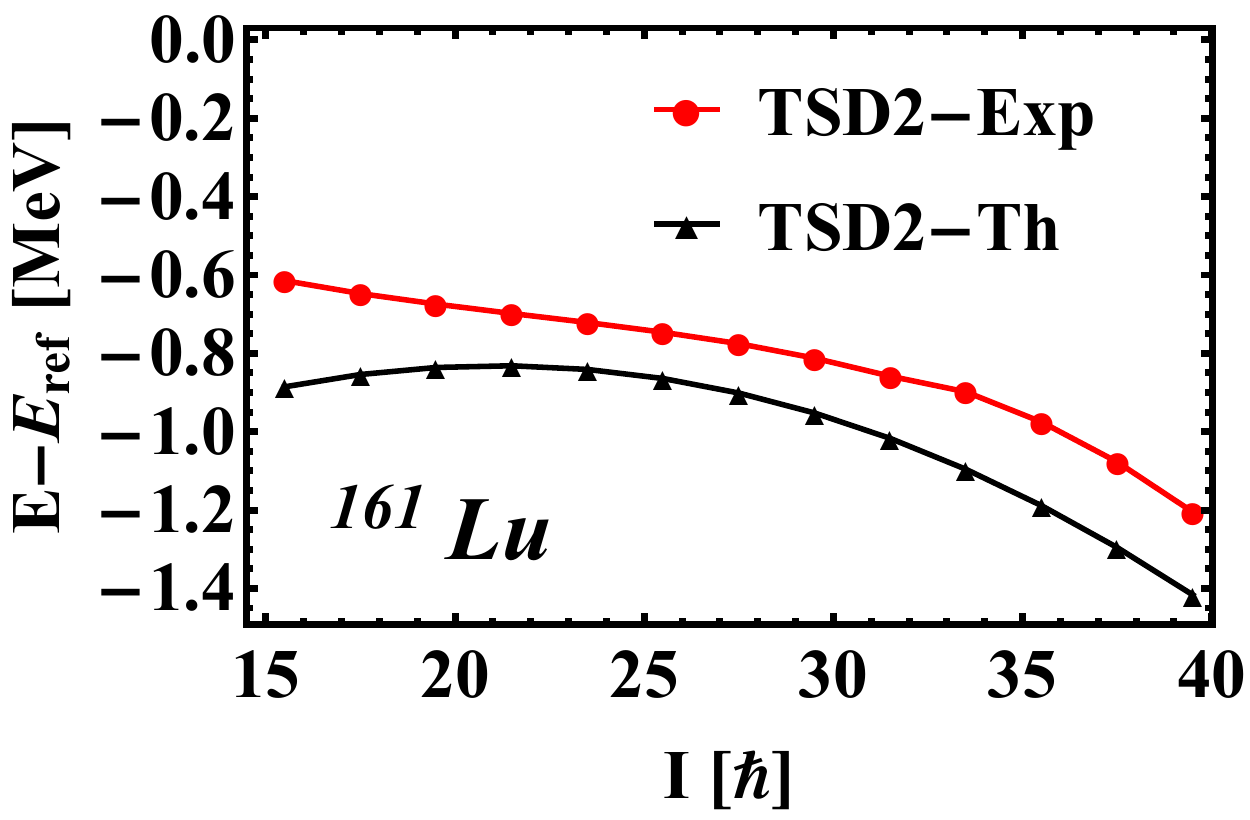}
\caption{(Color online)Theoretical and experimental excitation energies for TSD1 in $^{161}$Lu normalized to the energy of a rigid rotor with an effective moment of inertia, i.e. 
$E_{REF}=0.0075I(I+1)(MeV)$, are plotted as function of the angular momentum.}
\label{Fig.11}
\end{figure}

\begin{figure}[ht!]
\includegraphics[width=0.24\textwidth]{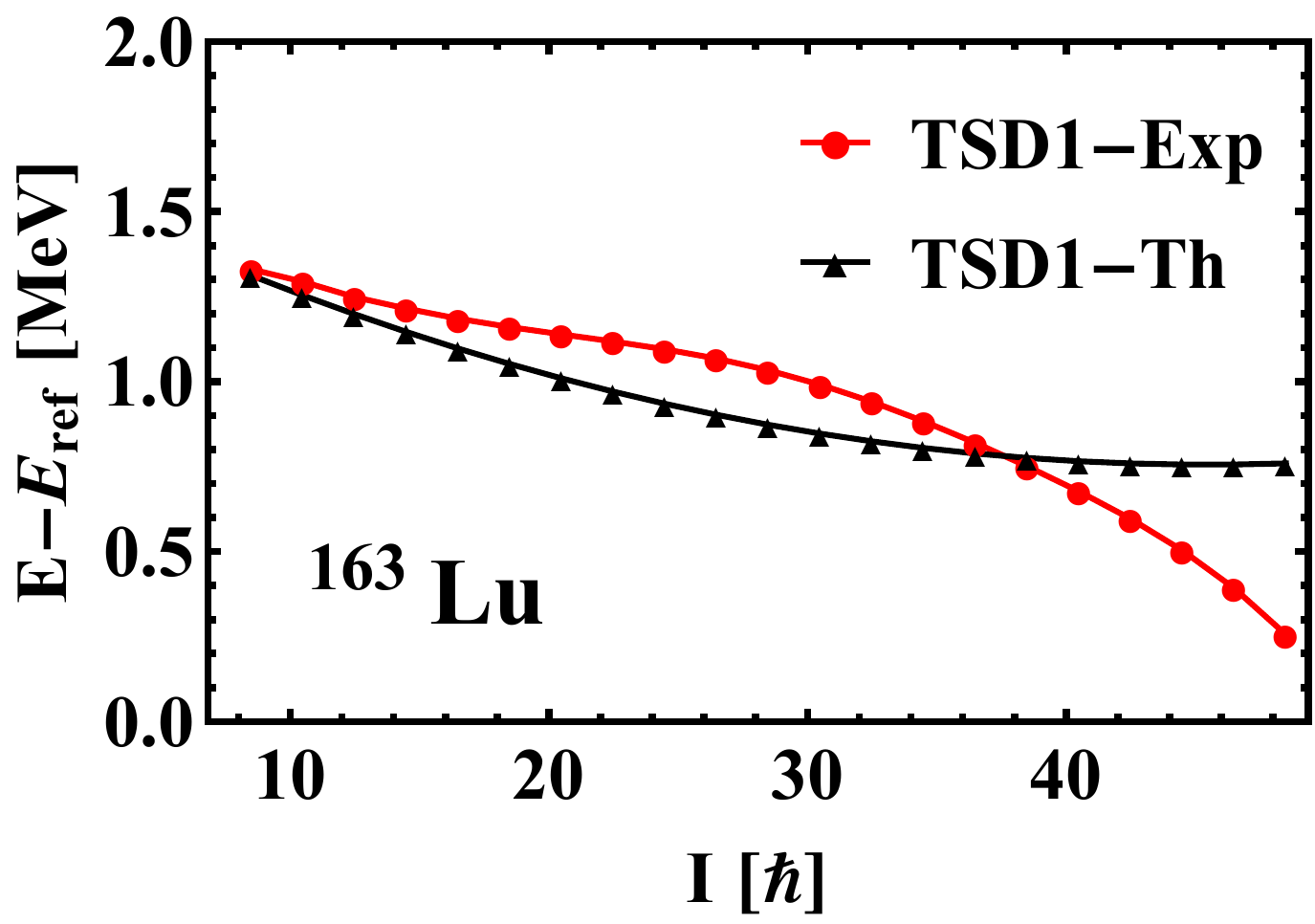}\includegraphics[width=0.24\textwidth]{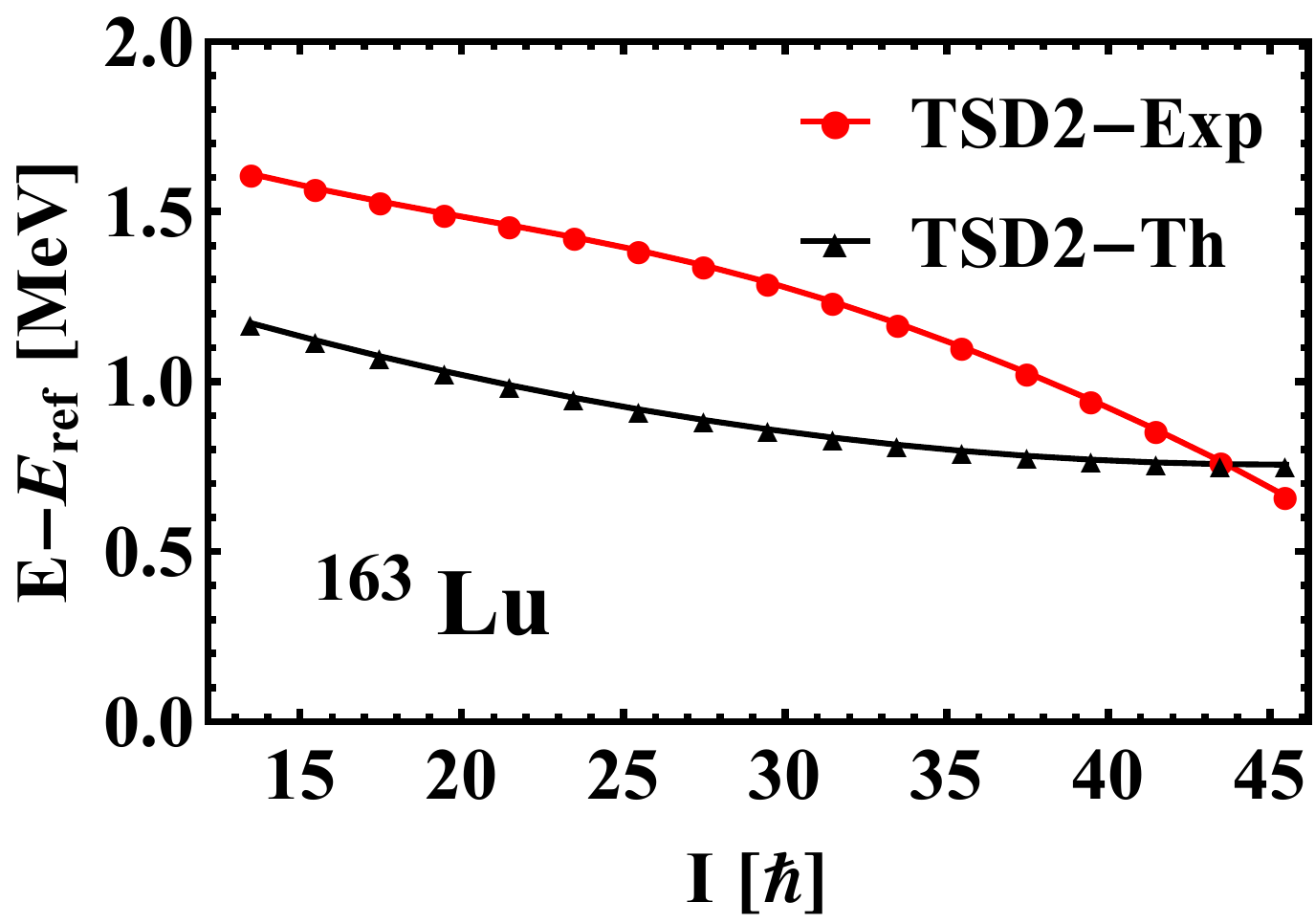}
\includegraphics[width=0.24\textwidth]{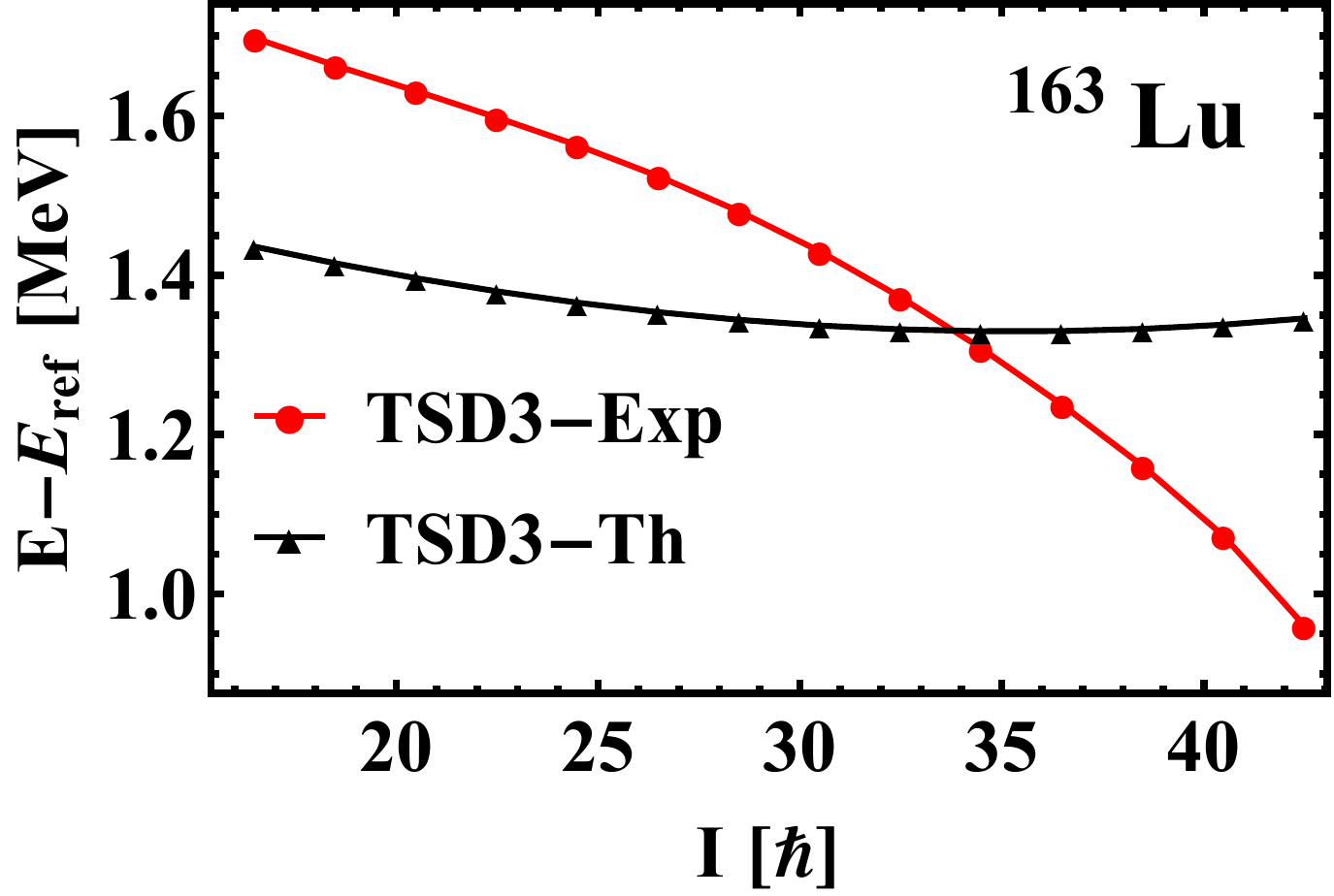}\includegraphics[width=0.24\textwidth]{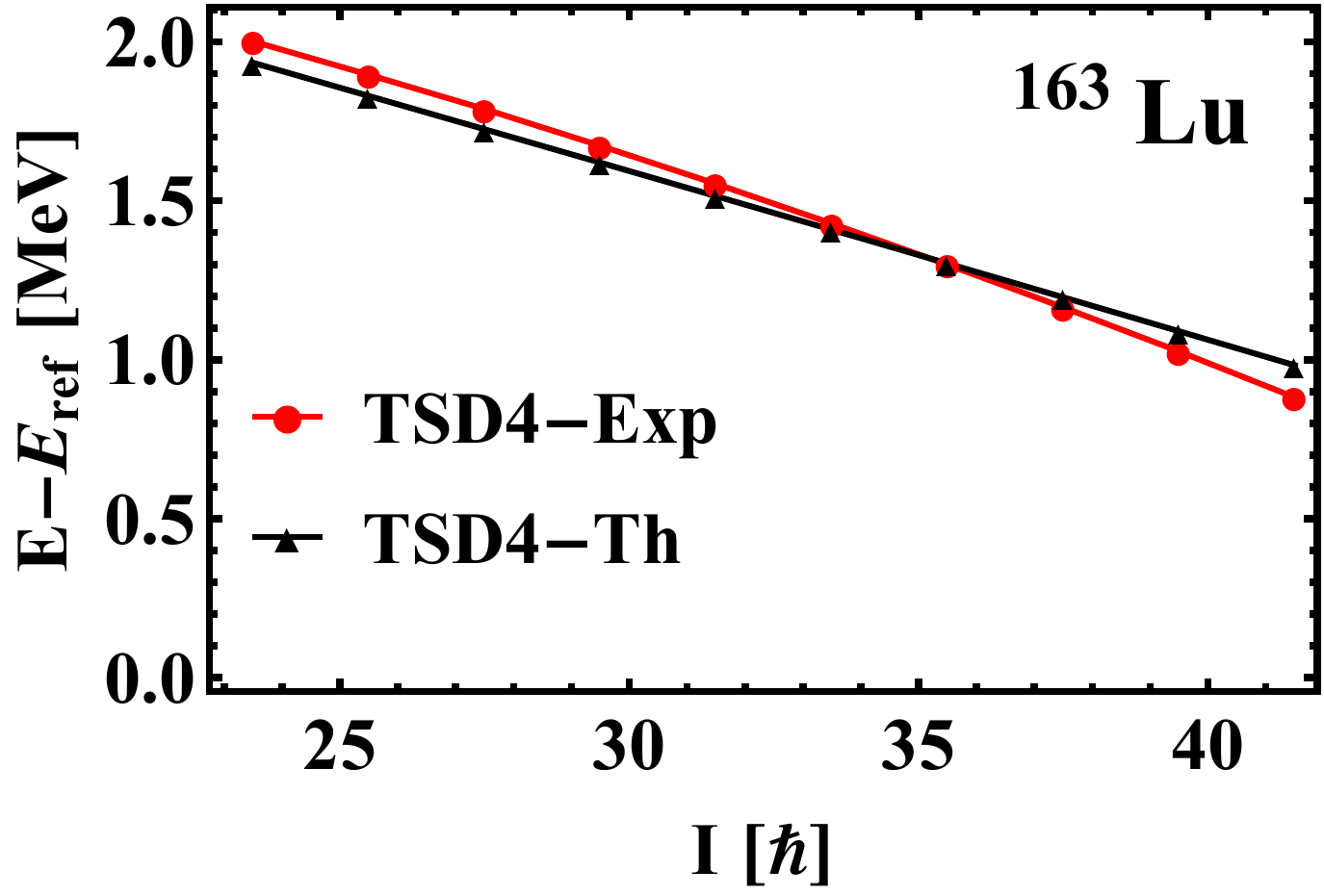}
\caption{(Color online)The same as in Fig. 11, but for the bands TSD1, TSD2, TSD3, and TSD4 of $^{163}$Lu.}
\label{Fig.12}
\end{figure}

\begin{figure}[ht!]
\includegraphics[width=0.25\textwidth]{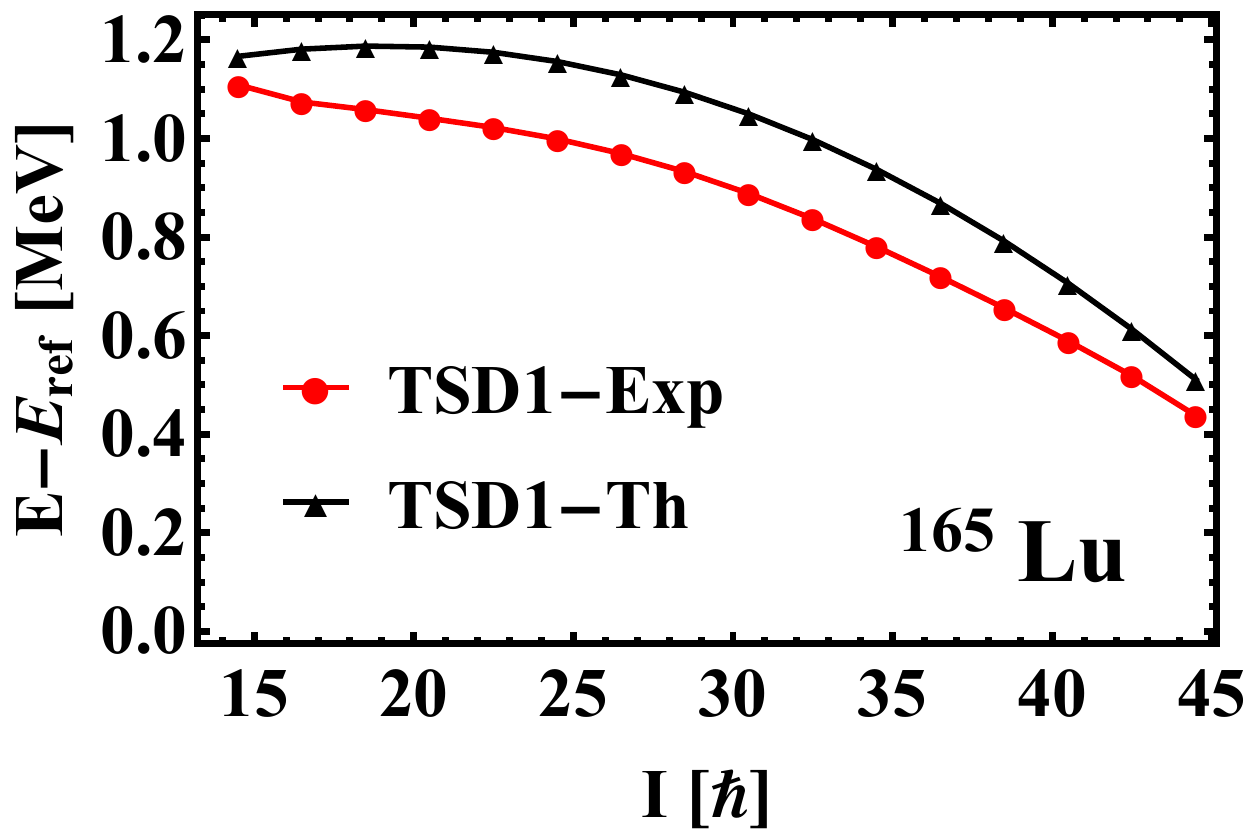}\includegraphics[width=0.25\textwidth]{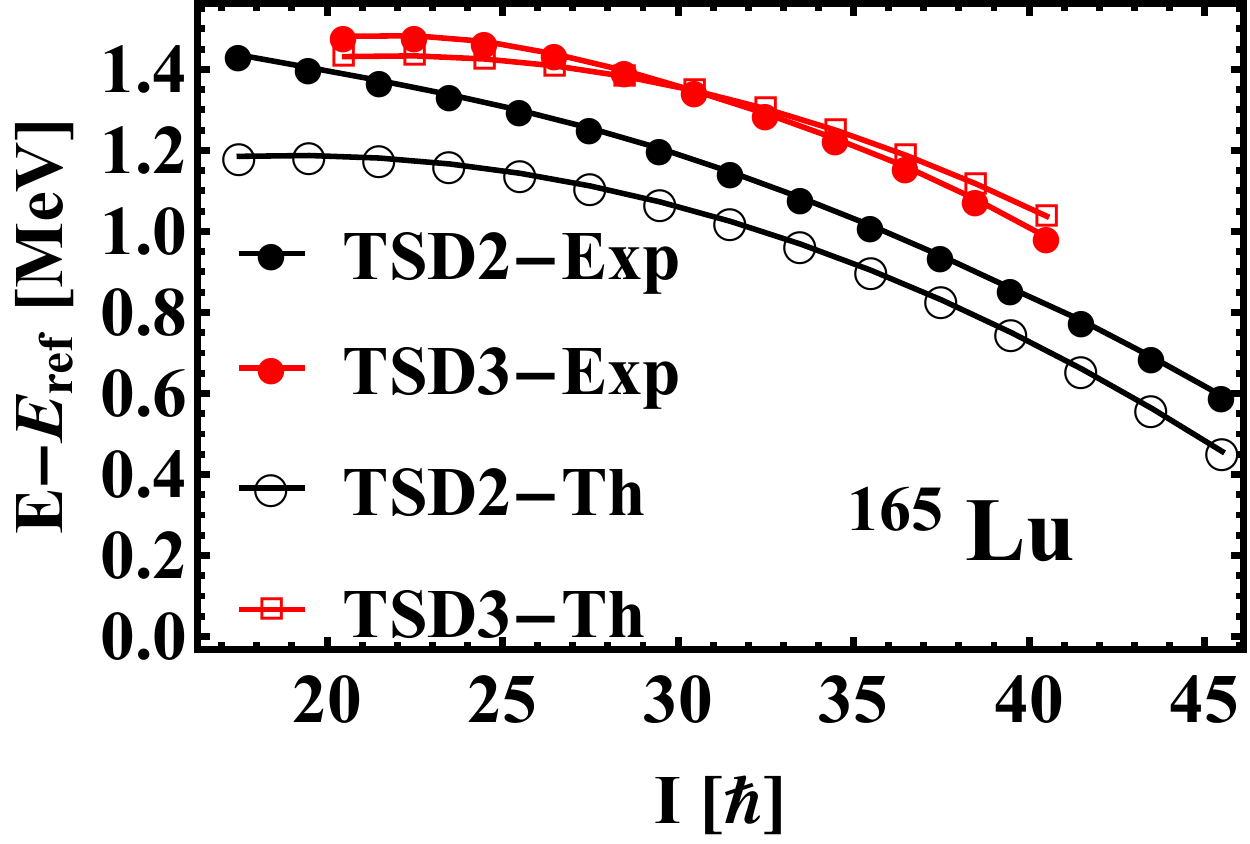}
\caption{(Color online)The same as in Fig. 11, but for $^{165}$Lu.}
\label{Fig.13}
\end{figure}

\begin{figure}[ht!]
\includegraphics[width=0.24\textwidth]{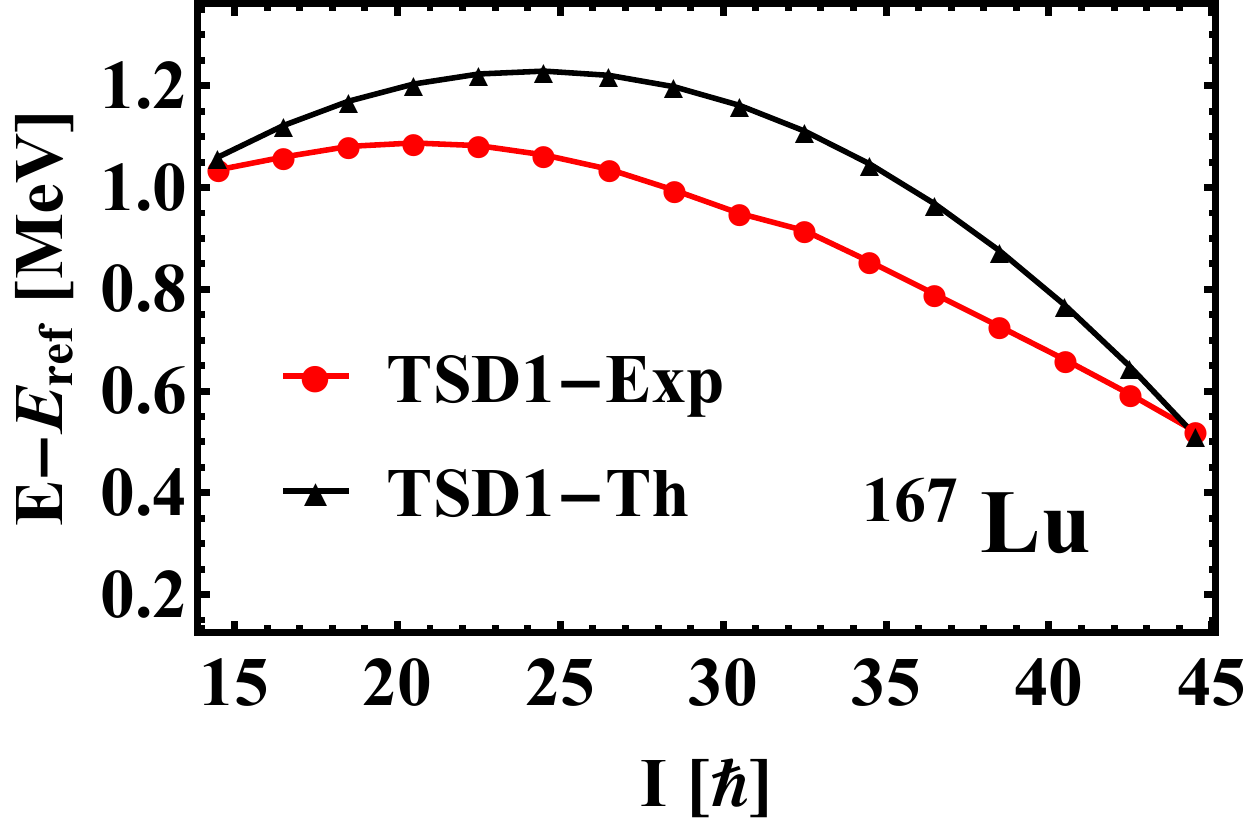}\includegraphics[width=0.24\textwidth]{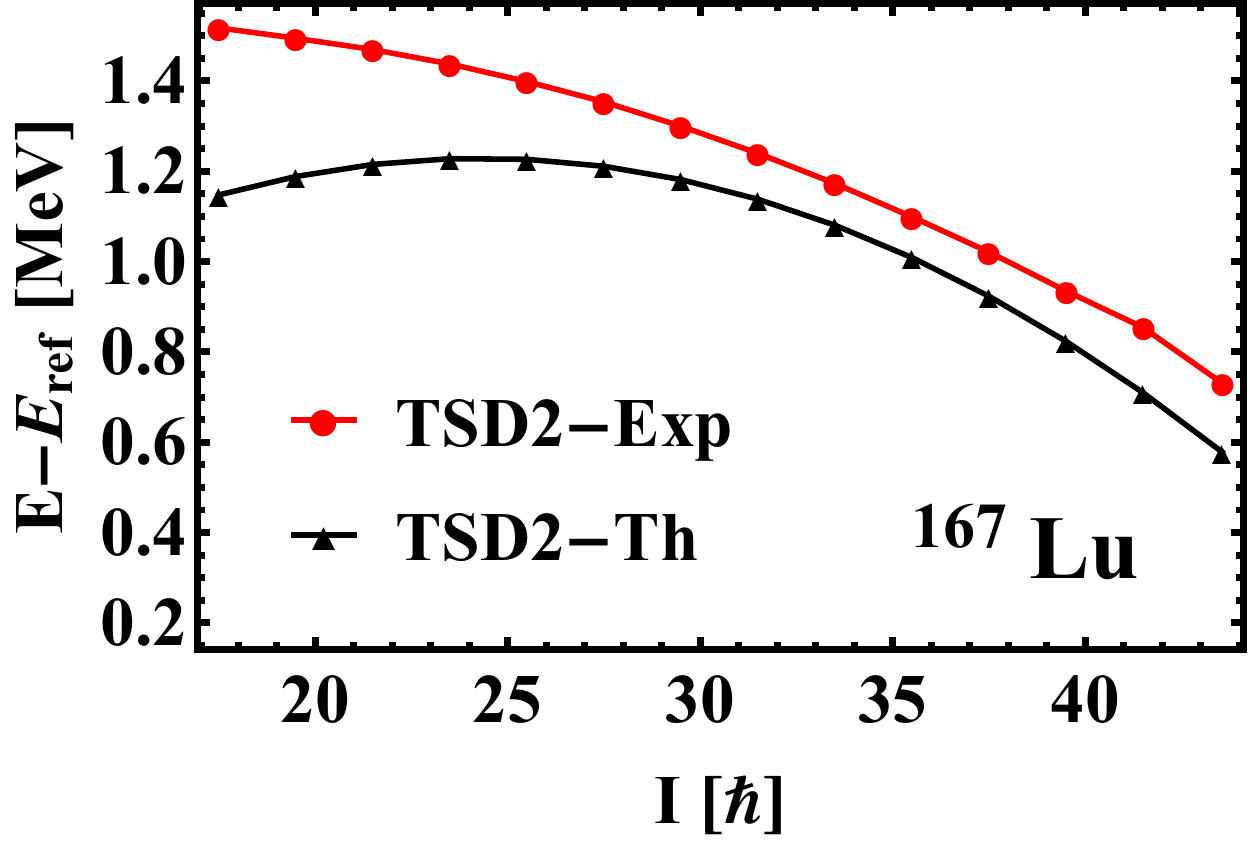}
\caption{(Color online)The same as in Fig.11, but for the bands TSD1 and TSD2 of $^{167}$Lu.}
\label{Fig.14}
\end{figure}

\subsection{The dynamic moment of inertia}
The dynamic MoI is a  magnitude sensitive to the rotation frequency variation. Its behavior is shown in Figs. 15-18 for both theoretical and experimental TSD bands, in all four considered isotopes. While in the extreme limits of the $\omega$ interval, the experimental dynamic moment of inertia depends on the energy spacings, inducing a staggering or a sharp increase, in the complementary interval this is almost constant for all three bands. In $^{161}$Lu, the sharp increase of the dynamic moment of inertia is caused by the alignment of the odd-proton angular momentum while for the heavier isotopes the staggering, seen in low spin part of the spectrum, reflects an interaction with the states from the neighboring normal deformed bands. Also the strong variation seen for large frequency might be also attributed to the alignment of the $i_{13/2}$ proton.
On the other hand, the theoretical dynamic moment of inertia is a constant function of $\omega$, which reflects a linear dependence of $\omega$ on the angular momentum I, and, moreover, a similar slope for this dependence in the three bands. Concluding, apart from the staggering of a few states placed at the beginning and at the end of the $\omega$ interval, the results of our calculations agree with the experimental data. The agreement quality is better than that presented in Ref.\cite{Rad018} for $^{165,167}$Lu (see Figs. 17 and 18 from here and 8 and 14 from there) by a different approach.

\begin{figure}[ht!]
\includegraphics[width=0.25\textwidth]{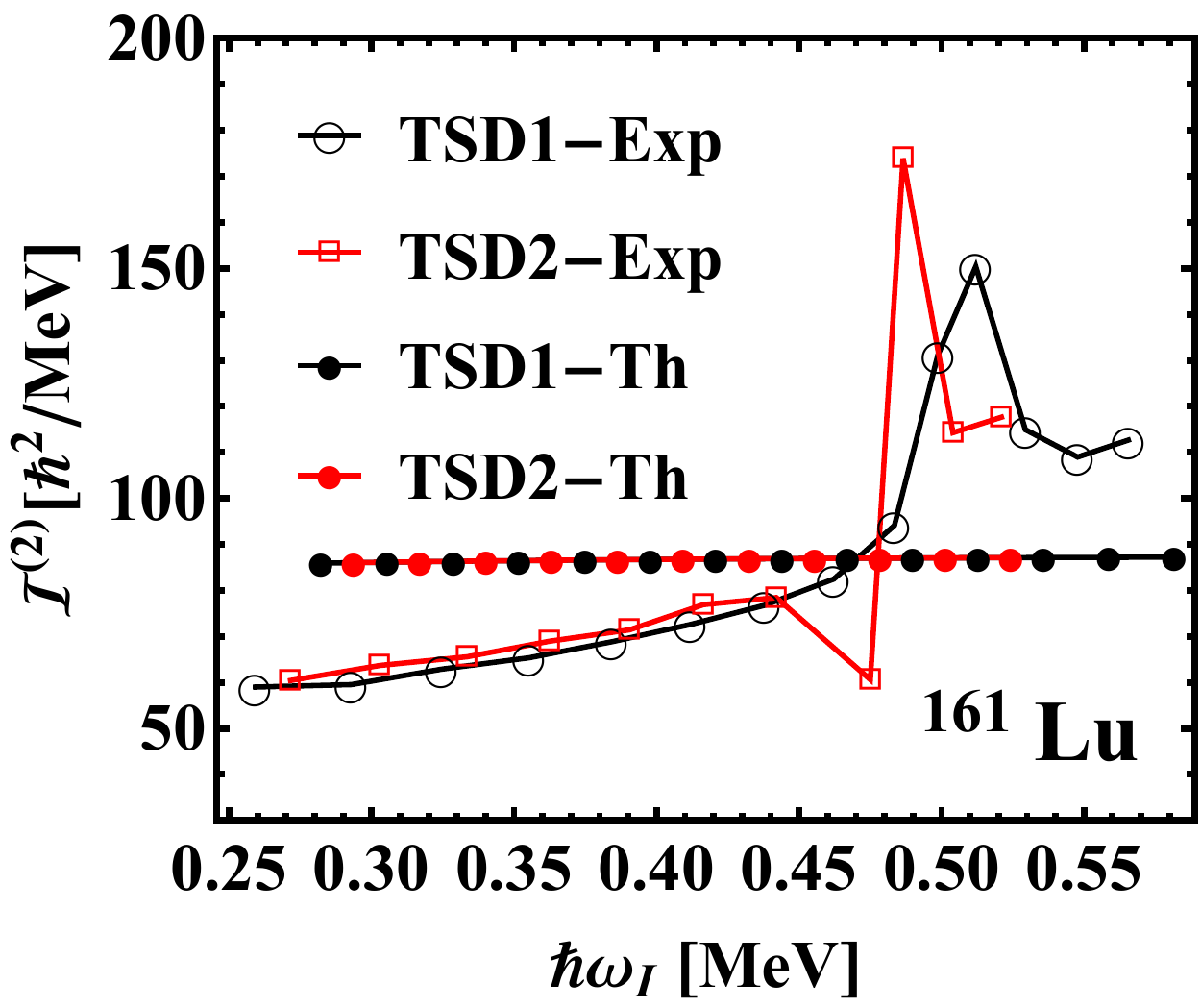}
\caption{(Color online)Results for the dynamic MoI's in the bands TSD1, TSD2 of $^{161}$Lu are compared with the corresponding experimental data taken from Ref.\cite{Bring}.}
\label{Fig.15}
\end{figure}

\begin{figure}[ht!]
\includegraphics[width=0.24\textwidth]{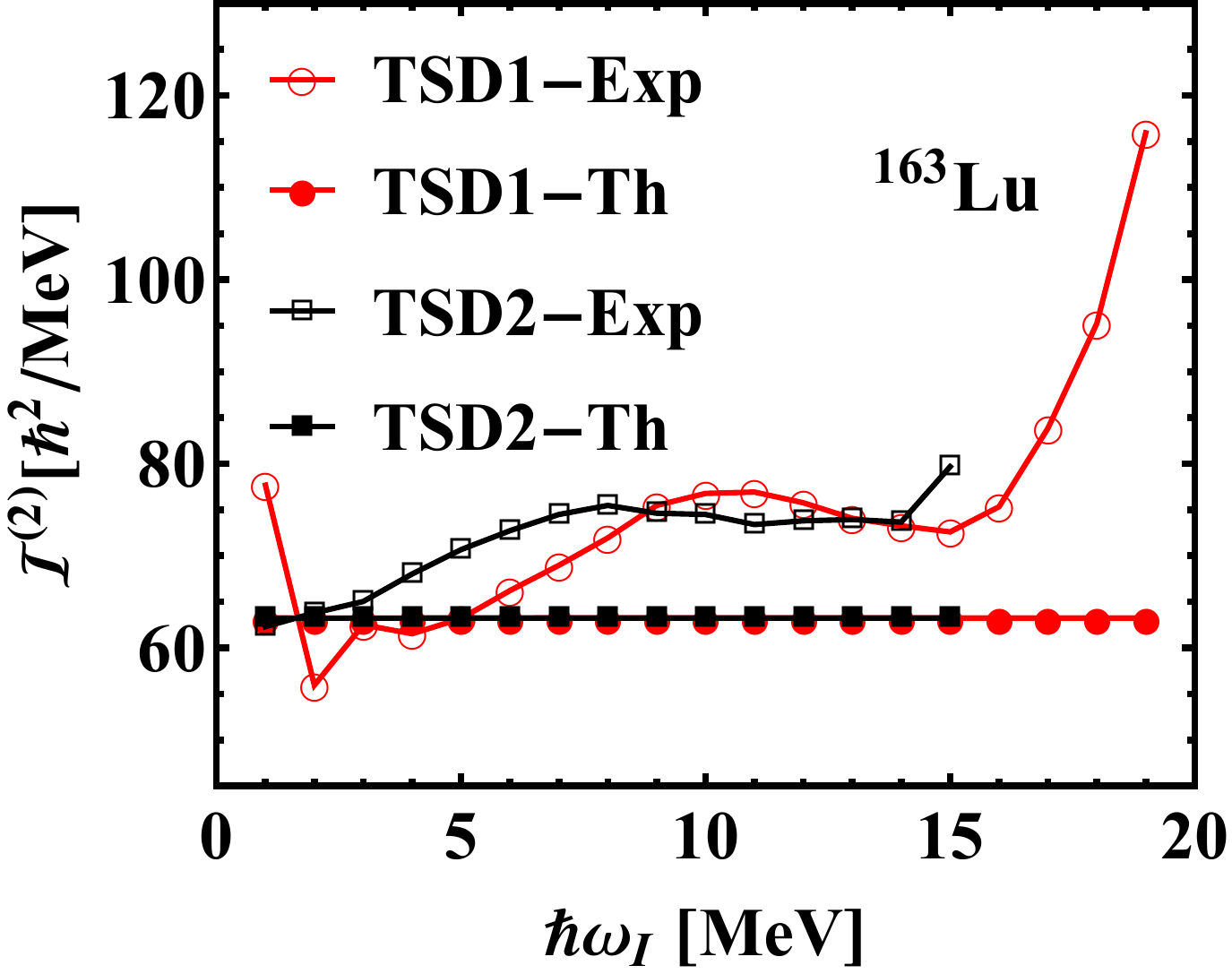}\includegraphics[width=0.24\textwidth]{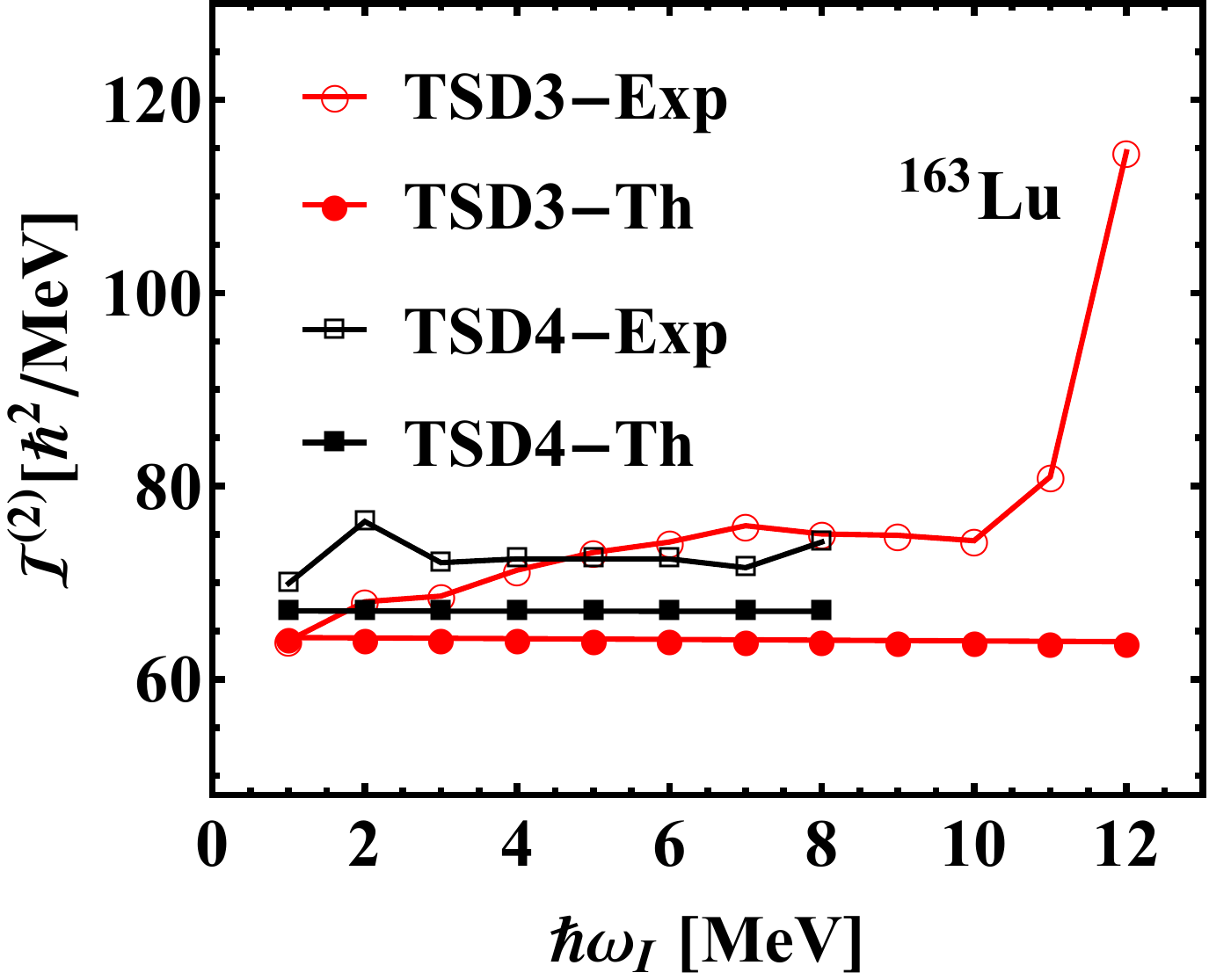}
\caption{(Color online)Results for the dynamic MoI's in the bands TSD1, TSD2, TSD3 and TSD4 of $^{163}$Lu, are compared with the corresponding experimental data taken from Ref.\cite{Jens1,Odeg}.}
\label{Fig.16}
\end{figure}

\begin{figure}[ht!]
\includegraphics[width=0.24\textwidth]{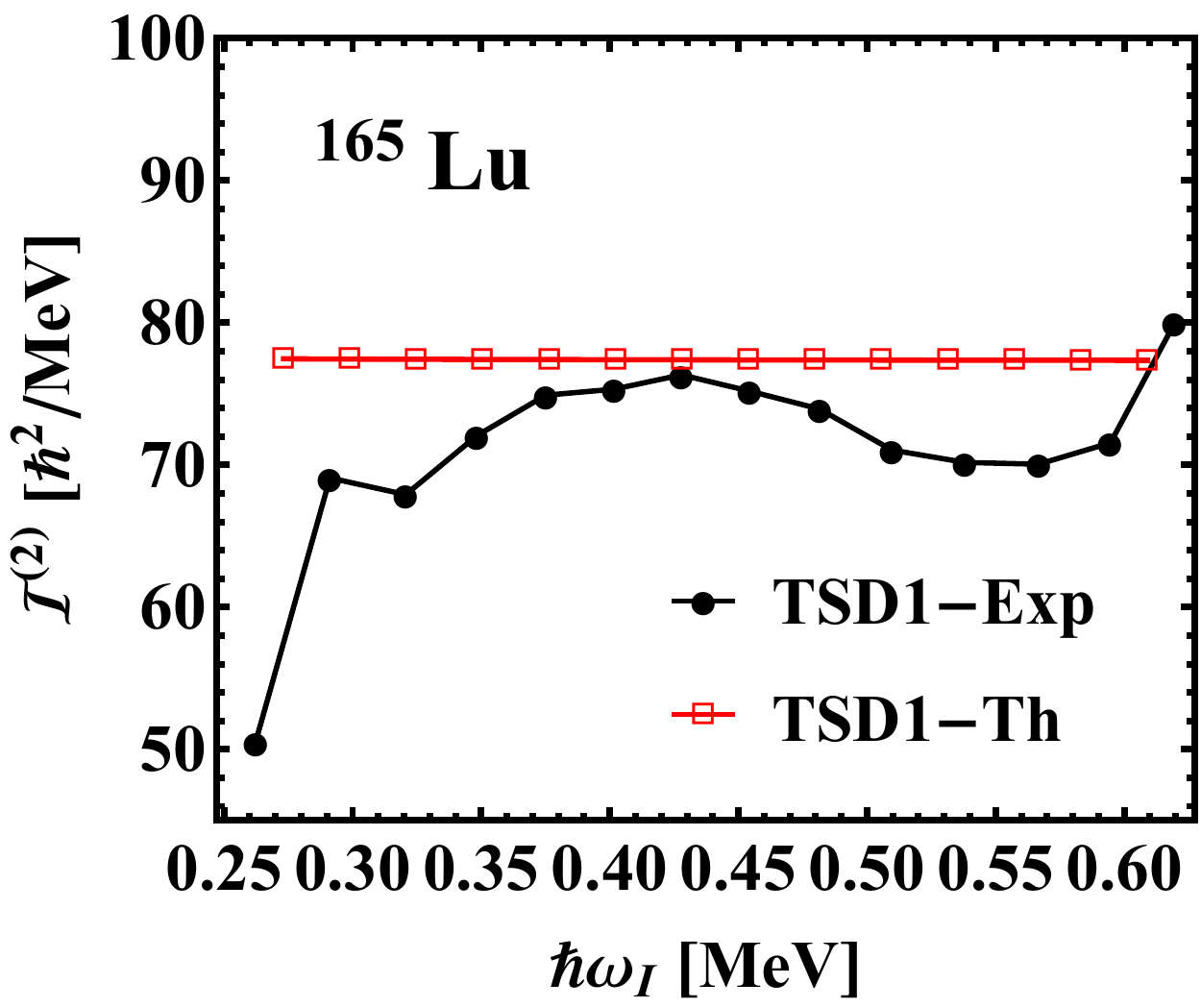}\includegraphics[width=0.24\textwidth]{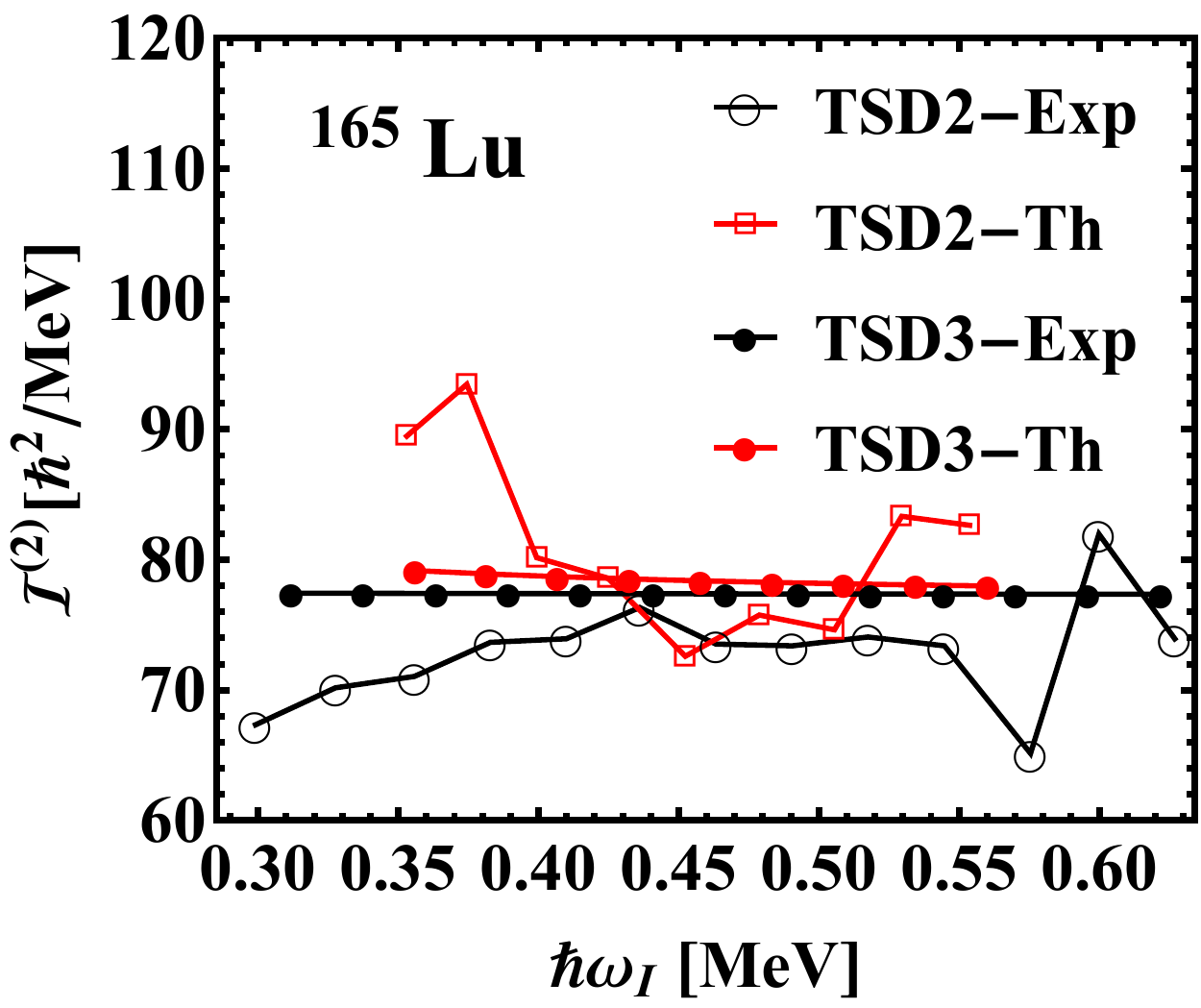}
\caption{(Color online)The calculated dynamic MoI's in the bands TSD1, TSD2 and TSD3 of $^{165}$Lu are compared with the corresponding experimental data taken from Ref.\cite{Scho}.}
\label{Fig.17}
\end{figure}

\begin{figure}
\includegraphics[width=0.25\textwidth]{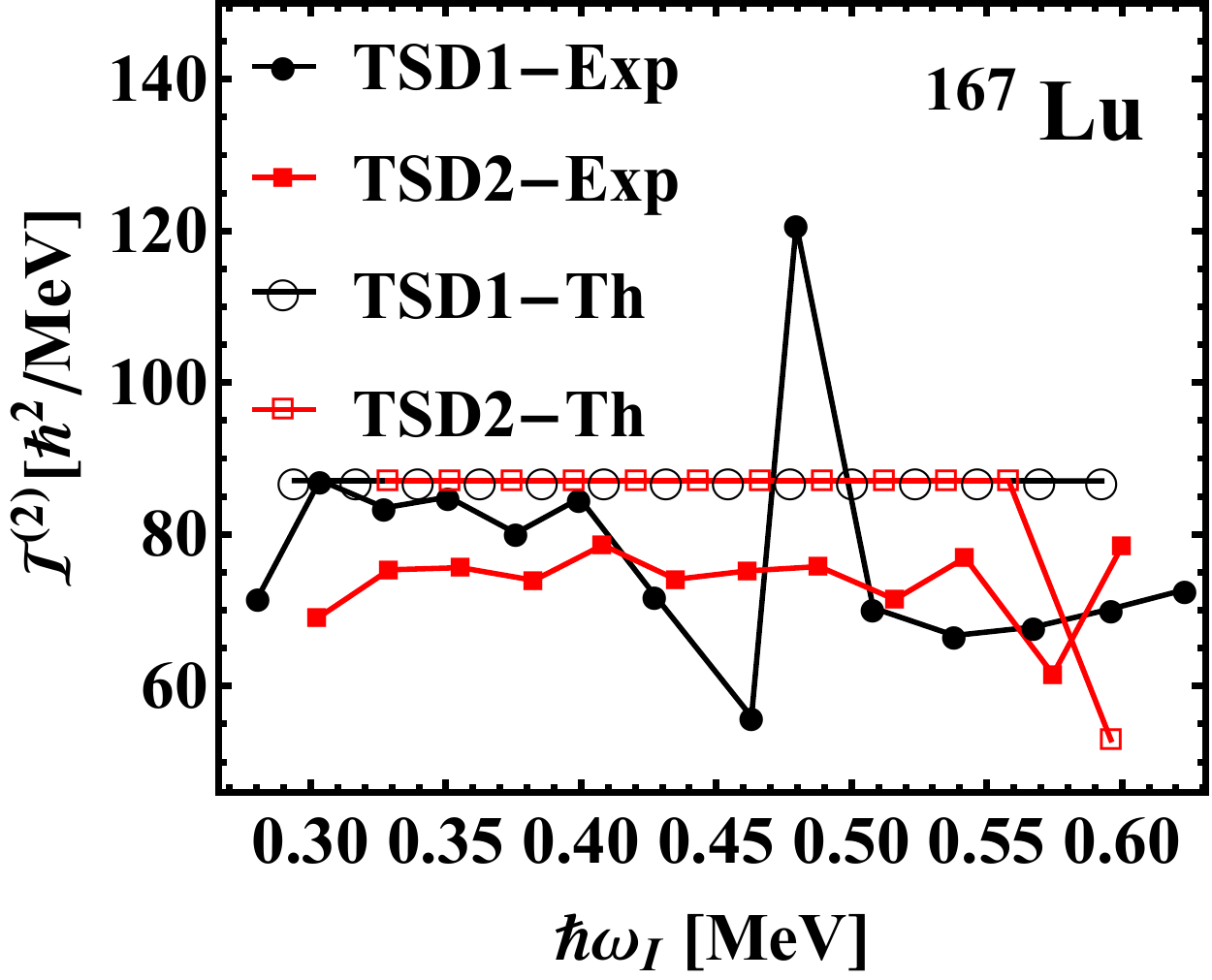}
\caption{(Color online)The calculated dynamic MoI's in the bands TSD1 and TSD2 of $^{167}$Lu are compared with the corresponding experimental data taken from Ref.\cite{Amro}.}
\label{Fig.18}
\end{figure}
\subsection{e.m. transitions}
To calculate the quadrupole electric transition probabilities we need the expression of the wave-functions describing the states involved in the given transition and the quadrupole transition operator. The available experimental data concern the states of TSD1 and TSD2. As we saw before, the level energies from these bands account for the wobbling motion through the zero point energy. Therefore the wave function is considered to be the one corresponding to the classical minimum energy, corrected by the first order expansion term, with the coordinates deviations from the minimum quantized. Thus, one arrived at the following wave function:
\begin{eqnarray}
&&\Phi^{(1)}_{IjM}={\bf N}_{Ij}\sum_{K,\Omega}C_{IK}C_{j\Omega}|IMK\rangle |j\Omega\rangle\left\{1+\right.\\
&&\left.\frac{i}{\sqrt{2}}\left[\left(\frac{K}{I}k+\frac{I-K}{k}\right)a^{\dagger}+
\left(\frac{\Omega}{j}k'+\frac{j-\Omega}{k'}\right)b^{\dagger}\right]\right\}|0\rangle_{I},\nonumber
\label{Phi}
\end{eqnarray}
with ${\bf N}_{Ij}$ standing for the normalization factor, and $|0\rangle_{I}$ for the vacuum state of the bosons $a^{\dagger}$ and $b^{\dagger}$ determined by the classical coordinates $\varphi$, 
$\psi$ and the corresponding conjugate momenta $r$ and $t$ through the canonical parameters $k$ and $k'$, which are analytically expressed in Appendix A.
 Expansion coefficients of the trial function corresponding to the minimum point, in terms of the normalized Wigner function, $C_{IK}$, were  analytically expressed in \cite{Rad018}:  
\begin{equation}
\text{C}_{IK}=\frac{1}{2^{I}}\left(\begin{matrix}2I\cr I-K\end{matrix}\right)^{1/2},\;\text{C}_{j\Omega}=\frac{1}{2^{j}}\left(\begin{matrix}2j\cr j-\Omega\end{matrix}\right)^{1/2}.
\label{CIK}
\end{equation}
The electric quadrupole transition operator is defined by:
\begin{equation}
{\cal M}(E2,\mu)=\left[Q_0D^2_{\mu0}-Q_2(D^2_{\mu2}+D^2_{\mu-2})\right]
+e\sum_{\nu=-2}^{2}D^2_{\mu\nu}Y_{2\nu}r^2,
\end{equation}
with $Q_0$ and $Q_2$ taken as free parameters which are to be fixed by fitting two intra-band transitions for TSD1 and TSD2. The results are shown in Fig.19, where one remarks on the change of the 
$Q_0$ and $-Q_2$ ordering at A=163. Again this might be a signal for a phase transition.
Note that MoI's are free parameters, that is, no option for  their nature, rigid or hydrodynamic, is adopted. To be consistent with this picture, the strengths  $Q_0$ and $Q_2$ were also considered as free parameters. However, this is not consistent with the structure of the single particle potential, which considers the collective quadrupole operator as emerging from the hydrodynamic model. These are fixed by fitting the B(E2) values for one intra-band (TSD1) and one inter-band  ($TSD2\to TSD1$) transition.  The remaining B(E2) transitions and the quadrupole transition moments, listed in Tables II-V, are free of any adjustable parameter.
Results for the B(E2) values are compared with the corresponding data in Tables III-VI.
\begin{figure}[t!]
\includegraphics[width=0.25\textwidth]{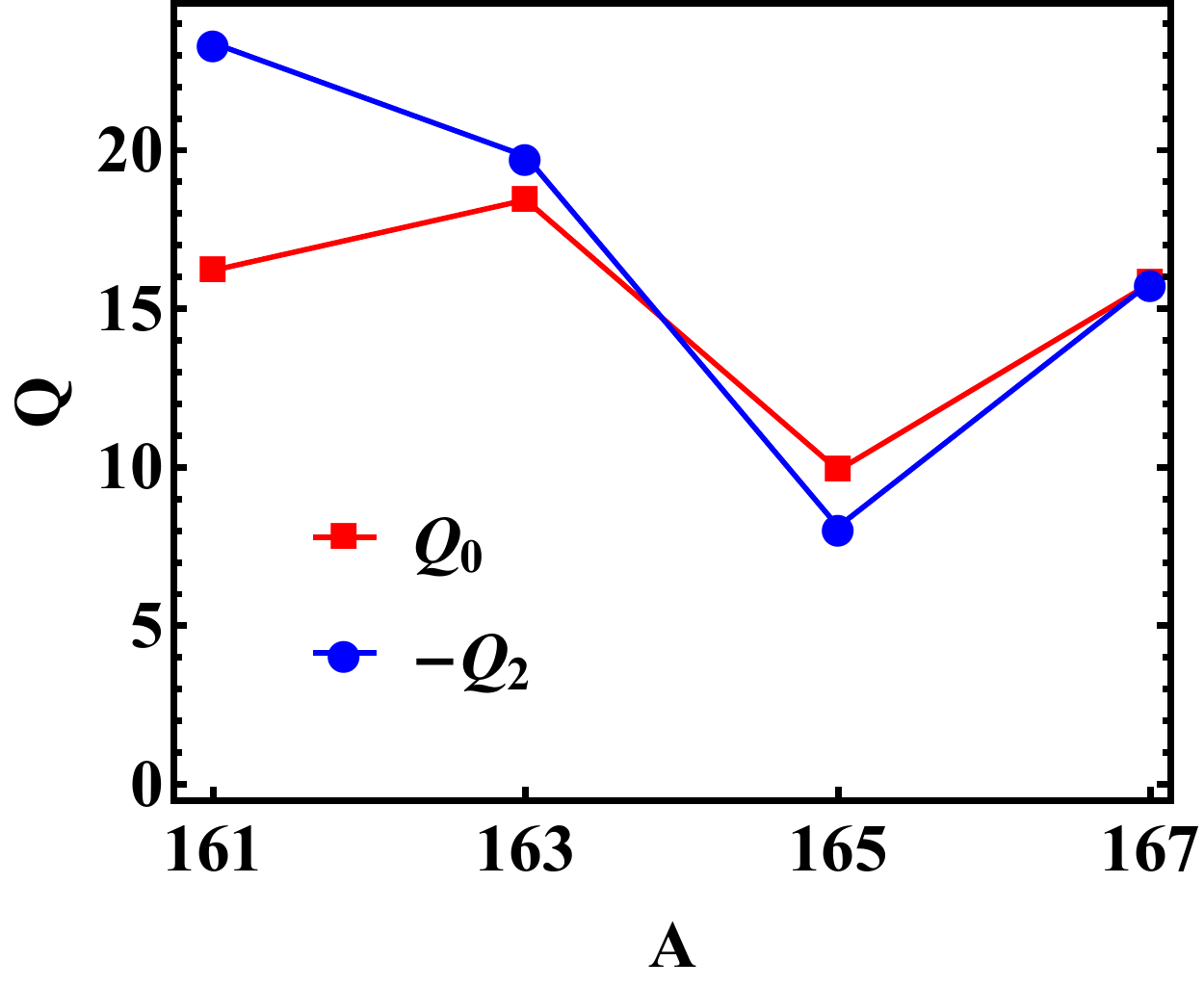}
\caption{(Color online)The parameters $Q_0$ and $Q_2$ involved in the quadrupole transition operator are given in units of $eb$, as function of the mass number A.}
\label{Fig.19}
\end{figure}

As mentioned before, the resulting strength $Q_0$ and $Q_2$ 
are not consistent with the structure of the single particle potential. Indeed, from their ratio one can extract the triaxial shape parameter $\gamma$ assuming,
 that $-Q_2/Q_0=\tan (\gamma)/\sqrt{2}$, valid in the hydrodynamic model, also holds within the present approach. In this way one determines $\gamma$ to be close to $60^{0}$ which is quite different from the values presented in Table I.
This does not surprise us since a similar situation is met in the Bohr-Mottelson model when the nuclear deformation $\beta$ is alternatively calculated by fitting the energy of the first $2^+$ state
and by fixing the B(E2) value corresponding to the transition $2^+\to 0^+$ \cite{David}. We tried to conciliate between the two options and first fixed $\gamma$ from the ratio $-Q_2/Q_0$. and then performed a least square fit for MoI's and V. The best fit was touched from the parameters listed in Table II. The quality of the energy  fits obtained with the two scenarios are similar  so that there is no need to remake the plots of subsections C, D and E. Although now there is a consistency of the electric transitions and the single particle potential one gets a disagreement with the microscopic descriptions, which however predict for $\gamma$ the values given in Table I. 
\begin{widetext}
\begin{table}[h!]
\begin{tabular}{|c|c|c|c|c|c|c|c|c|c|}
\hline
isotope&j& bands& ${\cal I}_1[\hbar^2/MeV]$& ${\cal I}_2[\hbar^2/MeV]$ &${\cal I}_3[\hbar^2/MeV]$& V[MeV] &$\gamma $ [degrees]&nr states&r.m.s.[MeV]\\
\hline
$^{161}$Lu&13/2&TSD1,TSD2     &77&3 &39 &0.3&64&29&0.185\\
$^{163}$Lu&13/2&TSD1,TSD2,TSD3&73.7  &  65.9  &  2.9  & 3.6 &  57&52&0.122\\
          &9/2&TSD4           &74    &  16& 2.1  & 1.7&  57&10&0.004\\
$^{165}$Lu&13/2&TSD1,TSD2,TSD3&78&18&4&1.2&49&42&0.125\\
$^{167}$Lu&13/2&TSD1,TSD2&85&2&61&0.5&55&30&0.165\\
\hline
\end{tabular}
\caption{The MoI's, the strength of the single particle potential (V), and  the triaxial parameter ($\gamma$) as provided by the adopted fitting procedure. The angles $\gamma$ are fixed from the strength of the quadrupole transition operator, $Q_0$ and $Q_{2}$ }
\label{Table 2}
\end{table}
\end{widetext}
The magnetic transition operator used in our calculations is:
\begin{equation}
 {\cal M}(M1,\mu)=\sqrt{\frac{3}{4\pi}}\mu_N\sum_{\nu=0,\pm 1}\left[g_RR_{\nu}+qg_jj_{\nu}\right]D^1_{\mu\nu},
\end{equation}
with $R_{\nu}$ denoting the components of the core's angular momentum with the corresponding gyromagnetic factor, $g_R=Z/A$, while $g_j$ is the free gyromagnetic factor for the single proton angular momentum $j(=13/2)$, which was quenched by a factor q=0.43 in order to account for the polarization effects not included in $g_j$. 
This factor takes care of the interaction of the odd-proton orbit with the currents distributed inside the core as well as of the internal structure of the proton, which may also influence its magnetic moment \cite{Iudi}.

To evaluate the transition matrix elements of the core's angular momentum, the  involved states are written in the form:
\begin{equation}
\Psi_{IM}=\frac{1}{\sqrt{2j+1}}\sum_{M_R \Omega K} C^{RjI}_{M_{R} \Omega M}C_{RK}|RM_{R}K\rangle|j\Omega\rangle .
\end{equation}
Again, the expansion coefficients of the core's wave function  in the basis of the normalized Wigner function are denoted by $C_{RK}$, and have the expression given by Eq. (\ref{CIK}).
Results for the relevant $B(M1)$ values of the inter-band transitions as well as for the mixing rations are collected in Table 3. Here the coordinate fluctuations around their minima are ignored since their contribution is negligible.
{\scriptsize
\begin{widetext}
\begin{table}
\begin{tabular}{|c|c|cc|cc|c|c|cc|cc|}
\hline
&&\multicolumn{2}{c|}{$B(E2;I^+\to (I-2)^+)$} &\multicolumn{2}{c|}{$Q_I$}&&&\multicolumn{2}{c|}{$B(E2;I^+\to (I-2)^+)$} &\multicolumn{2}{c|}{$Q_I$}\\
&&\multicolumn{2}{c|}{$[e^2b^2]$}    &\multicolumn{2}{c|}{$[b]$}&&&\multicolumn{2}{c|}{$[e^2b^2]$}    &\multicolumn{2}{c|}{$[b]$}  \\
\hline
TSD1&$I^{\pi}$&Th.  &  Exp.& Th. &Exp. &TSD2&$I^{\pi}$            &   Th.       &   Exp.   &  Th.   &  Exp.\\
\hline
&$\frac{41}{2}^+$&2.80&        &8.89&                       &&$\frac{47}{2}^+$&2.84&               &8.92&                \\
&$\frac{45}{2}^+$&2.83&        &8.91&                      &&$\frac{51}{2}^+$&2.86&                &8.93&                   \\
&$\frac{49}{2}^+$&2.85&        &8.92&                      &&$\frac{55}{2}^+$&2.88&                &8.95&                      \\
$^{161}$Lu&$\frac{53}{2}^+$&2.87&&8.94&                    &&$\frac{59}{2}^+$&2.89&                &8.96&                       \\
&$\frac{57}{2}^+$&2.88&         &8.95&                     &&$\frac{63}{2}^+$&2.54&                &8.97&                     \\
&$\frac{61}{2}^+$&2.90&         &8.96&                     &&$\frac{67}{2}^+$&2.51&                &8.98&                     \\
&$\frac{65}{2}^+$&2.91&         &8.97&                      &&$\frac{71}{2}^+$&2.49&               &8.98&                     \\
&$\frac{69}{2}^+$&2.92&         &8.98&                      &&               &    &                      &         &                  \\
\hline
&$\frac{41}{2}^+$&2.80&3.45$^{+0.80}_{-0.69}$&8.89&9.93$^{+1.14}_{-0.99}$&&$\frac{47}{2}^+$&2.71&2.56$^{+0.57}_{-0.44}$&8.71&8.51$^{+0.95}_{-0.73}$\\
&$\frac{45}{2}^+$&2.74&3.07$^{+0.48}_{-0.43}$&8.77&9.34$^{+0.72}_{-0.65}$&&$\frac{51}{2}^+$&2.66&2.67$^{+0.41}_{-0.33}$&8.62&8.67$^{+0.66}_{-0.53}$\\
&$\frac{49}{2}^+$&2.69&2.45$^{+0.28}_{-0.25}$&8.66&8.32$^{+0.47}_{-0.42}$&&$\frac{55}{2}^+$&2.62&2.81$^{+0.53}_{-0.41}$&8.53&8.88$^{+0.83}_{-0.64}$\\
$^{163}$Lu&$\frac{53}{2}^+$&2.64&2.84$^{+0.24}_{-0.22}$&8.57&8.93$^{+0.38}_{-0.35}$&&$\frac{59}{2}^+$&2.58&2.19$^{+0.94}_{-0.65}$&8.46&7.82$^{+1.66}_{-1.15}$\\
&$\frac{57}{2}^+$&2.60&2.50$^{+0.32}_{-0.29}$&8.50&8.37$^{+0.54}_{-0.49}$&&$\frac{63}{2}^+$&2.54&2.25$^{+0.75}_{-0.48}$&8.39&7.91$^{+1.32}_{-0.84}$\\
&$\frac{61}{2}^+$&2.56&1.99$^{+0.26}_{-0.23}$&8.43&7.45$^{+0.49}_{-0.43}$&&$\frac{67}{2}^+$&2.51&1.60$^{+0.52}_{-0.37}$&8.34&6.66$^{+1.09}_{-0.76}$\\
&$\frac{65}{2}^+$&2.53&1.95$^{+0.44}_{-0.30}$&8.36&7.37$^{+0.82}_{-0.57}$&&$\frac{71}{2}^+$&2.49&1.61$^{+0.82}_{-0.49}$&8.28&6.68$^{+1.70}_{-1.02}$\\
&$\frac{69}{2}^+$&2.50&2.10$^{+0.80}_{-0.48}$&8.31&7.63$^{+1.46}_{-0.88}$ &&               &    &                      &         &                  \\
\hline
&$\frac{41}{2}^+$&3.63&&10.12&&&$\frac{47}{2}^+$&3.68&&10.15&\\
&$\frac{45}{2}^+$&3.66&&10.14&&&$\frac{51}{2}^+$&3.71&&10.17&\\
&$\frac{49}{2}^+$&3.69&&10.16&&&$\frac{55}{2}^+$&3.73&&10.19&\\
$^{165}$Lu&$\frac{53}{2}^+$&3.72&&10.18&&&$\frac{59}{2}^+$&3.75&&10.20&\\
&$\frac{57}{2}^+$&3.77&&10.19&&&$\frac{63}{2}^+$&3.77&&10.21&\\
&$\frac{61}{2}^+$&3.76&&10.21&&&$\frac{67}{2}^+$&3.78&&10.22&\\
&$\frac{65}{2}^+$&3.77&&10.22&&&$\frac{71}{2}^+$&3.79&&10.23&\\
&$\frac{69}{2}^+$&3.79&&10.23&&&               &    &                      &         &                  \\
\hline
&$\frac{41}{2}^+$&2.80&&8.90&&&$\frac{47}{2}^+$&2.84&&8.92&\\
&$\frac{45}{2}^+$&2.83&&8.91&&&$\frac{51}{2}^+$&2.86&&8.94&\\
&$\frac{49}{2}^+$&2.85&&8.92&&&$\frac{55}{2}^+$&2.88&&8.95&\\
$^{167}$Lu&$\frac{53}{2}^+$&2.87&&8.94&&&$\frac{59}{2}^+$&2.89&&8.96&\\
&$\frac{57}{2}^+$&2.88&&8.95&&&$\frac{63}{2}^+$&2.90&&8.97&\\
&$\frac{61}{2}^+$&2.90&&8.96&&&$\frac{67}{2}^+$&2.92&&8.98&\\
&$\frac{65}{2}^+$&2.91&&8.97&&&$\frac{71}{2}^+$&2.93&&8.98&\\
&$\frac{69}{2}^+$&2.92&&8.98&&&               &    &                      &         &                  \\
\hline
\end{tabular}
\caption{The E2 intra-band transitions $I\to (I-2)$ for TSD1 and TSD2 bands. Also, the transition quadrupole moments, defined as in Ref.\cite{Hage1}, are given. Theoretical results (Th.) are compared with the corresponding experimental data (Exp.) taken from Ref. \cite{Gorg}. B(E2) values are given in units of $e^2b^2$,
while the quadrupole transition moment in $b$.}
\label{Table 3}
\end{table}
\end{widetext}}
\clearpage

{\scriptsize
\begin{table}
\begin{tabular}{|c|c|cc|cc|cc|}
\hline
&&\multicolumn{2}{c|}{$B(E2)[e^2b^2]$} &\multicolumn{2}{c|}{$B(M1)[\mu_N^2]$}&\multicolumn{2}{c|}{$\delta_{I\to(I-1)}$}\\
&&\multicolumn{2}{c|}{$I^+\to (I-1)^+$}&\multicolumn{2}{c|}{$I^+\to (I-1)^+$}&\multicolumn{2}{c|}{$[MeV.fm]$}  \\
&$I^{\pi}$&Th.  &  Exp.& Th. &Exp. &Th.&Exp.\\
\hline
&$\frac{47}{2}^+$&0.54&&0.018&&-1.55&\\
&$\frac{51}{2}^+$&0.47&&0.018&&-1.56&\\
$^{161}$Lu&$\frac{55}{2}^+$&0.42&&0.019&&-1.57&\\
&$\frac{59}{2}^+$&0.37&&0.019&&-1.58&\\
&$\frac{63}{2}^+$&0.33&&0.020&&-1.59&\\
\hline
&$\frac{47}{2}^+$&0.54&0.54$^{+0.13}_{-0.11}$&0.017&0.017$^{+0.006}_{-0.005}$&-1.55&-3.1$^{+0.36}_{-0.44}$\\
&$\frac{51}{2}^+$&0.49&0.54$^{+0.09}_{-0.08}$&0.018&0.017$^{+0.005}_{-0.005}$&-1.58&-3.1$\pm 0.4$$^{a)}$\\
$^{163}$Lu&$\frac{55}{2}^+$&0.44&0.70$^{+0.18}_{-0.15}$&0.019&0.024$^{+0.008}_{-0.007}$&-1.61&-3.1$\pm 0.4$$^{a)}$\\
&$\frac{59}{2}^+$&0.34&0.65$^{+0.34}_{-0.26}$&0.019&0.023$^{+0.013}_{-0.011}$&-1.64&-3.1$\pm 0.4$$^{a)}$\\
&$\frac{63}{2}^+$&0.36&0.66$^{+0.29}_{-0.24}$&0.020&0.024$^{+0.012}_{-0.010}$&-1.66&\\
\hline
&$\frac{47}{2}^+$&0.37&&0.018&&-1.32&\\
&$\frac{51}{2}^+$&0.34&&0.018&&-1.34&\\
$^{165}$Lu&$\frac{55}{2}^+$&0.32&&0.019&&-1.36&\\
&$\frac{59}{2}^+$&0.29&&0.019&&-1.38&\\
&$\frac{63}{2}^+$&0.27&&0.020&&-1.40&\\
\hline
&$\frac{39}{2}^+$&0.66&&0.016&&-1.67&$-3.1^{+1.1}_{-3.4}$\\
&$\frac{47}{2}^+$&0.54&&0.018&&-1.65&$-5.1^{-1.6}_{-2.5}$\\
&$\frac{51}{2}^+$&0.49&&0.018&&-1.65&$-3.9^{+2.7}_{-8.4}$\\
$^{167}$Lu&$\frac{55}{2}^+$&0.45&&0.019&&-1.65&\\
&$\frac{59}{2}^+$&0.41&&0.019&&-1.65&\\
&$\frac{63}{2}^+$&0.38&&0.020&&-1.65&\\
\hline
\end{tabular}
\caption{The B(E2) and B(M1) values for the transitions from TSD2 to TSD1. Mixing ratios are also mentioned. Theoretical results (Th.) are compared with the corresponding experimental (Exp.) data taken from Ref.\cite{Gorg,Jens01}. Data labeled by $^{a)}$ are from Ref.\cite{Reich}.}
\label{Tabel 4}
\end{table}}

{\scriptsize
\begin{table}
\begin{tabular}{|c|c|cc|}
\hline
&&\multicolumn{2}{c|}{$B(E2)_{out}/B(E2)_{in}$} \\
&I$^{\pi}$&Th.& Exp. \\
\hline
&$\frac{31}{2}$              & 0.29           &0.21$\pm 0.11$ \\
&$\frac{35}{2}$              & 0.26           &0.22$\pm 0.02$  \\
&$\frac{39}{2}$              & 0.24           &0.21$\pm 0.02$  \\
&$\frac{43}{2}$              & 0.22           &0.22$\pm 0.02$  \\
$^{163}$Lu&$\frac{47}{2}$              & 0.20           &0.21$\pm 0.03$  \\
&$\frac{51}{2}$              & 0.18           &0.21$\pm 0.02$  \\
&$\frac{55}{2}$              & 0.17           &0.26$\pm 0.05$  \\
&$\frac{59}{2}$              & 0.15           &0.30$\pm 0.09$  \\
&$\frac{63}{2}$              & 0.14           &0.30$\pm 0.11$  \\
\hline
&$\frac{39}{2}^{+}$          &0.12&0.17$\pm$0.05\\
$^{165}$Lu&$\frac{43}{2}^{+}$&0.11&0.16$\pm$0.03\\
&$\frac{47}{2}^{+}          $&0.10&0.22$\pm$0.08\\
\hline
&$\frac{39}{2}^{+}          $&0.24&0.23$^{+0.02}_{-0.05}$\\
$^{167}$Lu&$\frac{43}{2}^{+}$&0.21&0.26$^{+0.03}_{-0.04}$\\
&$\frac{47}{2}^{+}          $&0.19&0.27$^{+0.02}_{-0.10}$\\
\hline
\end{tabular}
\caption{Branching ratios of some  states from the band TSD2. Experimental data are from Refs.\cite{Hage,Scho,Amro,Jens,Jensen}}
\label{Table 5}
\end{table}}

{\scriptsize
\begin{table}
\begin{tabular}{|c|c|cc|}
\hline
&&\multicolumn{2}{c|}{$B(M1)/B(E2)_{in}[10^2\frac{\mu_N^2}{e^2b^2}]$} \\
&I$^{\pi}$&Th.& Exp. \\
\hline
&$\frac{35}{2}$              & 0.502           &0.439$^{+0.082}_{-0.076}$ \\
&$\frac{39}{2}$              & 0.560           &0.447$^{+0.077}_{-0.078}$  \\
$^{163}$Lu&$\frac{43}{2}$    & 0.608           &0.509$^{+0.088}_{-0.086}$  \\
&$\frac{47}{2}$              & 0.650           &0.498$^{+0.091}_{-0.084}$  \\
&$\frac{51}{2}$              & 0.685          &0.709$^{+0.182}_{-0.196}$  \\
\hline
\end{tabular}
\caption{Results for the ratio  $B(M1)/B(E2)_{in}$ are compared with the corresponding experimental data \cite{Jens}for a few TSD2 levels from  $^{163}$Lu.}
\label{Table 6}
\end{table}}

One specific feature for the wobbling motion consists of a strong $E2$ transition from the TSD2 to the TSD1 bands. This is reflected by the relative large values of the branching ratios characterizing the states from TSD2. This is confirmed by Table 4, where the calculated branching ratios are compared with the corresponding experimental data. Also, the computed ratio $B(M1)/B(E2)_{in}$ are in good agreement with the experimental data in $^{163}$Lu. Another specific wobbling feature is the large transition quadrupole moment, as shown in Table II. From there one can see a very good agreement of our calculation results for $^{163}$Lu, and the coresponding data.
Concluding the application part of the present paper, one may say that the proposed semi-phenomenological approach seems to be an efficient tool to account for the main features of electromagnetic properties  of the even-odd $Lu$-isotopes.
\renewcommand{\theequation}{4.\arabic{equation}}
\setcounter{equation}{0}
\label{sec:level4}

\section{Phase diagram}
Here we shall consider the phase diagram associated with the classical energy function ${\cal H}$, for a given total angular momentum. From the equations of motion written in the Hamilton canonical form, it results that the angles play the role of the classical generalized coordinate, while  the variables $r$ and $t$ are the corresponding conjugate momenta. In virtue of this we may denote, more suggestively, the canonical conjugate coordinates as:
\begin{equation}
q_1=\varphi,\; q_2=\psi,\; p_1=r,\;p_2=t.
\end{equation}
The critical manifolds associated to the classical energy function are determined from the equation:
\begin{equation}
\det \left(\frac{\partial ^2{\cal H}}{\partial (q_i)^{k}\partial (p_j)^{l}}\right)=0;\;i,j,=1,2; k,l=0,1,2;k+l=2.
\end{equation}
After some algebraic manipulations,  the above equation leads to:
\begin{equation}
C=0,
\label{Ceq0}
\end{equation}
where C has the expression from Eq.(\ref{BandC}). From (\ref{equOm}) it is obvious that for this value of C the lower solution is vanishing. Thus, Eq.(\ref{Ceq0}) defines a Goldstone mode which suggests a transition to a new nuclear phase \cite{Gold}. Eq.(\ref{Ceq0}) splits to the following two equations:
\begin{eqnarray}
z&=& f_1(x),\nonumber\\
z&=& f_2(x,y),
\label{f1andf2}
\end{eqnarray}
with
{\scriptsize
\begin{eqnarray}
\hspace*{-0.4cm}f_1(x)&=&\left[\left(1-4(I-j)^2\right)x^2+\left(4I^2+4j^2-8Ij+2j+2I-2\right)x\right.\nonumber\\
&-&\left.(2I+2j-1)\right]\left[G_1\left((2I-2j-1)x-(2I-1)\right)\right]^{-1},\nonumber\\
\hspace*{-0.4cm}f_2(x,y)&=&\left[\left(1-4(I-j)^2\right)x^2+\left(1-2(I+j)\right)y^2+2\left(2(I-j)^2\right.\right.\nonumber\\
&+&\left.\left.(I+j)+1\right)xy\right]\left[G_2\left((2I-2j-1)x-(2I-1)y\right)\right]^{-1}.
\end{eqnarray}}
Here the following notations were used:
\begin{equation}
x=\frac{A_1}{A_3},\;\;y=\frac{A_2}{A_3},\;\;z=\frac{V}{A_3}.
\label{xandy}
\end{equation}
For a fixed $\gamma (=17^{o})$, the equations (\ref{f1andf2}) represent two singular surfaces, having the asymptotic planes:
\begin{equation}
x=\frac{2I-1}{2I-2j-1},\;\;y=\frac{2I-2j-1}{2I-1}x.
\label{planesxandy}
\end{equation}

On the other hand, we recall \cite{Rad018} that the wobbling frequencies are obtainable by a quadratic  expansion of the energy function around the minimum point, which results in getting a Hamiltonian for two coupled oscillators. Quantizing the independent oscillators, the coupling term is diagonalized through a canonical transformation. Thus, the same frequencies as given by 
Eq.(\ref{equOm}) are obtained. The frequencies for the uncoupled oscillators are real, provided the following restrictions hold (see Appendix A):
\begin{eqnarray}
\hspace*{-0.2cm}&&S_{Ij}A_1<A_2<A_3\;\; {\rm{or}}\;\;S_{Ij}A_1<A_3<A_2,\nonumber\\
\hspace*{-0.2cm}&&A_3>T_{Ij}A_1-\frac{G_1V}{2j-1},\;\;A_2>T_{Ij}A_1-\frac{G_2V}{2j-1},
\label{inequaA123}
\end{eqnarray}
with
\begin{equation}
S_{Ij}=\frac{2I-1-2j}{2I-1},\;\;T_{Ij}=\frac{2j-1-2I}{2j-1},
\end{equation}
and $G_1, G_2$ defined, as in Appendix A, through Eq.(\ref{STG1G2}). If $V>0$, the inequalities from the second row of (\ref{inequaA123}) are always satisfied.
The intervals given in the first line of (\ref{inequaA123}) together with the surfaces (\ref{f1andf2}) define, in the parameter space, sectors bordered by separatrices which, in fact, determine the nuclear phases.
Pictorially, the phase diagram is presented in Fig. 20, where the separatrices are shown for a given angular momentum, $I=45/2$. Therein, the planes $x=y,\;x=0,\;y=0$ are also shown; they are associated with the axial symmetric cases,  and therefore are forbidden. The planes asymptotic to the surfaces (\ref{f1andf2}) are shown too. 
\begin{figure}[ht!]
\hspace*{-0.4cm}\includegraphics[width=0.55\textwidth]{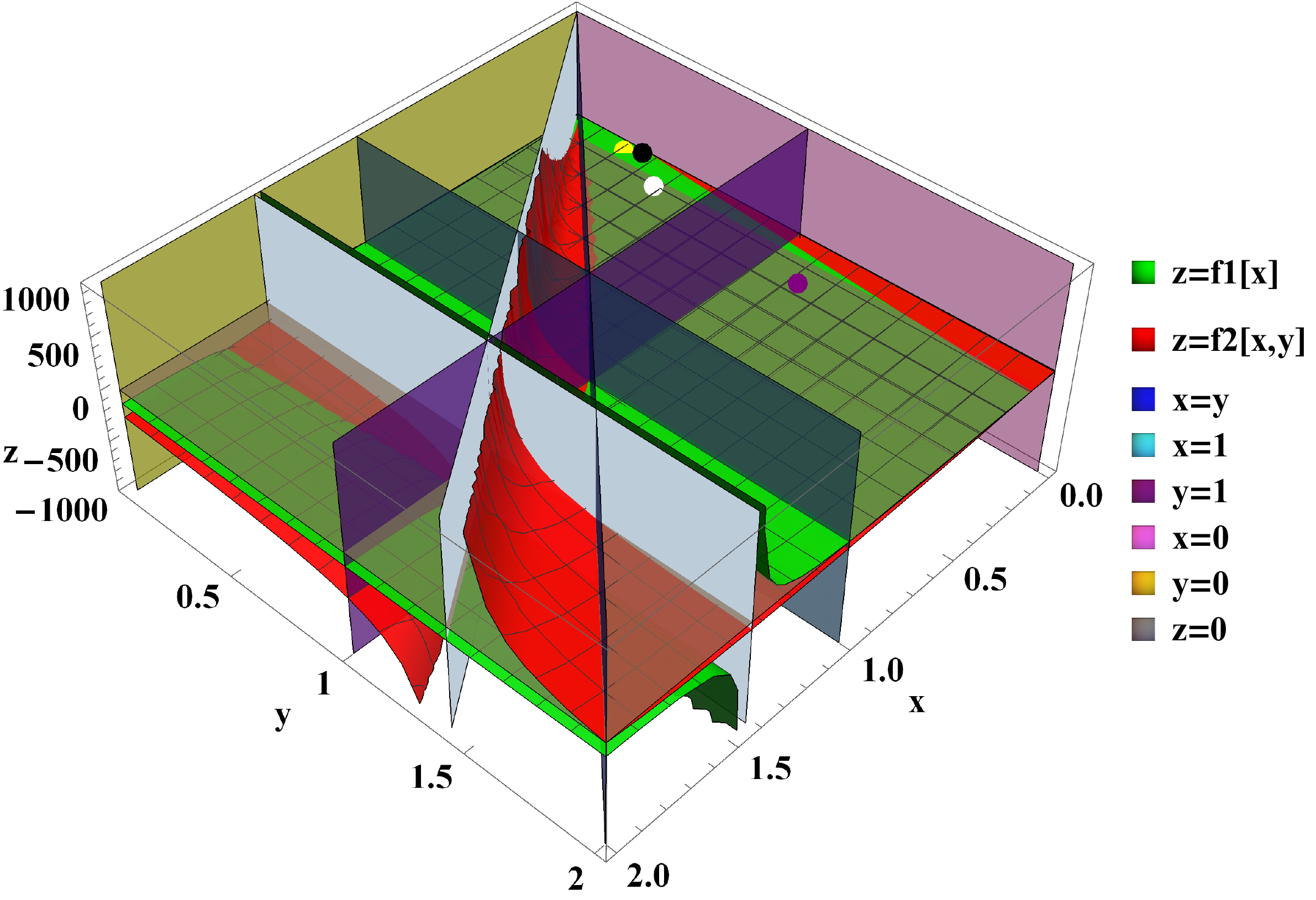}
\caption{(Color online)The phase diagram for a j-particle-triaxial rotor coupling Hamiltonian with j=13/2 and I=45/2.The coordinates x,y and z are a-dimensional.}
\label{Fig.20}
\end{figure}
Some separatrices depend on $\gamma$ through the functions $G_1$ and $G_2$.  Fig.20 is associated to $^{163}$Lu, while for the other isotopes the pictures look similarly, but the mentioned separatrices are slightly modified. 
For the fixed MoI's and V shown in Table I, the coordinates $(x,y,z)$ corresponding to the four isotopes respectively, are represented by small circles of different colors: purple ($^{161}$Lu)
, white ($^{163}$Lu), yellow ($^{165}$Lu) and black ($^{167}$Lu). In the case of $^{161}$Lu, the coordinate y is too large and therefore drops out the range shown in Fig.20. In order to keep it inside the figure we modified y to y-7. Even so, the purple circle falls in an adjacent phase, which is consistent with Fig. 1 suggesting that this isotope belongs to a different nuclear phase. Inside a given phase the classical Hamiltonian has specific stationary points. If one of these is a minimum, then the classical trajectories surround it with a certain time period. If the point in the parameter space approaches the separatrices, the period tends to infinity 
\cite{Rad98}. When $V>0$, ${\bf j}$ is always oriented along the short axis, that is the one-axis,  and the region where ${\cal I}_2>{\cal I}_1>{\cal I}_3$ is the phase where the transversal wobbling may take place. More specifically, this region is bounded by four planes, one being the diagonal plane, one is given by the second Eq. (\ref{planesxandy}), one is the plane $x=1$, and the fourth one is the plane $y=1$. There are other two planes bordering the phase of interest defined in Appendix B in Eq.(\ref{restr}). For this region we have to depict the minimum of ${\cal H}$, if that exists. If ${\cal H}$ exhibits, indeed, a minimum in the considered sector for $\gamma$, i.e. $[0^0,60^0]$, further, the frequencies describing the small oscillations around the found minimum are to be determined. Actually, the mentioned project is partially already accomplished in Appendix B. {\it The results from there confirm the existence of a transversal mode but for  ideal restrictions \cite{Frau}, while within the Holstein-Primakoff description, the minimum for energy surface reflecting a transversal wobbling regime does not exist\cite{Tan017}, if one keeps all energy terms. Therefore, there is no contradiction between the two formalisms \cite{Frau018,Tana018}, since they deal with different Hamiltonians. Moreover, if in each of the two formalisms as well as in the present one, one keeps all terms from the starting Hamiltonian, it seems that no solution for the transversal wobbling exists.} Although in our case the transversal mode is determined by a part of the starting Hamiltonian, for the time being we cannot say whether this ideal picture is preserved when the remaining interaction is accounted for or is totally spoiled by the Coriolis interaction. An indirect answer to this question is, actually, provided by the wobbling energy behavior as function of spin, this being considered as a signature for the wobbling character. Contrary to what is stated in Ref.\cite{Frau}, in the present approach the monotony of the experimental and theoretical curves are the same, namely both are increasing functions of the angular momentum. This, in fact, confirms that the considered isotopes are longitudinal wobblers.

It is worth noting that for $z<0$, the
three-axis, the long one, is energetically favored in aligning j. Therefore, another region where the transversal wobbling motion may show up, is bordered by the planes $x=0,\;y=1$, by the asymptotic plane for the surface $z=f_1(x)$, and below  the surface $z=f_2(x,y)$. In the region between the two surfaces $z=f_1(x)$ and $z=f_2(x,y)$,
the motion of the odd system is not allowed. Indeed, there $C<0$ and consequently the phonon lower frequency becomes imaginary.
\begin{widetext}
\begin{figure}[ht!]
\hspace*{-0.4cm}\includegraphics[width=0.5\textwidth]{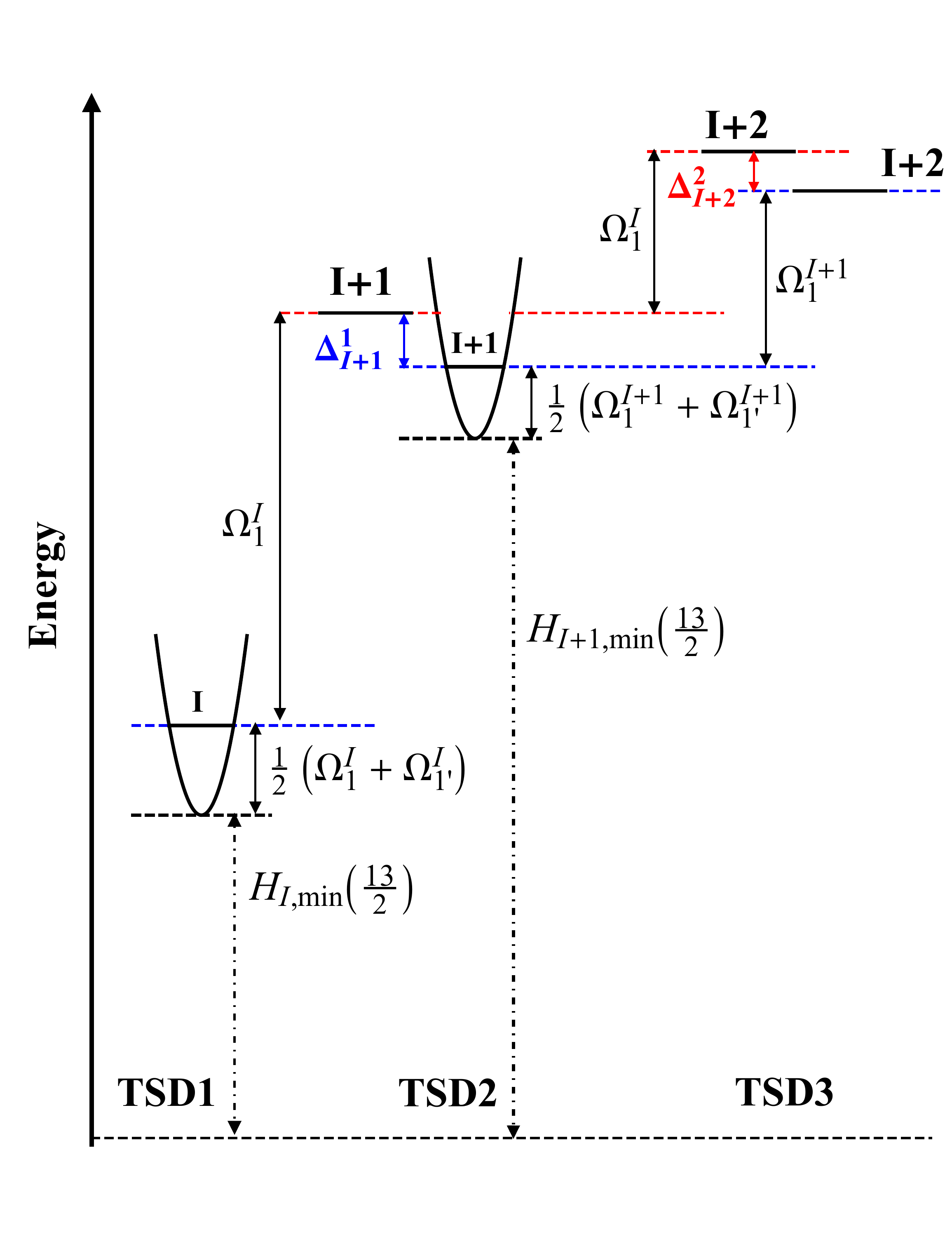}\includegraphics[width=0.4\textwidth]{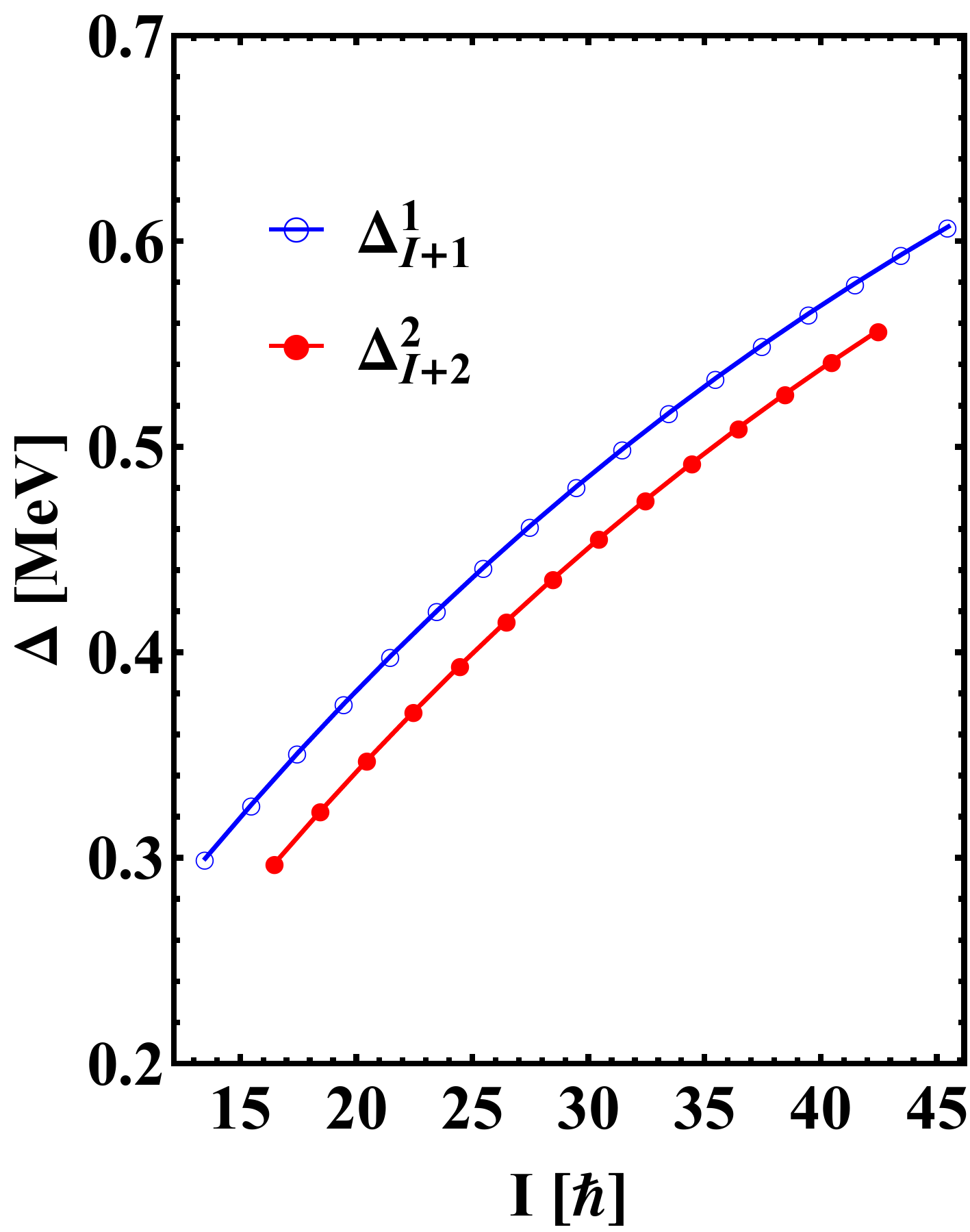}
\caption{(Color online) Left panel:Suggestive comparison of the I-state energies, taken in arbitrary units, obtained with the actual and previous approaches. Right panel: The energy difference for the states of the same angular momenta in the bands TSD2 ($\Delta_{I+1}^{1}$) and TSD3 ($\Delta_{I+2}^{2}$) within the two descriptions, respectively.}
\label{Fig.21}
\end{figure}
\end{widetext}
\renewcommand{\theequation}{5.\arabic{equation}}
\setcounter{equation}{0}
\label{sec:level5}
\section{Conclusion}
The main results of our investigation are: i) The classical equations of motion are obtained through a time dependent variational principle with a trial function which is a product of two coherent states, one associated to the core and one to the odd nucleon. The coherent states depend on two complex parameters depending on time which play the role of classical canonical conjugate coordinates. The classical equations are brought to the Hamilton form by a suitable transformation. Linearizing the equations around the coordinates which make the energy function minimum, one determines a dispersion equation for two normal wobbling modes. The corresponding frequencies are used to define the rotational bands; they depend on five parameters, which are fixed so that the experimental excitation energies in the bands TSD1,TSD2,TSD3 and TSD4 are at best reproduced. In such a way the model parameters, the MoI's and the strength of the particle-core interaction, are determined. We underline the fact that here the bands TSD1,TSD2 and TSD4 are collections of energy levels describing the ground states corresponding to distinct sets of angular momenta. By contrast, the band TSD3 has a one phonon wobbling character and is obtained by exciting each of the TSD2 states by a specific one phonon operator.
ii)The e.m. transition probabilities are calculated  and compared with the available experimental data. The numerical analysis shows a very good agreement between our results and the corresponding experimental data. iii) Other observables calculated, with the result compared to the corresponding data are: the alignment, the relative energy with respect to a reference energy, the dynamic moment of inertia. The result's analysis concludes on the wobbling character of the studied magnitudes.
iv) In the parameter space one defines manifolds bordered by separatrices defined by equating the associated Hessian to zero. One identifies the regions where a longitudinal or a transversal wobbling mode may show up. Also, the regions where the system motion is forbidden, are pointed out.
v) In Appendix B it is shown that the transversal wobbling motion is possible under some important restrictions and is incompatible with realistic Hamiltonians, where the Coriolis interaction prevails over the particle-deformation coupling. In that respect, the present paper confirms the result of \cite{Tan017}, claiming that, for a realistic Hamiltonian, the minimum of the energy function, specific for a transversal wobbling, does not exist and therefore no solution for such a motion is possible. In the present paper, this conclusion is strengthen by the wobbling energy calculations showing that both the theoretical and experimental curves have an increasing behavior with respect to the angular momentum, which might indicate that the $^{161,163,165,167}$Lu isotopes are longitudinal wobblers.

Note that in the present paper the moments of inertia are fixed by a fitting procedure, which accounts for the global structure of the considered bands. In that respect, the obtained MoI's have an effective character  including the effect coming from the particle-core coupling, which leads to a longitudinal wobbling regime where the whole system rotates around the short axis. But this is the final picture and therefore we cannot state that this emerges from a transversal wobbling motion where the odd particle coupling to the core deformation prevails over the Coriolis-like interaction. To make the transition from the transversal to the longitudinal wobbling regime explicit, one needs to infer a spin dependence for the MoI's, that might be determined by the renormalization effect due to the particle-core interaction. However, such a feature cannot be touched within the present approach, since, indeed, the linear terms in the core's angular momentum are not used to renormalize the MoI's, in an approximative way, but treated semiclassically on equal footing with the quadratic ones.

Finally, a short comment on the two band definitions is in order.
In the previous approaches, where the bands TSD2 and TSD3 bands have a one and two phonon character, there is an inconsistency in the treatment of the three bands. While TSD1 states are determined variationally, the others are obtained by adding one and two phonon excitations on the top of the former levels \cite{IHam,Raduta84}. We asked ourselves whether it is possible that at least one of the two bands be variationally obtained. We want to speculate on the fact that by diagonalizing the rigid rotor Hamiltonian one finds out an even-odd degeneracy. Even for the liquid drop with large deformation an even-odd staggering like $(2^+,3^+), (4^+,5^+), (6^+,7^+)$,...for the $\gamma$ band is currently seen, with the pair members emerging from vibrational states with number of phonons differing by one unit. On the other hand, by acting with a phonon operator on a TSD1 state the angular momentum for the mother state is increased by one unit. Thus one could try to approximate
the state $I$ from the TSD2 band in the old picture by the one obtained variationally with I=R+j, but having the core angular momentum given by R=1,3,5,....The mechanism of creating the states of TSD2 and TSD3 bands in the old and current picture is schematically suggested in Fig. 21, the left panel. The difference in energy for the levels $I+1$ and $I+2$ obtained in the old and current approach respectively, are denoted by $\Delta^{1}_{I+1}$ and $\Delta^{1}_{I+2}$. The magnitude of these differences is relatively small, as shown in the right panel of Fig. 21. Note that both energies are calculated with the moments of inertia of Table I. The agreement between the two sets of energies would be much increased if in the old formalism a separate least square fit  is performed. Obviously the output parameters would be different from those from Table I, but this is the consequence of using different approximations in the two approaches. Anyway, the result for $\Delta$'s confirms that the adopted approximation has a realistic character.

Within the ultimate cranking model (\cite{Jens}) the authors studied the possibility that the TSD2 band be associated to the unfavored signature of the $\pi i_{13/2}$ orbital. Since the signature splitting in the deepest minimum is large ($\ge 1 MeV$), it is most likely that this signature belongs to a local minimum showing up at a smaller deformation but a larger triaxiality. Due to this fact the highly excited signature partner band has properties different from those of TSD1, and also contrasts the measured features of TSD2. On the other hand in \cite{Tana3} the invariance properties of the triaxial rotor Hamiltonian and its eigen-functions with respect to the $D2$ group of transformations are studied, in order to efficiently use the Holstein-Primakoff boson expansion method.

Let us also discuss, now, the present results in terms of signature. We recall that signature is a quantum number related to the invariance of the wave function of an axially symmetric deformed nucleus with respect to a rotation by $\pi$ around an axis perpendicular to the symmetry axis. Due to the D2 group invariance of the triaxial rotor Hamiltonian, this property can be extended also to the non-axially symmetric nuclei. As we shall see in the next paragraph, the signature concept is valid also within the classical picture associated to a triaxial nucleus.

Indeed, in the case of triaxial nuclei there is no symmetry axis; however, it is well established that the system rotates around the short axis, that is the one-axis, labeled as the x axis. By convention, we shall call the z-axis as the symmetry axis  which is perpendicular on the rotation axis. In this case it is convenient to choose the x-axis as the  quantization axis. Then, the lowering angular momentum operator is defined as:
\begin{equation}
\hat{I}_-=\hat{I}_y+i\hat{I}_z;\;\;\hat{j}_-=\hat{j}_y+i\hat{j}_z.
\end{equation}
Acting on the trial function $\Psi_{Ij;M}(\rho,\varphi;t,\psi)$, defined by Eq.(2.4), with the rotation operator $\hat{R}_x(\pi)=e^{-i\pi\hat{I}_x}\otimes e^{-i\pi\hat{j}_x}$, one obtains:
\begin{eqnarray}
&&\hat{R}_x(\pi)\Psi_{Ij;M}(\rho,\varphi;t,\psi)=e^{-z\hat{I}_-}e^{-s\hat{j}_-}|IMI\rangle|j,j\rangle e^{-i\pi(I+j)}\nonumber\\
&&=\Psi_{Ij;M}(\rho,\pi+\varphi;t,\pi+\psi)(-1)^{I+j}.
\end{eqnarray}
Note that the energy function ${\cal H}$ (2.8) is invariant to the change of variables:
\begin{equation}
(\rho,\varphi;t,\psi)\to (\rho,\pi+\varphi;t,\pi+\psi).
\end{equation}
Moreover, the wave function is invariant against such a transformation, otherwise the trial function would be degenerate. Consequently, the eigenvalue of $\hat{R}_x(\pi)$ is $r_x(\pi)=e^{-i\pi(I+j)}$, hence the signature is either $1$ or $0$.
Hence, the angular momentum $I+j$ may take values from two sets: $1(mod\;2)$ and  $0(mod\;2)$.The mentioned signatures and eigenvalues are associated to the even system. From here, one extracts the signatures for the odd system to be $+\frac{1}{2}$ and $-\frac{1}{2}$, while the corresponding eigenvalues are $i$ and $-i$, respectively. The angular momenta, for the two signatures, form the sets $\frac{13}{2}, \frac{17}{2}, \frac{21}{2},...$ and $\frac{15}{2}, \frac{19}{2}, \frac{23}{2},...$,respectively. Thus, in the present formalism the band TSD1 is  characterized by the signature
$+\frac{1}{2}$ (favored), while the band TSD2 seems to be of signature $-\frac{1}{2}$ (unfavored).
The argument given in Ref.\cite{Jens} saying that the signature split is so large that it is unlikely that the higher band belongs to the deepest potential well, which results in having for TSD2 properties which differ from those of TSD1. 

In the present case such an argument does not hold, since the potential well is very deep (see Refs.\cite{Rad017,Rad018}) so that the potential barrier prevents the TSD2 states to share the secondary minimum through a tunneling effect. Concluding, the  TSD1 and TSD2 bands are signature partners.

We may conclude that the present formalism provides an alternative interpretation for the excited wobbling bands. The corresponding numerical results are in good agreement with the experimental data. The agreement quality is close and better than that yielded by the previous semi-classical approach where the bands TSD2, TSD3 and TSD4 have  one,  two and three phonon character, respectively.

{\bf Acknowledgment.} This work was supported by the Romanian Ministry of Research and Innovation through the project PN19060101/2019

\renewcommand{\theequation}{A.\arabic{equation}}
\setcounter{equation}{0}
\label{sec:level6}

\subsection{Appendix A}
From the equations of motion of the classical coordinates, one readily finds that the function ${\cal H}$ is a constant of motion, i.e. $\stackrel{\bullet}{\cal H}=0$. This equation defines a surface,  called as equi-energy surface, ${\cal H}=const.$ This result appears to be a consequence of the fact that the equations of motion emerge from a variational principle. Also, one notes that the stationary coordinates, having vanishing time derivatives, are stationary points for the equi-energy surface. There are several stationary points, among which some are minima, as suggested by the sign of the associated Hessian. For example, one minimum is achieved in the point.
$(r,\varphi;t, \psi)=(I,0;j,0)$

Expanding the classical energy function around the minimum point and denoting the deviations from the minimum by prime letters, one obtains:
\begin{eqnarray}
&&{\cal H}={\cal H}_{min}+\frac{1}{I}\left[(2I-1)(A_3-A_1)+2jA_1\right]\frac{r^{\prime 2}}{2}\nonumber\\
        &+&I\left[(2I-1)(A_2-A_1)+2jA_1\right]\frac{\varphi^{\prime 2}}{2}\nonumber\\
        &+&\frac{1}{j}\left[(2j-1)(A_3-A_1)+2IA_1\right.\nonumber\\
        &+&\left.V\frac{2j-1}{j(j+1)}2\sqrt{3}\sin(\gamma+\frac{\pi}{3})\right]\frac{t^{\prime 2}}{2}\\
        &+&j\left[(2j-1)(A_2-A_1)+2IA_1\right.\nonumber\\
        &+&\left.V\frac{2j-1}{j(j+1)}2\sqrt{3}\sin\gamma \right]\frac{\psi^{\prime 2}}{2}
        -2A_3r^{\prime}t^{\prime}-2IjA_2\varphi^{\prime}\psi^{\prime}.\nonumber
\end{eqnarray}
Neglecting for the moment the coupling terms, one obtains that the classical energy is the sum of two independent oscillators whose frequencies are:
{\scriptsize
\begin{eqnarray}
&&\omega_1=\left[\left((2I-1)(A_3-A_1)+2jA_1\right)\left((2I-1)(A_2-A_1)+2jA_1\right)\right]^{1/2},\nonumber\\
&&\omega_2=\left[(2j-1)(A_3-A_1)+2IA_1+V\frac{2j-1}{j(j+1)}\sqrt{3}\left(\sqrt{3}\cos\gamma+\sin\gamma\right)\right]^{1/2}\nonumber\\
&&\times\left[(2j-1)(A_2-A_1)+2IA_1+V\frac{2j-1}{j(j+1)}2\sqrt{3}\sin\gamma \right]^{1/2}.
\end{eqnarray}}
In order to have real solutions for the two frequencies, the MoI parameters and the single particle potential strength V must fulfill some restrictions:
\begin{eqnarray}
&&S_{Ij}A_1<A_2<A_3,\; {\rm or}\; \;S_{Ij}A_1<A_3<A_2,\nonumber\\
&&A_3>T_{Ij}A_1-\frac{G_1V}{2j-1},\;\;A_2>T_{Ij}A_1-\frac{G_2V}{2j-1},\hskip0.4cm
\label{ineqTij}
\end{eqnarray}
where the following notations have been used:
\begin{eqnarray}
&&S_{Ij}=\frac{2I-1-2j}{2I-1},\;\;T_{Ij}=\frac{2j-1-2I}{2j-1},\label{STG1G2}\\
&&G_1=\frac{2j-1}{j(j+1)}2\sqrt{3}\sin(\gamma+\frac{\pi}{3}),\;\;G_2=\frac{2j-1}{j(j+1)}2\sqrt{3}\sin\gamma .\nonumber
\end{eqnarray}
Note that the inequalities of the second line from (\ref{ineqTij}) are always satisfied and consequently $\omega_2$ is real, irrespective of the positive value of V. 
To treat the coupling term involved in the energy function it is useful to  quantize the phase space coordinates:
\begin{eqnarray}
&&\varphi\to\hat{q};\;\;r\to \hat{p};\;\; [\hat{q},\hat{p}]=i,\nonumber\\
&&\psi\to\hat{q}_1;\;\;t\to \hat{p}_1;\;\; [\hat{q}_1,\hat{p}_1]=i.
\end{eqnarray}
We associate to the two oscillations, defined above, the creation/annihilation operators:
\begin{eqnarray}
&&\hat{q}=\frac{1}{\sqrt{2}k}(a^{\dagger}+a),\;\;\hat{p}=\frac{ik}{\sqrt{2}}(a^{\dagger}-a),\nonumber\\
&&\hat{q}_1=\frac{1}{\sqrt{2}k'}(b^{\dagger}+b),\;\;\hat{p}_1=\frac{ik'}{\sqrt{2}}(b^{\dagger}-b).
\end{eqnarray}
The transformation relating the coordinates and the conjugate momenta with the operators $a^{\dagger}, a$ and $b^{\dagger}, b$ is canonical irrespective of the constants $k$ and $k'$. 
These were fixed such that the quantized form of the two oscillators Hamiltonian does not comprise cross terms like $a^{\dagger 2}+a^2$ and $b^{\dagger 2}+b^2$. The result is:
{\scriptsize
\begin{eqnarray}
&&k=\left[\frac{(2I-1)(A_2-A_1)+2jA_1}{(2I-1)(A_3-A_1)+2jA_1}I^2\right]^{1/4},\\
&&k'=\left[\left((2j-1)(A_2-A_1)+2IA_1+V\frac{2j-1}{j(j+1)}2\sqrt{3}\sin\gamma\right)j^2\right]^{1/4}\nonumber\\
&&\times\left[(2j-1)(A_3-A_1)+2IA_1+V\frac{2j-1}{j(j+1)}2\sqrt{3}\sin(\gamma+\frac{\pi}{3})\right]^{-1/4}.\nonumber
\end{eqnarray}}
In the new representation, the quantized Hamilton operator looks like:
\begin{eqnarray}
\hat{H}&=&{\cal H}_{min}+\omega_1(a^{\dagger}a+\frac{1}{2})+\omega_2(b^{\dagger}b+\frac{1}{2})\nonumber\\
       &+&A_3kk'(a^{\dagger}b^{\dagger}+ba-a^{\dagger}b-b^{\dagger}a)\nonumber\\
       &-&IjA_2\frac{1}{kk'}(a^{\dagger}b^{\dagger}+ba+a^{\dagger}b+b^{\dagger}a).
\end{eqnarray}
The off-diagonal terms will be treated by the equation of motion method. Thus, we have:
\begin{eqnarray}
&&[\hat{H},a^{\dagger}]=\omega_1a^{\dagger}+A_3kk'(b-b^{\dagger})-IjA_2\frac{1}{kk'}(b+b^{\dagger}),\nonumber\\
&&[\hat{H},b^{\dagger}]=\omega_2b^{\dagger}+A_3kk'(a-a^{\dagger})-IjA_2\frac{1}{kk'}(a+a^{\dagger}),\nonumber\\
&&[\hat{H},a]=-\omega_1a-A_3kk'(b^{\dagger}-b)+IjA_2\frac{1}{kk'}(b^{\dagger}+b),\nonumber\\
&&[\hat{H},b]=-\omega_2b-A_3kk'(a^{\dagger}-a)+IjA_2\frac{1}{kk'}(a^{\dagger}+a),\nonumber\\
\end{eqnarray}
This is a linear system of equations, which can be analytically solved. Indeed, one defines the phonon operator
\begin{equation}
\Gamma^{\dagger}=X_1a^{\dagger}+X_2b^{\dagger}-Y_1a-Y_2b,
\end{equation}
with the amplitudes $X_1, X_2, Y_1, Y_2$ fixed such that the following restrictions are satisfied:
\begin{equation}
\left[H,\Gamma^{\dagger}\right]=\Omega \Gamma^{\dagger},\;\;\left[\Gamma,\Gamma^{\dagger}\right]=1.
\end{equation}
The first restriction provides a homogeneous linear system of equations for the unknown amplitudes. The compatibility condition for this system leads to the equation defining the phonon energy $\Omega$,

\begin{equation}
\Omega^4+B'\Omega^2+C'=0,
\label{ecOm}
\end{equation}
with the coefficients $B'$ and $C'$ having the expressions:
\begin{eqnarray}
B'&=&-\left(\omega_1^2+\omega_2^2+8A_2A_3Ij\right),\\
C'&=&\omega_1^2\omega_2^2-4\left(A_3^2k^2k'^2+I^2j^2\frac{A_2^2}{k^2k'^2}\right)\omega_1\omega_2+16A_2^2A_3^2I^2j^2.\nonumber
\end{eqnarray}
By elementary algebraic manipulation, one finds that $B'=B,\;C'=C$. There exists an interval for variable $\gamma$ where Eq.(\ref{ecOm}) admits two positive solutions, which were used to define the level energies of the bands TSD.

\renewcommand{\theequation}{B.\arabic{equation}}
\setcounter{equation}{0}
\label{sec:level7}
\section{Appendix B}
Here we present the results concerning other two stationary points which might be also minima for the equi-energy surface.
\subsection{The case $\varphi=\frac{\pi}{2}, \psi=\frac{\pi}{2},\;r=I,\;t=j.$}
One can check that this  is a stationary point for the equations of motion. In this point the energy is:
\begin{equation}
{\cal H}_{min}=\frac{I+j}{2}(A_1+A_3)+A_2(I-j)^2.
\end{equation}
Expanding ${\cal H}$ around the point mentioned above, one obtains a Hamiltonian describing two interacting oscillators with the frequencies:
\begin{widetext}
\begin{eqnarray}
&&\omega_1=\left[(2I-1)(A_3-A_2)+2jA_2\right]^{1/2}\left[(2I-1)(A_1-A_2)+2jA_2\right]^{1/2},\\
&&\omega_2=\left[(2j-1)(A_3-A_2)+2IA_2-\frac{2j-1}{j(j+1)}V2\sqrt{3}\sin(\gamma-\frac{\pi}{3})\right]^{1/2}
\left[(2j-1)(A_1-A_2)+2IA_2-\frac{2j-1}{j(j+1)}V2\sqrt{3}\sin\gamma\right]^{1/2}.\nonumber
\label{omeg12}
\end{eqnarray}
\end{widetext}
The quantized Hamiltonian is diagonalized through a canonical transformation which results in having two independent oscillators whose frequencies $\Omega$ satisfy the equation:
\begin{equation}
\Omega^4-B\Omega^2+C=0.
\label{OBC}
\end{equation} 
with the coefficients $B$ and $C$ given by:
\begin{widetext}
\begin{eqnarray}
&-&B=\left[(2I-2)(A_3-A_2)+2jA_2\right]\left[(2I-1)(A_1-A_2)+2jA_2\right]+8A_1A_3Ij\\
&+&\left[(2j-1)(A_3-A_2)+2IA_2-\frac{2j-1}{j(j+1)}V2\sqrt{3}\sin(\gamma-\frac{\pi}{3})\right]
\left[(2j-1)(A_1-A_2)+2IA_2-\frac{2j-1}{j(j+1)}V2\sqrt{3}\sin\gamma\right],\nonumber\\
&&C=\left\{\left[(2I-1)(A_1-A_2)+2jA_2\right]\left[(2j-1)(A_1-A_2)+2IA_2-\frac{2j-1}{j(j+1)}V2\sqrt{3}\sin\gamma\right]-4IjA_1^2\right\}\nonumber\\
&\times&\left\{\left[(2I-2)(A_3-A_2)+2jA_2\right]\left[(2j-1)(A_3-A_2)+2IA_2-\frac{2j-1}{j(j+1)}V2\sqrt{3}\sin(\gamma-\frac{\pi}{3})\right]-4IjA_3^2\right\}.
\label{BsiC}
\end{eqnarray}
\end{widetext}
It is worth noting that the expressions of $B$ and $C$ can be formally obtained from those from Appendix A by changing $A_1\to A_2$, $A_2\to A_1$, and $\gamma\to-\gamma$.

Inserting the mentioned coordinates in the expression of the classical angular momentum components, we obtain:
\begin{eqnarray}
&&I^{cl}_{1}\equiv\langle\Psi_{IMj}|\hat{I}_1|\Psi_{IMj}\rangle \left|_{\stackrel{(\varphi,r)=(\pi/2,I)}{(\psi,t)=(\pi/2,j)}}\right.=0,\nonumber\\
&&I^{cl}_{2}\equiv\langle\Psi_{IMj}|\hat{I}_2|\Psi_{IMj}\rangle\left|_{\stackrel{(\varphi,r)=(\pi/2,I)}{(\psi,t)=(\pi/2,j)}}\right.=-I,\nonumber\\
&&I^{cl}_{3}\equiv\langle\Psi_{IMj}|\hat{I}_3|\Psi_{IMj}\rangle\left|_{\stackrel{(\varphi,r)=(\pi/2,I)}{(\psi,t)=(\pi/2,j)}}\right.=0,\nonumber\\
&&j^{cl}_{1}\equiv\langle\Psi_{IMj}|\hat{j}_1|\Psi_{IMj}\rangle\left|_{\stackrel{(\varphi,r)=(\pi/2,I)}{(\psi,t)=(\pi/2,j)}}\right.=0,\nonumber\\
&&j^{cl}_{2}\equiv\langle\Psi_{IMj}|\hat{j}_2|\Psi_{IMj}\rangle\left|_{\stackrel{(\varphi,r)=(\pi/2,I)}{(\psi,t)=(\pi/2,j)}}\right.=-j,\nonumber\\
&&j^{cl}_{3}\equiv\langle\Psi_{IMj}|\hat{j}_3|\Psi_{IMj}\rangle\left|_{\stackrel{(\varphi,r)=(\pi/2,I)}{(\psi,t)=(\pi/2,j)}}\right.=0.
\end{eqnarray}
The two solutions \ref{omeg12} are real provided the following restrictions are fulfilled:
\begin{eqnarray}
&&A_1>S_{Ij}A_2,\;A_3>A_2,\;A_1>T_{Ij}A_2+\frac{2\sqrt{3}V}{j(j+1)}\sin\gamma,\nonumber\\
&&A_3>T_{Ij}A_2+\frac{2\sqrt{3}V}{j(j+1)}\sin(\gamma-\frac{\pi}{3}).
\label{A123}
\end{eqnarray}
If $C>0$, Eq. (\ref{OBC}) admits  real solutions, two positive and two negative. If C=0, two solutions are vanishing, one is positive, and one negative. 
In this case Eq.(\ref{A123}) defines several separatrices bordering the nuclear phases. These are obtained by modifying the separatrices defined in Appendix A, by changing $x\to y,\;y\to x$, and
$\gamma \to -\gamma$.   
If $C<0$, two solutions are imaginary, one solution is  positive, and one negative. 
\vspace*{0.5cm}
\subsection{The case $\varphi = \frac{\pi}{2},\;\psi=0,\; r= I,\; t=j$.}

For this case we followed the same algorithm  as in the previous subsection and obtained the results: The energy corresponding to the point, mentioned above, is: 
\begin{eqnarray}
&&{\cal H}_{0}=A_2I^2+A_1j^2\\  
&+&\frac{I}{2}(A_1+A_3)+\frac{j}{2}(A_2+A_3)-\frac{(2j-1)V}{j+1}\cos(\gamma-\frac{\pi}{3}).\nonumber
\end{eqnarray}
Expanding the classical energy function around the point mentioned above, it results a Hamiltonian of two coupled oscillators with the frequencies:
\begin{eqnarray}
&&\omega_1=(2I-1)\left[(A_3-A_2)(A_1-A_2)\right]^{1/2},\nonumber\\
&&\omega_2=(2j-1)\left[(A_3-A_1)+\frac{2\sqrt{3}V}{j(j+1)}\sin(\gamma+\frac{\pi}{3})\right]^{1/2}\nonumber\\
&\times&\left[(A_2-A_1)+\frac{2\sqrt{3}V}{j(j+1)}\sin\gamma\right]^{1/2}.
\end{eqnarray}
The coupling Hamiltonian is further diagonalized through a canonical transformation which leads to the following dispersion equation for the final
independent oscillations:
\begin{equation}
\Omega^4-B\Omega^2+C=0,
\label{OmBC}
\end{equation}
with the coefficients defined by:
\begin{widetext}
\begin{eqnarray}
&-&B=(2I-1)^2(A_3-A_2)(A_1-A_2)+(2j-1)^2\left[(A_3-A_1)+\frac{2\sqrt{3}V}{j(j+1)}\sin(\gamma+\frac{\pi}{3})\right]
\left[(A_2-A_1)+\frac{2\sqrt{3}V}{j(j+1)}\sin\gamma\right],\nonumber\\
&&C=\left\{(2I-1)(2j-1)(A_1-A_2)\left[(A_2-A_1)+\frac{2\sqrt{3}V}{j(j+1)}\sin\gamma\right]\right\}\nonumber\\
&\times&\left\{(2I-1)(2j-1)(A_3-A_2)\left[(A_3-A_1)+\frac{2\sqrt{3}V}{j(j+1)}\sin(\gamma+\frac{\pi}{3})\right]-4IjA_3^2\right\}.\nonumber\\
\end{eqnarray}
\end{widetext}
The solutions of Eq.(\ref{OmBC}) are all real if the following conditions are fulfilled:
\begin{eqnarray}
&&x>y,\;\;y<1,\;\;z>\frac{G_2}{2j-1}(x-y),\nonumber\\
&&z>\frac{4Ij}{(2I-1)G_1(1-y)}+\frac{2j-1}{G_1}(x-1).
\label{restr}
\end{eqnarray}
Replacing in the above relations the inequality sign by an equality sign, one obtains the equations defining the separatrices. These can be also obtained by equating the Hessian to zero, which results in having a vanishing value for the coefficient $C$. The components of the core and of the odd proton angular momenta are:
\begin{eqnarray}
&&I^{cl}_{1}=0,\;I^{cl}_{2}=-I,\;I^{cl}_{3}=0,\nonumber\\
&&j^{cl}_{1}=j,\;j^{cl}_{2}=0,\;j^{cl}_{3}=0.
\end{eqnarray}
Such a situation is met with the hydrodynamic model for the MoI parameters and the particle-core potential given by Eq.(2.1). In order to prove that, let us  write the particle-core potential in the form:
\begin{equation}
V_{pc}=\frac{2V}{j(j+1)}\left[a_1j_1^2+a_2j_2^2+a_3j_3^2\right],
\label{Vpc}
\end{equation}
with the notations:
\begin{equation}
a_1=-\cos(\gamma-\frac{\pi}{3}),\;a_2=-\cos(\gamma+\frac{\pi}{3}),\;a_3=\cos\gamma.  
\end{equation}
\begin{figure}[h!]
\includegraphics[width=0.25\textwidth,height=3.2cm]{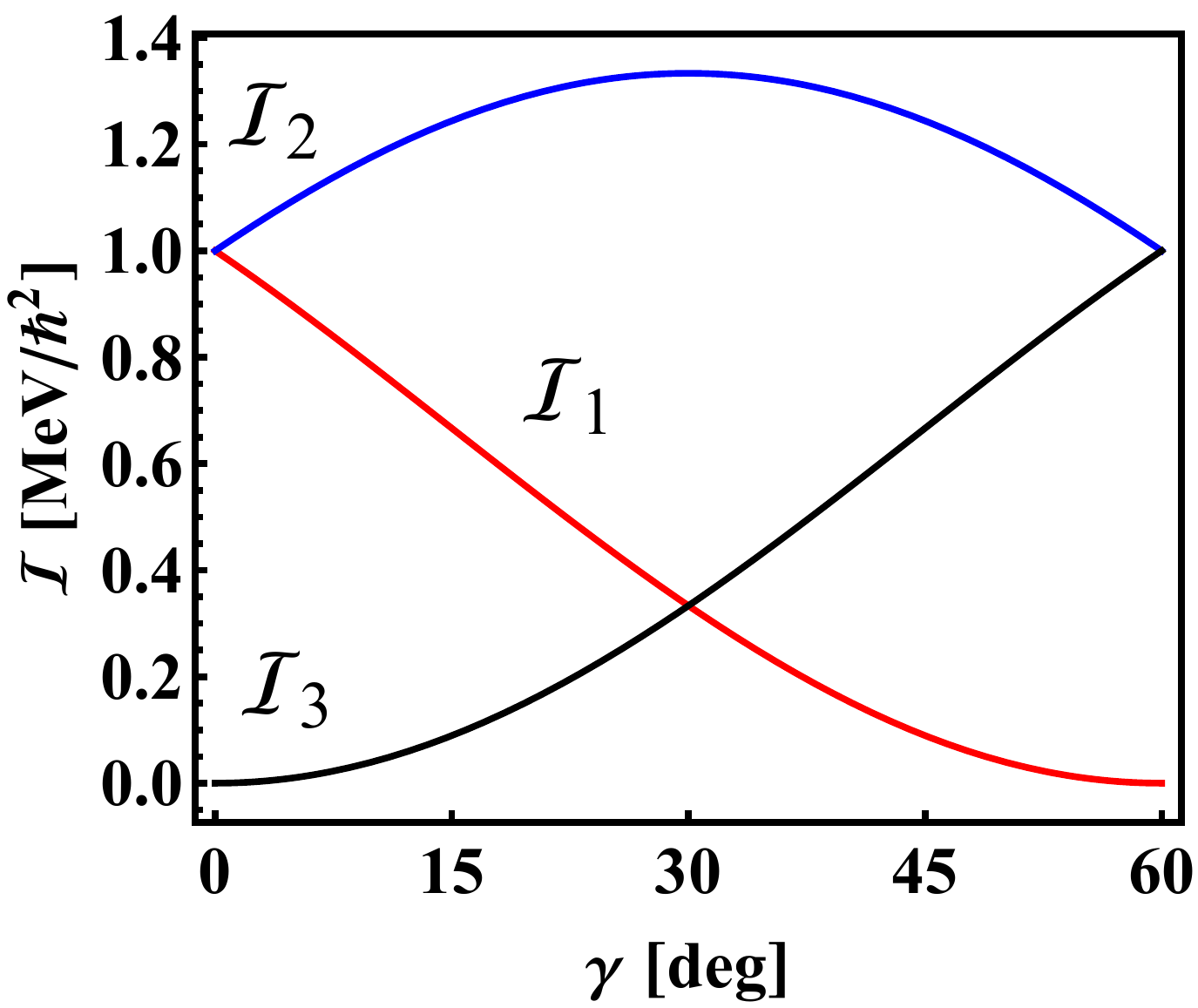}\includegraphics[width=0.25\textwidth,height=3cm]{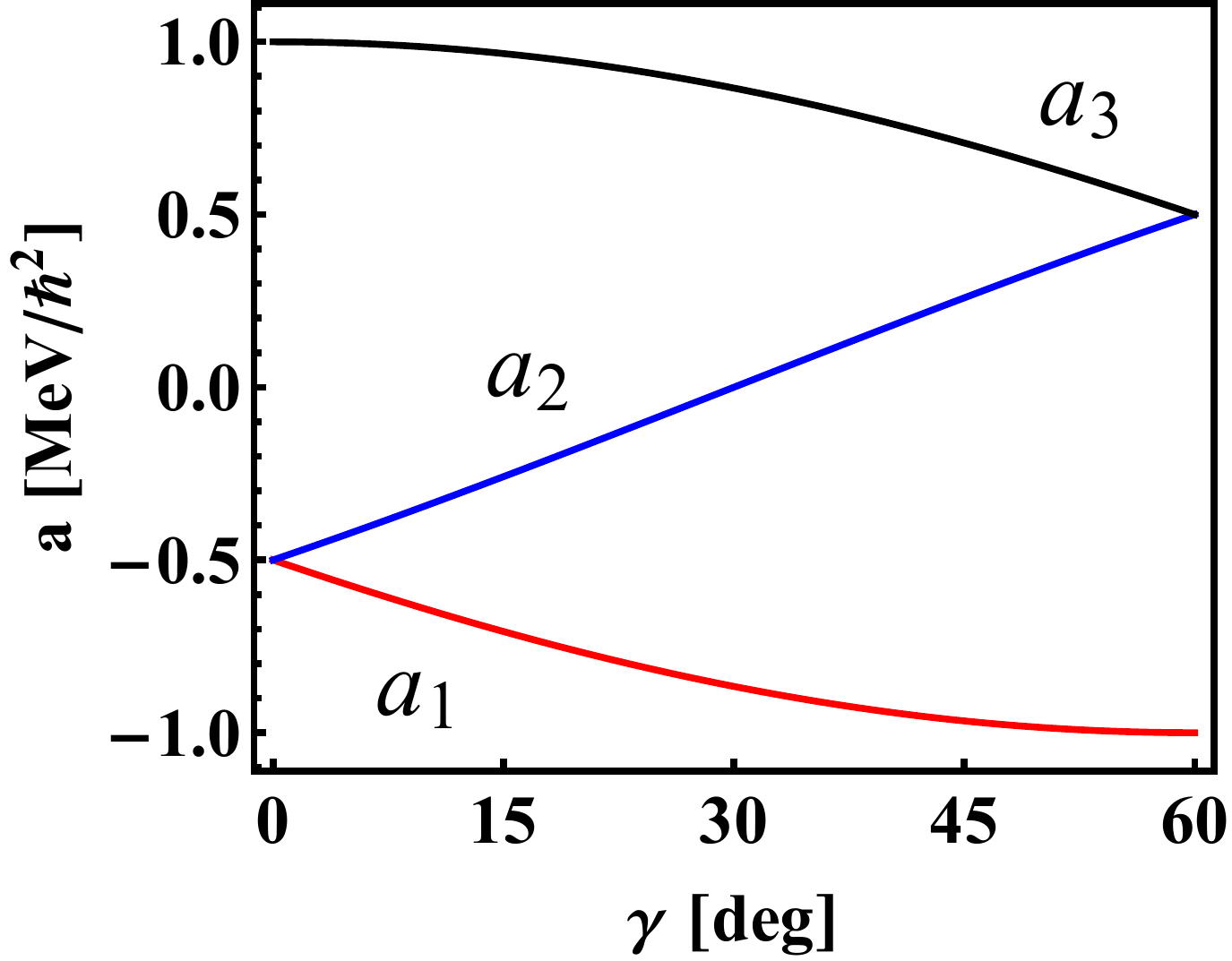}
\caption{(Color online) The hydrodynamic MoI(left panel) as well as the terms $a_i$ with i=1,2,3, involved in the single particle-core coupling potential $\ref{Vpc}$ (right panel) are plotted as function of $\gamma$, in the convention of negative gamma.}
\label{Fig.22}
\end{figure}
In Fig.22, the hydrodynamic MoI, and the particle-core attractive potential $V_{pc}$, for a  particle-like proton, are plotted as function of $\gamma$, in the interval $[0^0,60^0]$. We note that in the mentioned interval the maximal MoI is ${\cal J}_2$, while the term of $V_{pc}$ with the lowest energy is that multiplied by $a_1$. Consequently, the core is rotating around the middle axis,
 the two-axis, while the odd proton around the short axis, that is the one-axis. Thus, the situation described in the present subsection corresponds to a transversal wobbling of the even-odd system. This means that the odd proton is strongly coupled  to the core deformation. However, for a critical angular momentum the existence conditions (\ref{restr}) are no longer obeyed and the system passes from a transversal wobbling regime to a longitudinal wobbling one. This is caused by the Coriolis interaction which aligns the angular momentum of the odd proton to the principal axis to which the maximal MoI corresponds. Therefore, the transversal or a longitudinal wobbling motion is determined by which of the  interactions, with the core deformation or the Coriolis one, prevails. Some additional comments are necessary. The point considered is not really a stationary point, since there the time derivatives are:
\begin{eqnarray}
&&\stackrel{\bullet}{\varphi}=0,\;\stackrel{\bullet}{\psi}=0,\nonumber\\
&&\stackrel{\bullet}{r}=-2IjA_1,\;\stackrel{\bullet}{t}=2IjA_2.
\end{eqnarray}
Therefore, the considered point is not the ground state for the whole Hamiltonian, but it is for a part of it consisting of the sum of the two oscillators of energies $\omega_1$ and $\omega_2$. In
order to find the ground state of the whole Hamiltonian, the conjugate coordinates corresponding to the $\omega_1$ and $\omega_2$ are to be  mixed up through a canonical transformation of an RPA   (random phase approximation) type. If such a transformation exists, one obtains two wobbling phonons corresponding to  the frequencies $\Omega_1$ and $\Omega_2$, respectively. The newly determined representation defines the true ground state which, however, might become unstable due to the Coriolis interaction. Such an instability reclaims a redefining of a new stable ground state which  
is associated to the longitudinal wobbling motion. Note that for a rigid coupling, the coordinates $t, \psi$ disappear, and the stationary point $(r=I,\varphi=\alpha)$, with $\alpha$ defined by
\begin{equation}
\cos\alpha=\frac{2j}{2I-1}\frac{A_1}{A_1-A_2},
\end{equation}
is a minimum point for the energy function which results that the instability of the ground state is avoided. Note that $\alpha\ne\frac{\pi}{2}$, and only for $I\gg j$ one may approximate 
$\alpha\approx \frac{\pi}{2}$. We may conclude that even for a rigid coupling of the odd proton along the short axis, the transversal wobbling mode may show up only in the limit of a very large I. Moreover, the rigid coupling infers the fact that the Coriolis coupling terms determined by the core components corresponding to the middle and long axes is ignored. In a similar manner, in the present formalism
the transversal wobbling appears with the price of ignoring important terms which leads to an energy function describing two independent oscillators. In this picture, the collective wobbling mode is determined exclusively by the core. Switching on the ignored interaction new wobbling frequencies are obtained and the transversal picture is gradually blurred. Concluding, the transversal wobbling situation appears to be specific to ideal restrictions which abusively neglect some of the Coriolis coupling terms. Actually, the rigid coupling means that the initial Hamiltonian is truncated to a sum of two terms one describing the triaxial rotor-core and one linear in $I_1$ which cranks the system to rotate around the one-axis. When this happens, the longitudinal wobbling
regime is achieved. 

 It is conspicuous that the scenario presented here points out that  the  picture where the transversal wobbling  shows up corresponds to ideal restrictions \cite{Frau}, while within the Holstein-Primakoff description, the minimum for energy surface reflecting a transversal wobbling regime does not exist\cite{Tan017} if one keeps all energy terms. Therefore, there is no contradiction between the two formalisms \cite{Frau018,Tana018}, since they deal with different Hamiltonians.

\end{document}